\documentclass[]{pasj02}
\Received{}
\Accepted{}

\usepackage[switch,mathlines]{lineno}

\begin{document}

\title{Suzaku observations of outskirts of nearby clusters and groups: I.  electron density and gas fraction to the virial radius}

 \author{%
 Kyoko \textsc{Matsushita}\altaffilmark{1}\orcid{0000-0003-2907-0902}\email{matusita@rs.tus.ac.jp},
 Marie \textsc{Kondo}\altaffilmark{2}\orcid{0009-0005-5685-1562},
 Kosuke \textsc{Sato}\altaffilmark{3}\orcid{0000-0001-5774-1633},
Toru \textsc{Sasaki}\altaffilmark{1},
 Nobuhiro \textsc{Okabe}\altaffilmark{4,5}\orcid{0000-0003-2898-0728},
Kotaro \textsc{Fukushima}\altaffilmark{6}\orcid{0000-0001-8055-7113}
}
\altaffiltext{1}{Department of Physics, Tokyo University of Science, 1-3 Kagurazaka, Shinjuku-ku, Tokyo 162-8601 Japan}

\altaffiltext{2}{Graduate School of Science and Engineering, Saitama University, 255 Shimo-Okubo, Sakura-ku, Saitama, Saitama 338-8570, Japan}
\altaffiltext{3}{Department of Astrophysics and Atmospheric Sciences, Kyoto Sangyo University, Kyoto 603-8555, Japan}
\altaffiltext{4}{Department of Physical Science, Hiroshima University, 1-3-1 Kagamiyama, Higashi-Hiroshima, Hiroshima 739-8526, Japan}
\altaffiltext{5}{Hirohima Astrophysical Science Center, Hiroshima University, Higashi-Hiroshima, Kagamiyama 1-3-1, 739-8526, Japan}
\altaffiltext{6}{Institute of Space and Astronautical Science, Japan Aerospace Exploration Agency, 3-1-1 Yoshinodai,
Chuo-ku, Sagamihara, Kanagawa 229-8510, Japan}

\maketitle

\begin{abstract}
We present an analysis of Suzaku observations of 14 nearby galaxy clusters and groups
 ($z<0.06$),  extending radial coverage out to the virial radius ($\sim  r_{200}$).
The sample spans a wide mass range, from $M_{500} \sim 2\times 10^{13}$ to $7\times 10^{14} M_\odot$, and includes well-studied systems such as Coma, Perseus, and Virgo.
We carefully modeled all background components,  including the soft X-ray foregrounds (the Local Hot Bubble, Milky Way Halo, and super-virial temperature components),  the cosmic X-ray background, and the non-X-ray background, and assessed their effects on the derived properties of the intracluster medium (ICM). 
We constructed radial profiles of emission measure, electron density, and temperature. 
 Temperatures decrease smoothly with radius, typically dropping to about one-third to half of their peak values near $r_{200}$.
For relaxed clusters, the emission measure profiles outside the core regions are well described by a $\beta$-model with $\beta \sim 0.6$--0.7, while groups show slightly flatter slopes of $\beta \sim 0.4$--0.65.
 Beyond $r_{2500}$, electron density profiles follow a power-law decline with a slope close to 2.
At $r_{500}$ and $r_{200}$, the electron density and the gas mass fraction show a tight correlation with the system mass, except for three clusters with bright subclusters.
In massive clusters, the gas fraction increases with radius and approaches the cosmic baryon fraction near $r_{200}$. In contrast,  lower-mass systems exhibit gas fractions of around 0.1 at $r_{200}$.
The observed mass dependence of gas fractions suggests that feedback and related processes play an increasingly important role toward the group scale, shaping the connection between baryons and dark matter halos.

\end{abstract}


\section{Introduction}
\label{sec:intro}

Clusters of galaxies are the largest gravitationally bound structures in the Universe.
According to the cold dark matter (CDM) paradigm,  clusters form through hierarchical merging and accretion of
smaller systems driven by the gravitational force of dark matter. 
The infalling gas is shock-heated to the virial temperature and compressed adiabatically, forming the intracluster medium (ICM), 
which fills the cluster potential wells.
The CDM framework predicts that massive halos form in a self-similar manner primarily governed by gravity. 
As a result, the ICM's thermodynamic properties, such as density, temperature, pressure, and entropy,
are expected to scale with cluster mass and redshift \citep{Voit2005}.
Since clusters form through the collapse of dark matter and baryons, their total baryon fraction (ICM plus stars) is expected to be close to the cosmic mean \citep{White1993}.
Deviations from these expectations are interpreted as evidence for additional physical processes, such as heating or cooling,  beyond gravitational collapse \citep{Ponman1999}.

High-resolution X-ray observatories such as Chandra, XMM-Newton, and the low-background satellite Suzaku, along with measurements of the Sunyaev-Zel'dovich (SZ) effect, have enabled detailed studies of the baryonic content of galaxy clusters out to and beyond $r_{500}$, the radius enclosing an overdensity of 500 times the critical density.
Out to $r_{500}$, the gas mass fraction generally increases with radius and reaches $\sim$0.1 for massive clusters, although this remains lower than the cosmic baryon fraction \citep{Vikhlinin2009}.
 Beyond $r_{500}$,  Suzaku observations occasionally reported gas mass fractions exceeding the cosmic mean, suggesting a possible overestimation of the gas density due to unresolved gas clumping  (e.g., \cite{Simionescu2011}, \cite{Simionescu2017}, \cite{Urban2014}). 
 Alternatively, nonthermal pressure support from bulk motions or turbulence may cause deviations from hydrostatic equilibrium,
  as gas fractions derived relative to weak-lensing masses are more consistent with the cosmic mean \citep{Kawaharada2010, Ichikawa2013}.
To mitigate such biases, \citet{Eckert2019, Eckert2022} combined XMM-Newton and Planck data in the X-COP project,  excluding potential clump candidates using the much better angular resolution of XMM. Their results showed that in most massive clusters, the baryon fraction profiles increase smoothly with radius and remain consistent with theoretical expectations. Possible deviations from hydrostatic equilibrium are also discussed in \citet{Eckert2019, Eckert2022}.

Galaxy groups, which reside in shallower gravitational potentials, exhibit systematically lower baryon fractions \citep{Sun2009, Sun2012}. These trends are likely caused by nongravitational processes such as AGN feedback, which efficiently prevent the accretion of baryons in low-mass systems \citep{Ponman1999, Sun2009}. 
With Suzaku observations, the enclosed gas mass fractions of the galaxy groups do not exceed the cosmic baryon fraction up to the virial radius
\citep{UGC, Antlia, MKW4}.

Accurate background modeling is crucial in studies of the faint outskirts of clusters.
Suzaku observations have revealed excess emission in the 0.7--1 keV band in spectra of some regions without clusters and bright X-ray sources,
in addition to the standard soft X-ray background components such as the Local Hot Bubble (LHB) and Milky Way Halo (MWH). 
This excess is often modeled with a 0.6--1.2 keV plasma,  exceeding the virial temperature of the Milky Way halo, and has been detected in at least one third of the sky
\citep{Yoshino2009, Henley2013, Nakashima2018, Gupta2021, Gupta2023, Ueda2022, Sugiyama2023}.
 HaloSat observations also confirmed the presence of a similar super-virial temperature component over 85\% of the sky at $|b| > 30^\circ$ \citep{Bluem2022}.
In particular, \citet{Sugiyama2023} analyzed this supervirial temperature component (hereafter referred to as the “HG component”) and showed that its emission measure can vary by at least an order of magnitude,  with stronger emission typically observed at lower Galactic latitudes,
suggesting that at least part of this component may be related to stellar feedback.
This component can bias measurements of the ICM temperature and density in cluster outskirts.
Some Suzaku observations of cluster outskirts beyond the virial radius also included this additional component \citep{Ichikawa2013, Urban2014, Urban2017}, but a systematic and uniform analysis across datasets has not yet been conducted.

This paper presents a systematic analysis of Suzaku observations for 14 nearby clusters and groups, 
spanning a wide range of system masses and extending out to their virial radii. 
We carefully model the background, including the supervirial temperature component, the level of the cosmic X-ray background (CXB),  and the non-X-ray background (NXB), 
and assess their effects on derived ICM parameters. 
The present study focuses on the data analysis and density of the ICM and baryon content, while complementary results on Fe metallicity and Fe mass content, scaling relations of temperature, pressure, and entropy, and comparisons with stellar distributions will be presented in forthcoming companion papers (Matsushita et al., in preparation; Kondo et al., in preparation).
This paper is organized as follows.
In Section 2, we describe observations and data reduction.
We present spectral fitting results in Section 3.
We discuss the results in Section 4.
We adopt the solar abundance table of the proto-solar values by \citet{Lodders2009}.
We assume a $\Lambda$ CDM cosmology with $\Omega_{\rm m}=0.3$, $\Omega_\Lambda=0.7$, and  $H_0 = 70  \rm{km s^{-1} Mpc^{-1}}$.
The dimensionless Hubble parameter is defined as 
$h(z)=H(z)/H_0=\sqrt{E(z)^2}=\sqrt{\Omega_m(1+z)^3+\Omega_\Lambda}$, where $H(z)$ is the Hubble constant at redshift $z$.
Errors are reported at the 68\% confidence level unless otherwise stated.

\section{Observations and Data Reductions}
\label{sec:obs}

Using {\it Suzaku} archival data, we selected a sample of nearby ($z<0.06$) galaxy clusters and groups observed out to $\sim 2~ r_{500}$, where $r_{500}$ are obtained from our {Suzaku} data measurements (see Sec \ref{sec:r500}).
Due to short exposure times, we excluded the NGC 5129 and NGC 3402 groups, and
removed IC 1633 due to sparse sky coverage.
The final sample comprises 14 systems, spanning a wide range of halo masses, from poor groups to rich clusters.
Table \ref{tb:sample} lists the basic properties of these systems, and the Suzaku
 observation log is shown in Table \ref{tb:suzakuobslog} in Appendix \ref{sec:obslog}.
Figures \ref{fig:image} and \ref{fig:imageVirgo} show the XIS mosaic images.
We classify the Coma, Hydra-A, and Virgo clusters as "merging clusters" because they host very bright subclusters in the southwest, southeast, and southern directions, respectively, and others as "relaxed systems".
We also analyzed data from eight Lockman Hole observations with Suzaku to study the X-ray background, especially the CXB level. Their observation log is shown in Table \ref{tb:LHobs} in Appendix \ref{sec:obslog}.

\begin{table*}
	\caption{Basic properties of sample galaxy clusters and groups}
	\label{tb:sample}
	\begin{center}
		\begin{tabular}{llllllll}
			\hline
			Name & $z$\footnotemark[$*$] & $N_{H}$\footnotemark[$\dagger$] &  1$\arcmin$& (R.A., decl.)\footnotemark[$\ddagger$]   &   $F_{\rm th}$\footnotemark[\S]  \\ 
			&  &  $10^{20}$ cm$^2$ & kpc & J2000 &\\ \hline
			Coma          & 0.0231  &  0.85 & 28.0  & $\timeform{12h59m44s},\timeform{+27D56'50"}$ & 10   \\
			Perseus      & 0.0179  & 13.6  & 21.8  & $\timeform{03h19m48s},\timeform{+41D30'42"}$  & 10   \\
			Abell 2199    & 0.0305  &  0.89 & 36.7  & $\timeform{16h28m38s},\timeform{+39D33'06"}$  &  5  \\
			Abell 133    & 0.0566  &  1.57 & 65.9  & $\timeform{01h02m42s},\timeform{-21D52'25"}$  &  5    \\
			AWM7          & 0.0172  &  8.69 & 21.0  & $\timeform{02h54m32s},\timeform{+41D35'15"}$ &  5   \\
			Hydra A       & 0.0548  &  4.68 & 64.0  & $\timeform{09h18m06s},\timeform{-12D05'44"}$ &  5  \\
			UGC 03957    & 0.0341  &  4.27 & 40.8  & $\timeform{07h40m58s},\timeform{+55D25'38"}$ &  5  \\
			ESO 0306-017 & 0.0358  &  2.99 & 42.7  & $\timeform{05h40m07s},\timeform{-40D50'12"}$  &  5 \\
			Abell 262     & 0.0165  &  5.38 & 21.2  & $\timeform{01h52m46s},\timeform{+36D09'33"}$ &  5   \\
			Virgo        & 0.00436 &  1.96 &  4.26 & $\timeform{12h30m47s},\timeform{+12D23'28"}$ &  10 \\
			Antlia       & 0.00933 &  6.65 & 11.5  & $\timeform{10h30m04s},\timeform{-35D19'24"}$ &   5   \\
			MKW 4       & 0.0200  &  1.76 & 24.3  & $\timeform{12h04m27s},\timeform{+01D53'45"}$ & 5  \\
			NGC 1550     & 0.0124  & 10.2  & 15.2  & $\timeform{04h19m38s},\timeform{+02D24'36"}$ &  5   \\
			NGC 741     & 0.0185  &  4.37 & 22.6  & $\timeform{01h56m21s},\timeform{+05D37'44"}$ &  5  \\
			\hline
			\multicolumn{6}{@{}l@{}}{\hbox to 0pt{\parbox{150mm}{\footnotesize
						\par\noindent
						\footnotemark[$*$]     From NASA Extragalactic Database.   
						\par\noindent
						\footnotemark[$\dagger$]      The Galactic hydrogen column density \citep{Kalberla2005}.    
						\par\noindent
						\footnotemark[$\ddagger$]   The center of the annular region for the spectral analysis, the positions of the BCG galaxy of each cluster and group, 
						except for the Coma cluster and MKW 4 group, where the X-ray peaks are adopted.
						\par\noindent
						\footnotemark[$\S$] The threshold flux for excluding point sources.
						 The unit is 10$^{-14}$ erg s$^{-1}$ cm$^{-2}$ in 2.0-10.0 keV energy ranges with a power-law model of photon index $\Gamma=1.7$.
						\par\noindent
					}    \hss}}
		\end{tabular}
	\end{center}
\end{table*}

We analyzed the XIS data of the sample groups and clusters.
The XIS 0, and 3 are the front-illuminated (FI) sensors, while XIS 1 is the back-illuminated (BI) sensor,
 all operated in the normal clocking mode. 
We combined the data from the $3\times3$ and $5\times5$ editing modes for the spectral analysis.
We applied the standard data screening criteria\footnote{\ texttt {http://heasarc.nasa.gov/docs/Suzaku/processing/criteria\_xis.html}}.
In addition, we excluded time intervals with the geomagnetic cut-off rigidity (COR) below 6 GV, 
and when the Earth rim elevation angle, ELEVATION $<$10$^\circ$.
The analysis was performed using HEAsoft version 6.16.
Redistribution matrix files (RMFs) were generated by "xisrmfgen" ftools task. 
We generated ancillary response files (ARFs) radius using "xissimarfgen"  ftools task \citep{Ishisaki2007}, assuming a uniform emission within a 20$'$  radius.
The spectral fittings used XSPEC 12.13.1.

Using the ``wavdetect'' tool in CIAO\footnote{http://cxc.harvard.edu/ciao/}, 
we identified point-like sources in XIS images within the 0.5--2.0 keV and 2.0--5.0 keV ranges. 
We then fitted the spectra of these point source candidates with detection significance greater than $3 \sigma$ using a power-law model with a fixed photon index of $\Gamma=$1.7.  
Because detection sensitivity varies with exposure time, we applied a flux threshold ($F_{\rm th}$) specific to each object, as listed in Table \ref{tb:sample}.
As indicated by white circles in Figures \ref{fig:image}, and \ref{fig:imageVirgo},
we excluded circular regions around the point-like sources with a radius of 1.5$'$. 
Larger exclusion radii were used for brighter sources, as shown in the same figures.
Additionally, we excluded extended sources, including subclusters and background clusters.
In the case of the Coma cluster, we also excluded two X-ray sources associated with subhalos identified by weak-lensing observations \citep{Sasaki2015}.

To mitigate the impact of background variability, including contamination from solar wind charge exchange (SWCX), we extracted XIS light curves in the 0.5--2.0 keV band with 512 s time bins. Time intervals with count rates that exceeded the mean by more than 3$\sigma$ were excluded.
In some observations of the AWM7 cluster, Abell 262, the Antlia cluster, and the NGC 1550 group, the light curves exhibit strong flare-like variations. 
During these periods, the solar wind proton flux measured by the WIND/SWE instrument\footnote{http://web.mit.edu/afs/athena/org/s/space/www/wind.html} also shows pronounced flares, indicating a likely SWCX origin. Details of the SWCX filtering procedure and treatment in spectral analysis are described in the appendix \ref{sec:SWCX}.

\begin{figure*}[htpd]
	\begin{center}
		\includegraphics[width=0.31\textwidth,angle=0]{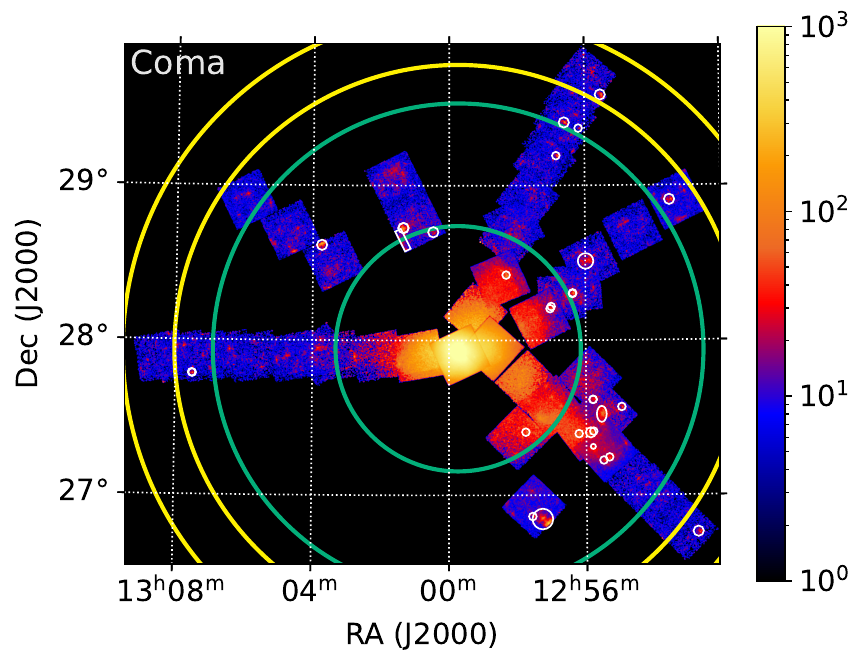}
		\includegraphics[width=0.29\textwidth,angle=0]{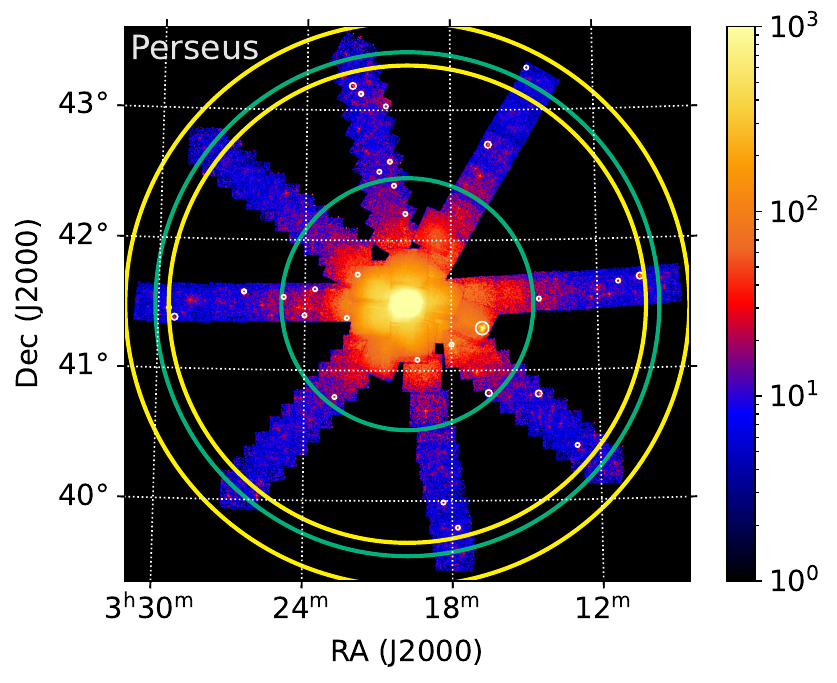}
		\includegraphics[width=0.38\textwidth,angle=0]{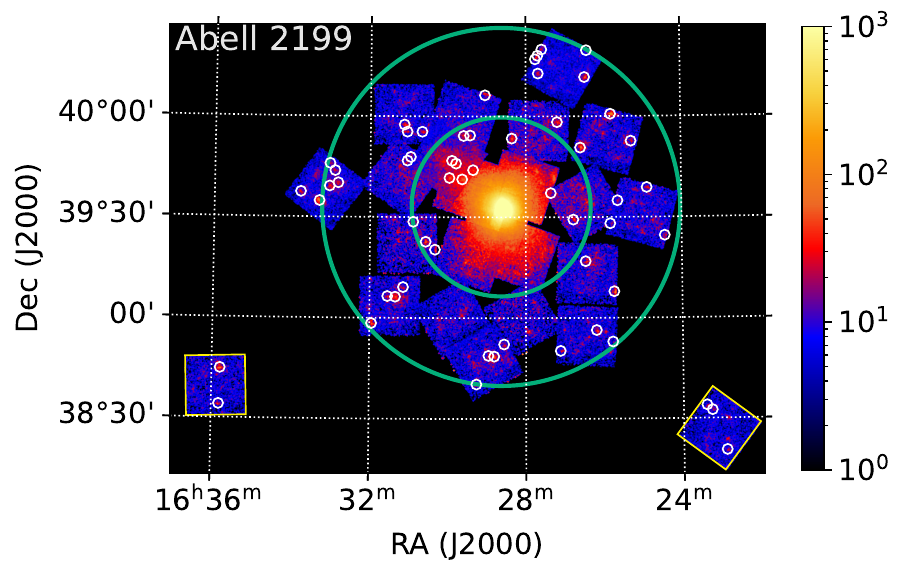}
		\includegraphics[width=0.33\textwidth,angle=0]{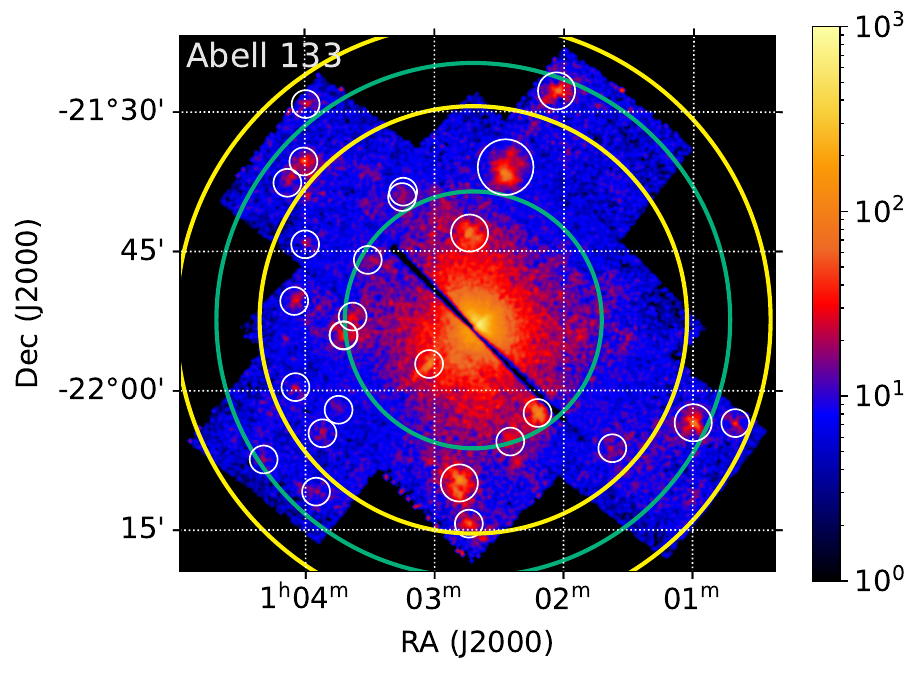}
		\includegraphics[width=0.36\textwidth,angle=0]{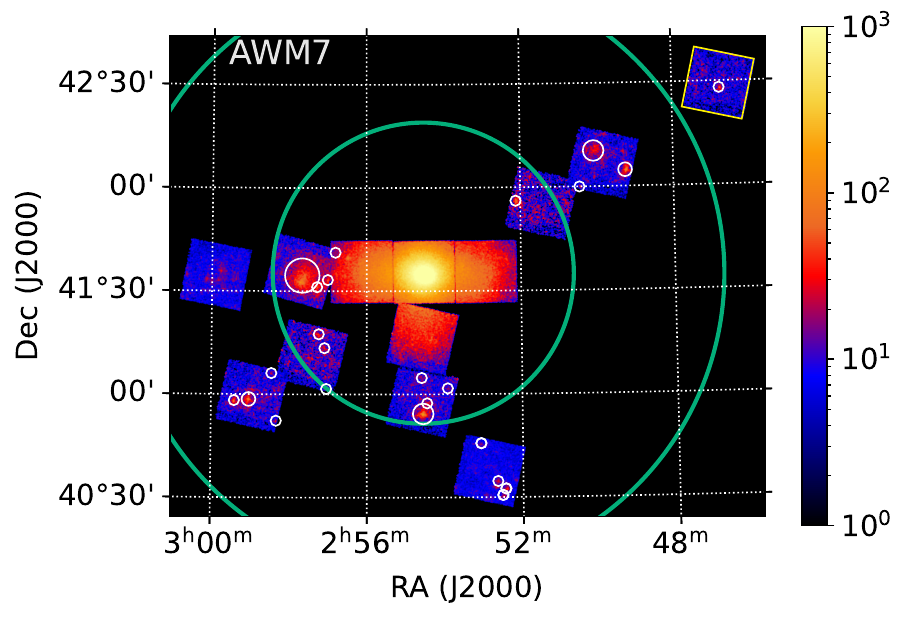}
		\includegraphics[width=0.28\textwidth,angle=0]{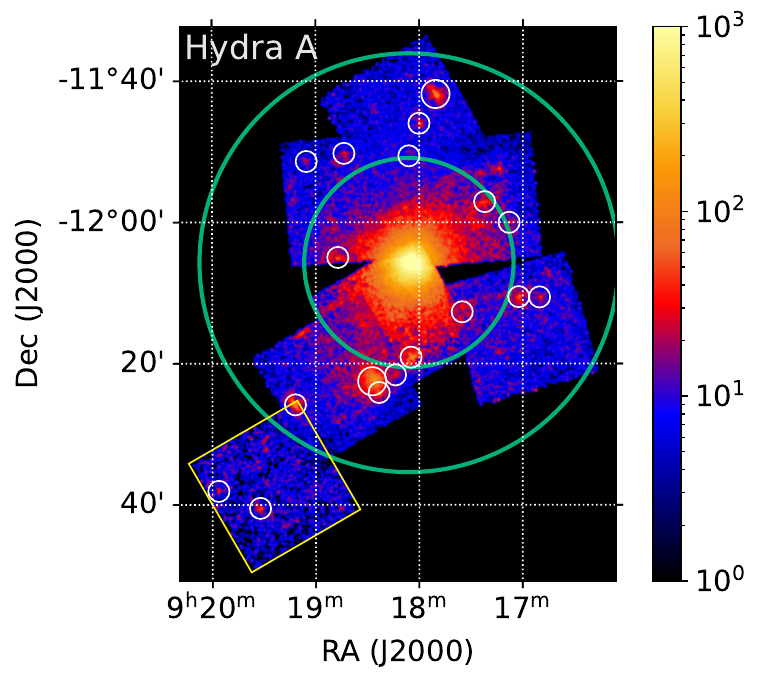}
        	\includegraphics[width=0.34\textwidth,angle=0]{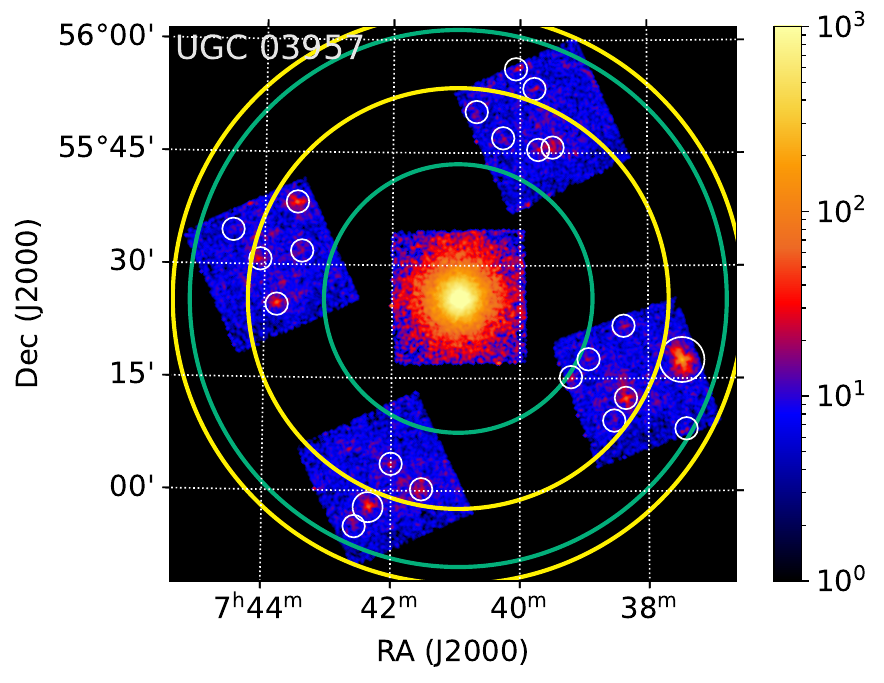}
		\includegraphics[width=0.25\textwidth,angle=0]{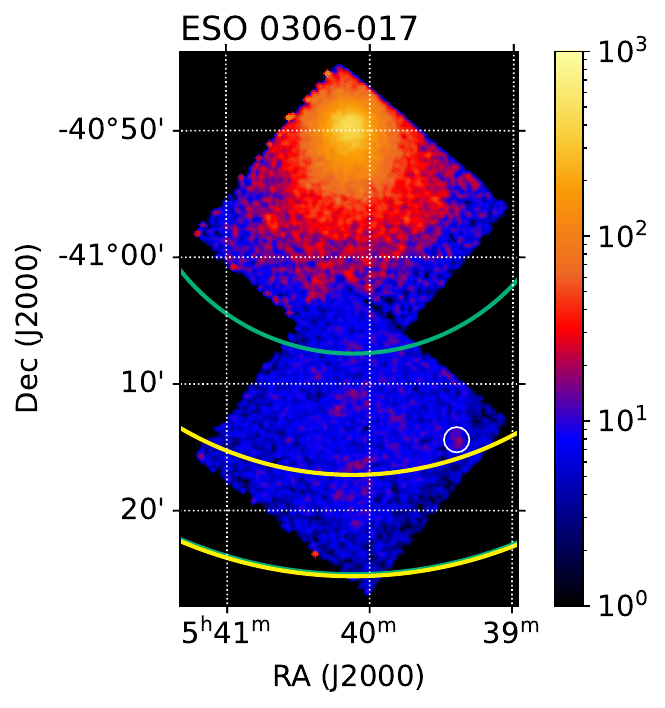}
		\includegraphics[width=0.4\textwidth,angle=0]{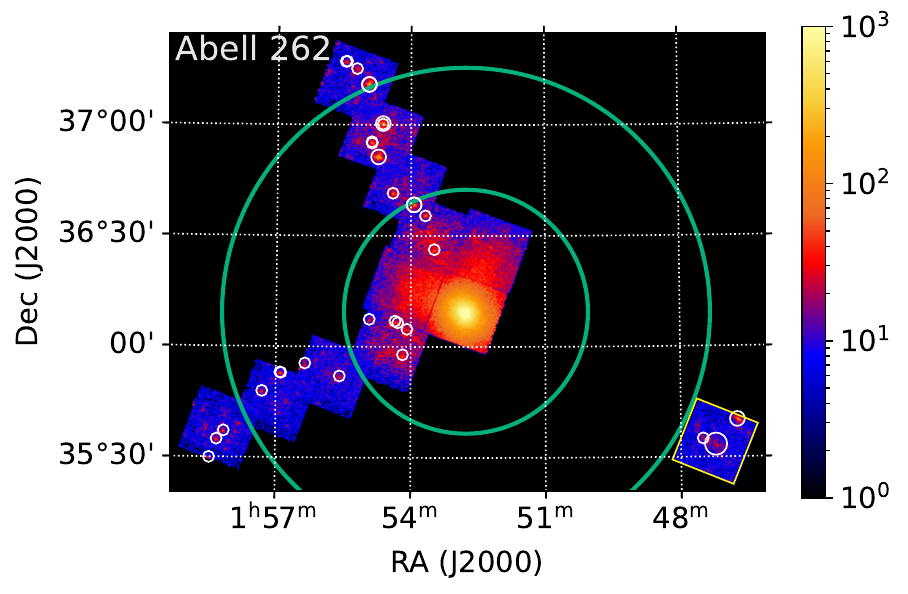}
        \includegraphics[width=0.7\textwidth,angle=0]{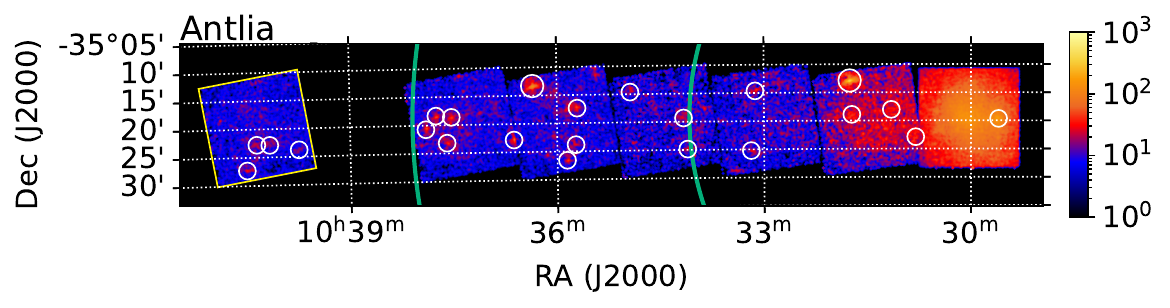}
        \includegraphics[width=0.28\textwidth,angle=0]{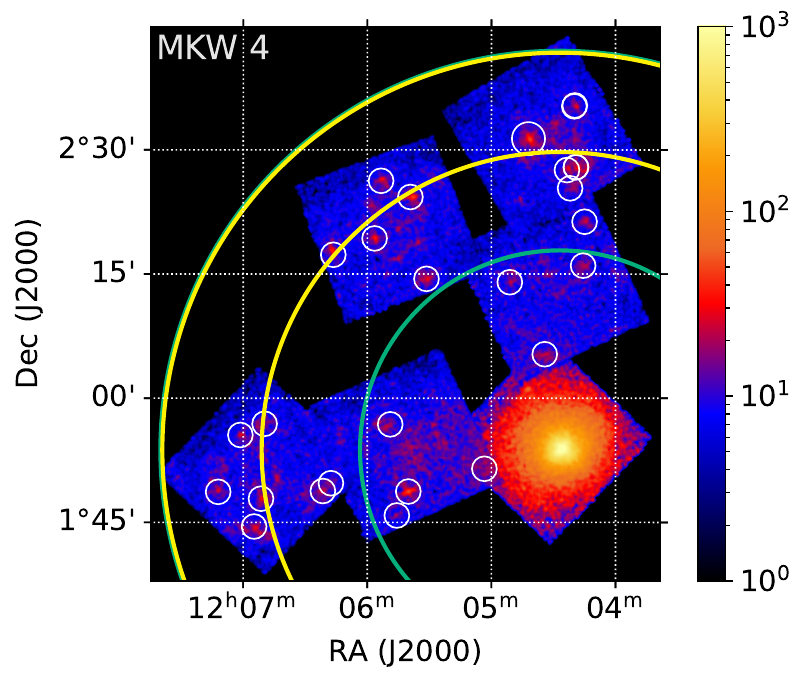}
         \includegraphics[width=0.5\textwidth,angle=0]{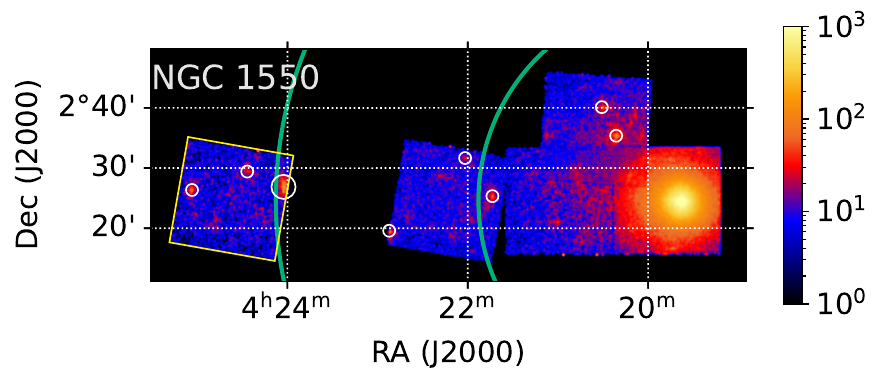}
         \includegraphics[width=0.28\textwidth,angle=0]{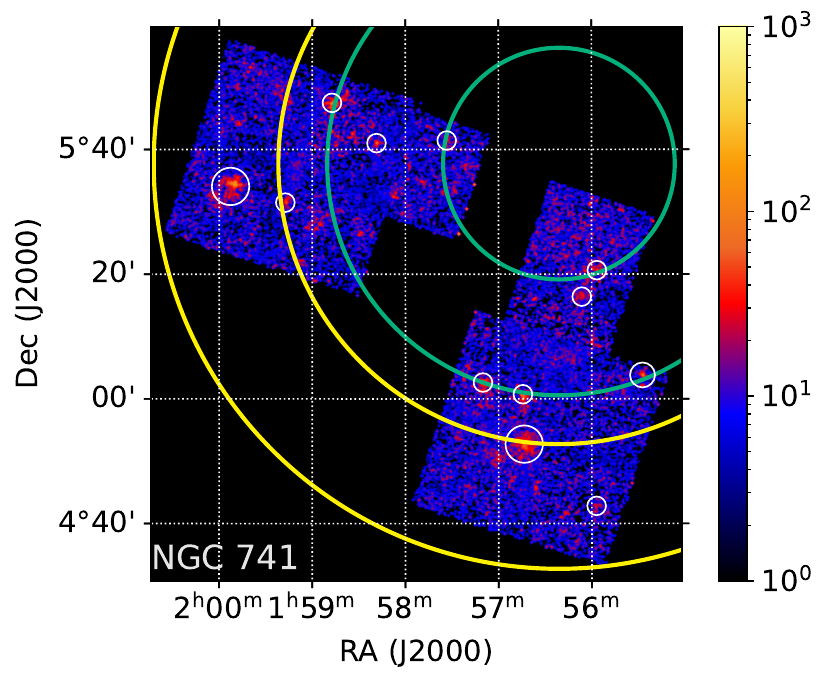}
		\caption{
			{\it Suzaku} XIS mosaic images in the 0.5--5.0 keV energy band.\@
			While the exposure time and instrumental background were corrected, the effect of the vignetting was not corrected. 
			The images were smoothed by a Gaussian of $\sigma=$16 pixels $\approx 17\arcsec$.  
			The numbers below the color bars have units of counts Ms$^{-1}$ pixel$^{-1}$.
			The green circles correspond  the $r_{500}$ and $2 r_{500}$, respectively. 
			The dashed circles indicate the point sources excluded from the analysis.
			The yellow boxes and annuli showed the background region where the background component parameters were estimated. 
			  {Alt text: Mosaic X-ray images of the sample except for the Virgo cluster observed with Suzaku XIS in the 0.5--5.0 keV band. The images are displayed in color, with the scale bar representing counts per pixel per megasecond. The horizontal and vertical axes show right ascension and declination.}
		}
		\label{fig:image}
	\end{center}
\end{figure*}

\begin{figure*}[htpd]
	\begin{center}
		\includegraphics[width=0.95\textwidth,angle=0 ]{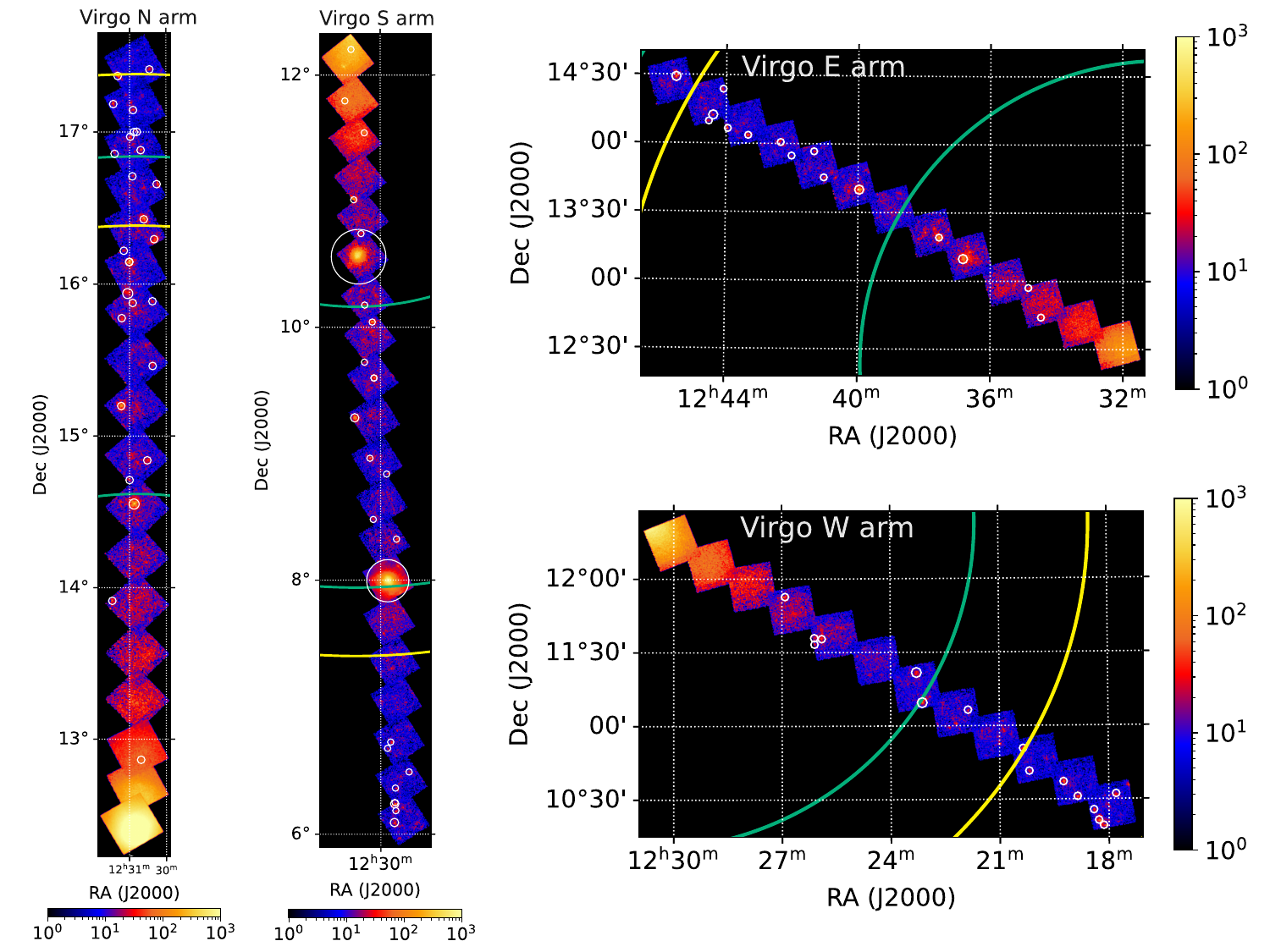}
		\caption{
			The same  figure as figure \ref{fig:image}, but (left) north arm, (middle) south arm, (right top) east arm, and (right bottom) west arm of the Virgo cluster. 
					{Alt text: The same as figure \ref{fig:image}, but showing the Virgo cluster.} 
		}
		\label{fig:imageVirgo}
	\end{center}
\end{figure*}

\section{Suzaku spectral analysis}
\label{sec:analysis}

\subsection{Extraction of spectra}

XIS spectra from each observation were extracted from
annular regions centered on the brightest cluster galaxy (BCG), except for the Coma cluster and MKW 4, where we used their X-ray peak as the center of the annuli.
We excluded regions within $30\arcsec$ from the edges of the field of view (FOV) to avoid vignetting and pointing uncertainties. 
Furthermore, we extracted spectra from the yellow
boxes or annuli shown in Figures \ref{fig:image} and \ref{fig:imageVirgo} (hereafter referred to as background regions).

The spectra of each XIS detector for a given annular region were combined. 
For the Coma and Hydra-A clusters, we divided the azimuthal directions into two sectors: one that includes the bright subcluster (southeast for the Hydra-A cluster and southwest for the Coma cluster), and another that is outside the sector.
We also created spectra for annular regions in each arm of the Coma and Virgo clusters.
We also avoided combining data from different observations of the Abell 262 cluster because some observations were severely affected by SWCX emissions (see Appendix \ref{sec:SWCX}); therefore, we did not use the three observations of the north-east arm (NE).
All spectra were binned to have at least one count per channel
and fitted using the C statistic \citep{Cash}.
We analyzed the 0.5--13 keV energy range for the FI detectors and 0.4--11.5 keV for the BI detector.
 
To estimate NXB contributions, we used the dark-earth database via ``xisnxbgen" ftools \citep{Tawa2008}.
The NXB spectra were combined using the same procedure as the corresponding spectra from the clusters
and then binned to have a minimum of 30 counts per channel.
Since the NXB spectra follow Gaussian statistics,   
we fitted them with a power law and multiple Gaussian
components using the $\chi^2$-statistic (see Appendix \ref{sec:nxb}), employing a diagonal RMF file.
Finally, we simultaneously fitted the sky spectra from the three XIS detectors, without subtracting the NXB, with astrophysical X-ray emission components and the NXB model (a power law plus Gaussians).

\subsection{Spectral fits of the background regions}

\begin{figure*}[tbhp]
	\begin{center}
	\includegraphics[width=0.23\textwidth,angle=0,clip]{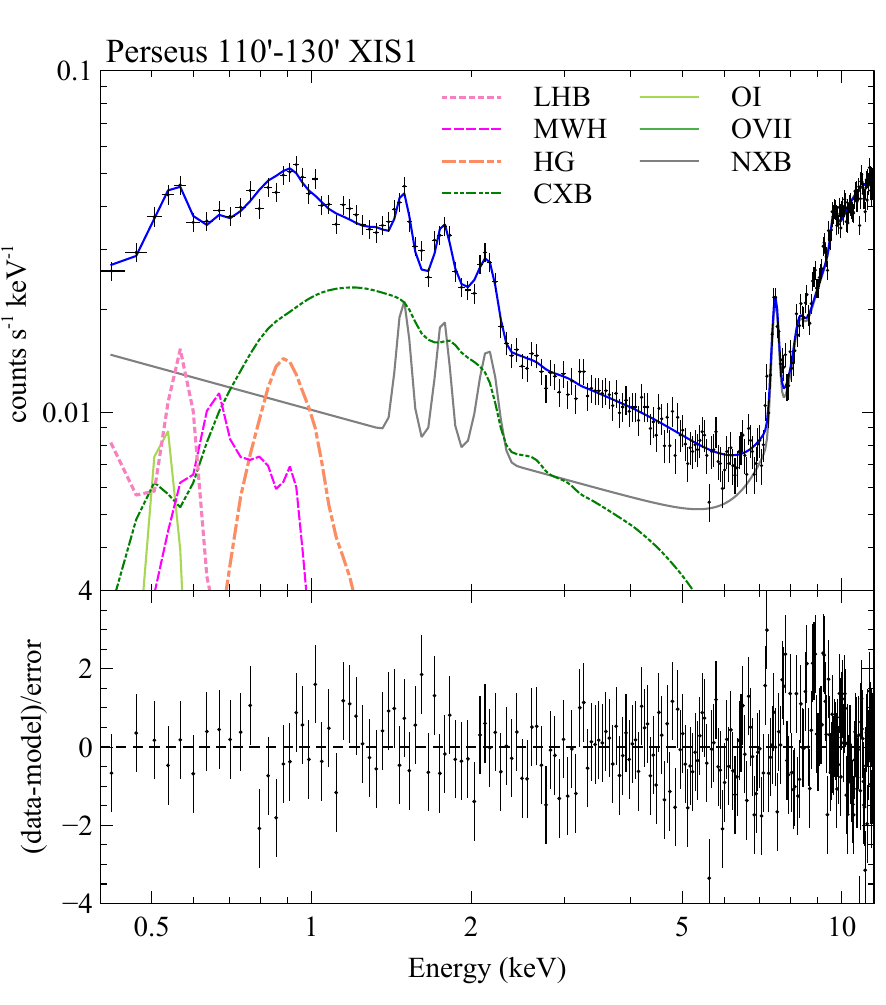}
	\includegraphics[width=0.23\textwidth,angle=0,clip]{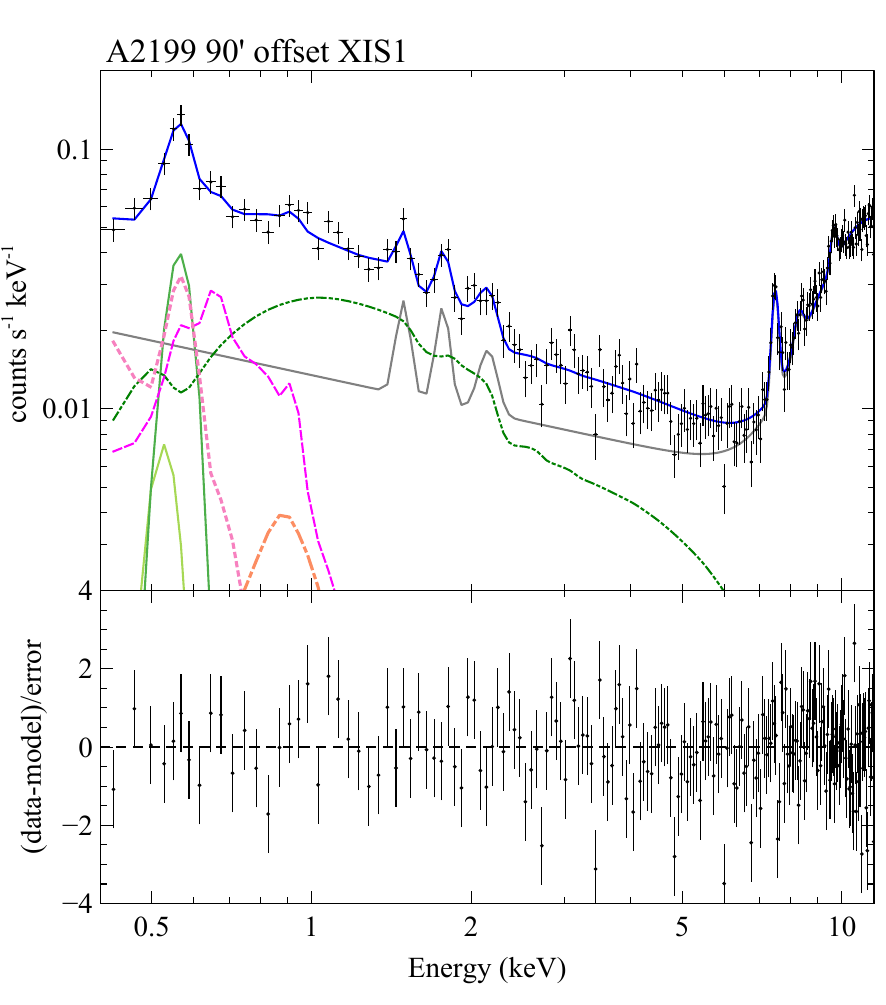}
	\includegraphics[width=0.23\textwidth,angle=0,clip]{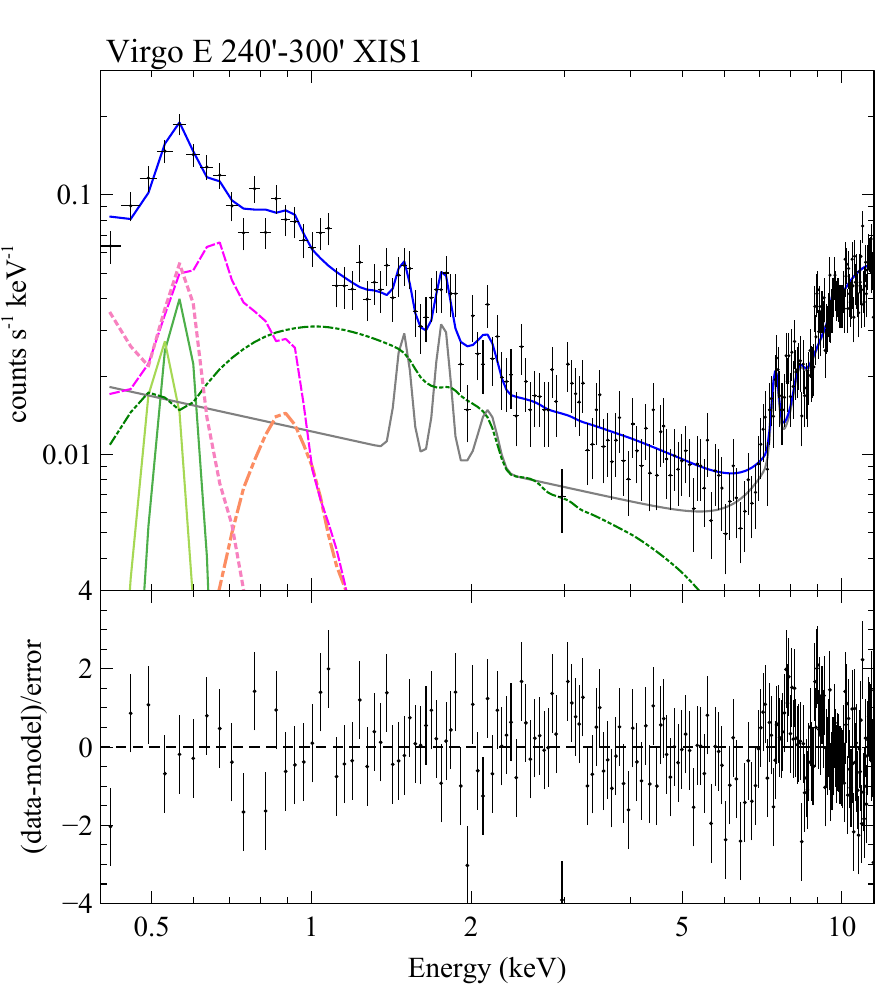}
	\includegraphics[width=0.23\textwidth,angle=0,clip]{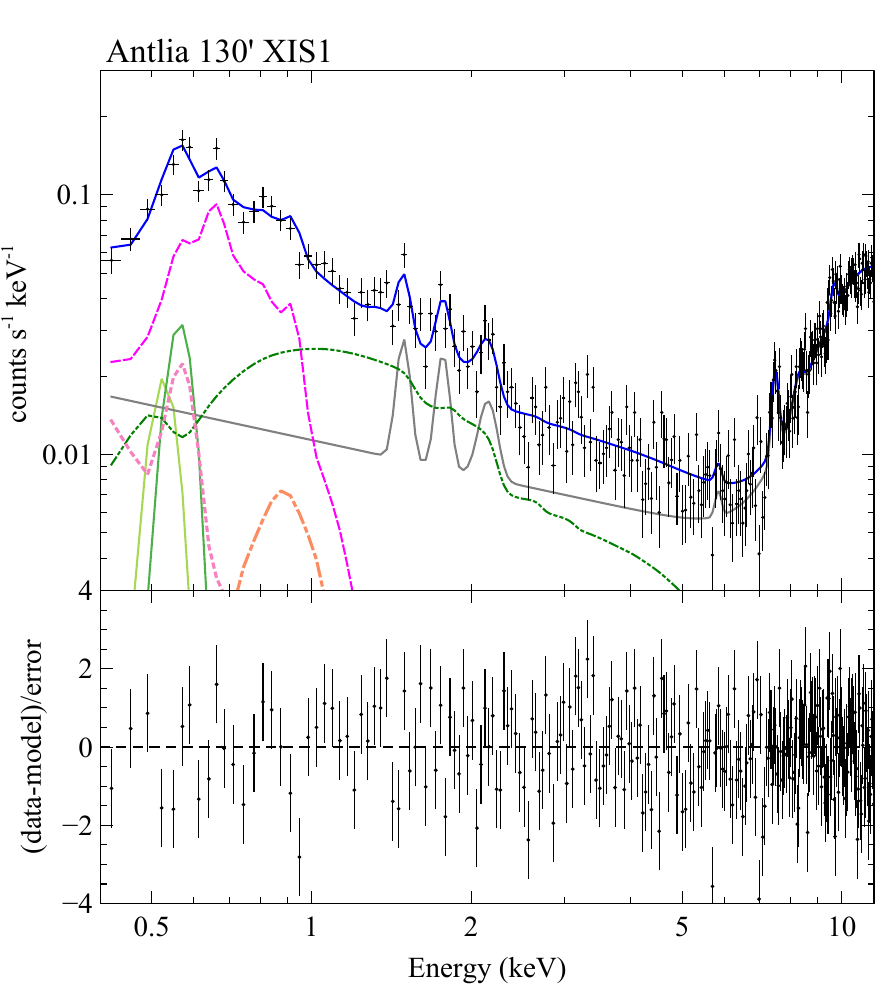}
		\caption{
			The representative XIS1 spectra of the background regions of the
			Perseus (110$'$--130$'$), A2199 (90$'$ offset pointings),    Virgo (240$'$--300$'$ east),
			and Antlia (130$'$ offset). 
			The contributions of the LXB (dotted lines), MWH (dashed lines), HG (dot-dashed lines), CXB (dot-dot-dashed lines), O\,\emissiontype{I} (light green solid lines), O\,\emissiontype{VII} (green solid lines), and NXB (gray solid lines) components are shown.
			The bottom panels show the residuals of the fit.
			{Alt text: Four subfigures, each consisting of two panels: the upper panels show the representative XIS1 spectra of the background regions and contributions of the spectral components with different colors, and the lower panels display the residuals.  } 
		}
		\label{fig:back}
	\end{center}
\end{figure*}

Accurate background and foreground estimation is crucial for studying faint X-ray emission in the outskirts of galaxy clusters and groups. 
Our background model consists of CXB, three collisional ionization equilibrium (CIE) plasmas representing LHB, MWH, and HG components,  
two Gaussians at 0.52 keV and 0.56 keV to model scattered O\,\emissiontype{I} emission \citep{Sekiya2014} and the O\,\emissiontype{VII} He$\alpha$  line, and the NXB components.
In our analysis, we model the CXB using a power law with a photon index of 1.4.
After 2011, the O\,\emissiontype{I} line (with a centroid at 0.525 keV), caused by solar X-ray fluorescence with neutral oxygen in the atmosphere of the Earth \citep{Sekiya2014}, has frequently contaminated the spectra.
Furthermore,  even after screening for SWCX emissions using light curves in the 0.5--2.0 keV,  the O\,\emissiontype{VII}  line, possibly originating from heliospheric SWCX,  often remains in the spectra
especially near the solar maximum and can lead to an underestimation of the MWH temperature \citep{Ueda2022}.
Outside the eRosita bubbles, using only the data near the solar minimum, the MWH temperatures are relatively uniform with a median value of 0.22 keV, 
which corresponds to the virial temperature of the Milky Way \citep{Ueda2022}.
We employ the APEC code \citep{Smith2001,Foster2012} with AtomDB 3.0.9 to model each CIE plasma.   
where the temperatures of the LHB, MWH, and HG components are fixed at 0.1 keV, 0.22 keV, and 0.8 keV, respectively, with solar abundances and zero redshift. The normalizations of the CIE components,  the two Gaussians, and CXB are allowed to vary. Galactic absorption $tbabs_{\rm GAL}$ was fixed for each object to the values listed in Tables \ref{tb:sample}.
  We assume that the HG component has a uniform surface brightness in the observed regions of each cluster or group, although spatial variation in its brightness beyond the virial radius of the Perseus cluster has been reported \citep{Matsushita2025}.
  The MWH, HG, and CXB components are modified by a photoelectric absorption model, $tbabs$, with the same Galactic $N_{\rm H}$ values.
We fitted the Suzaku spectra from the Lockman Hole and each background region using this model.
Additional details of the background modeling and fitting results are provided in Appendices \ref{sec:lockman} and \ref{sec:soft}.
As shown in figure \ref{fig:back}, the model successfully reproduces the observed spectra in these background fields, which span a wide range of Galactic
environment: located near (Perseus) and above (A2199) the Galactic disk, close to the eROSITA bubbles (Virgo), and even within a possible supernova remnant (Antlia).

\subsection{Spectral fit of the cluster regions}
\label{sec:fit}

\begin{figure*}[tbhp]
	\begin{center}
	\includegraphics[width=0.23\textwidth,angle=0,clip]{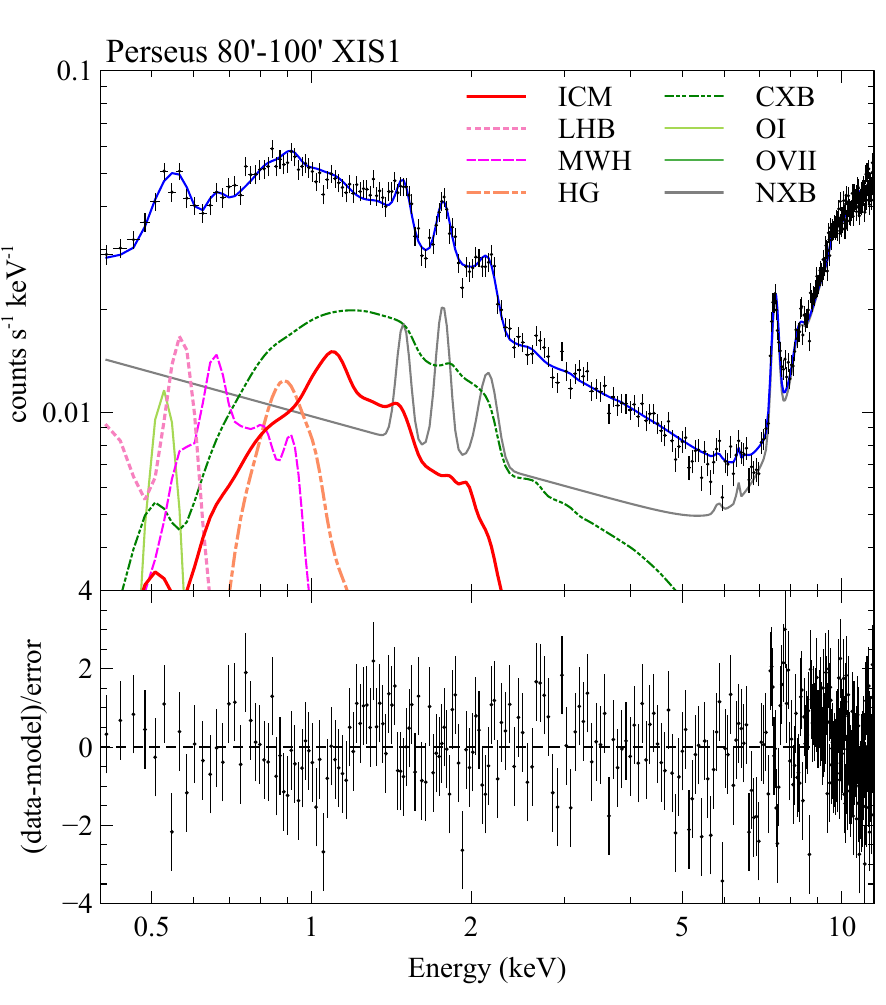}
	\includegraphics[width=0.23\textwidth,angle=0,clip]{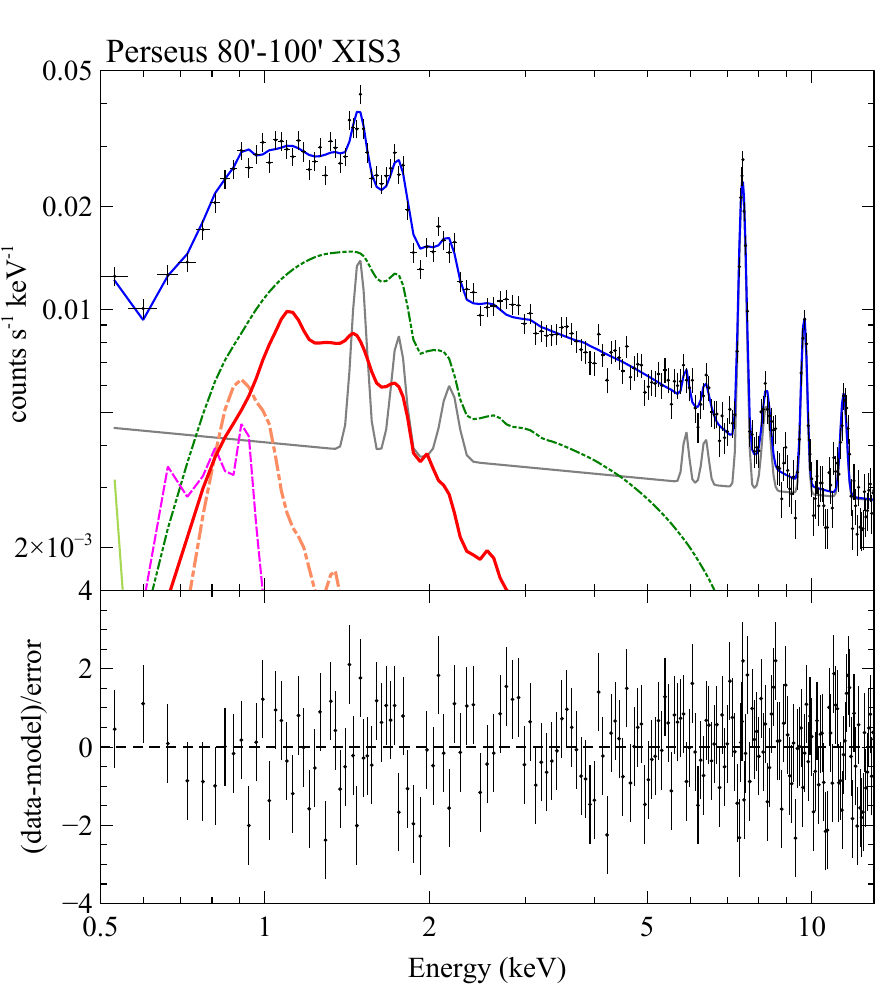}
		\includegraphics[width=0.23\textwidth,angle=0,clip]{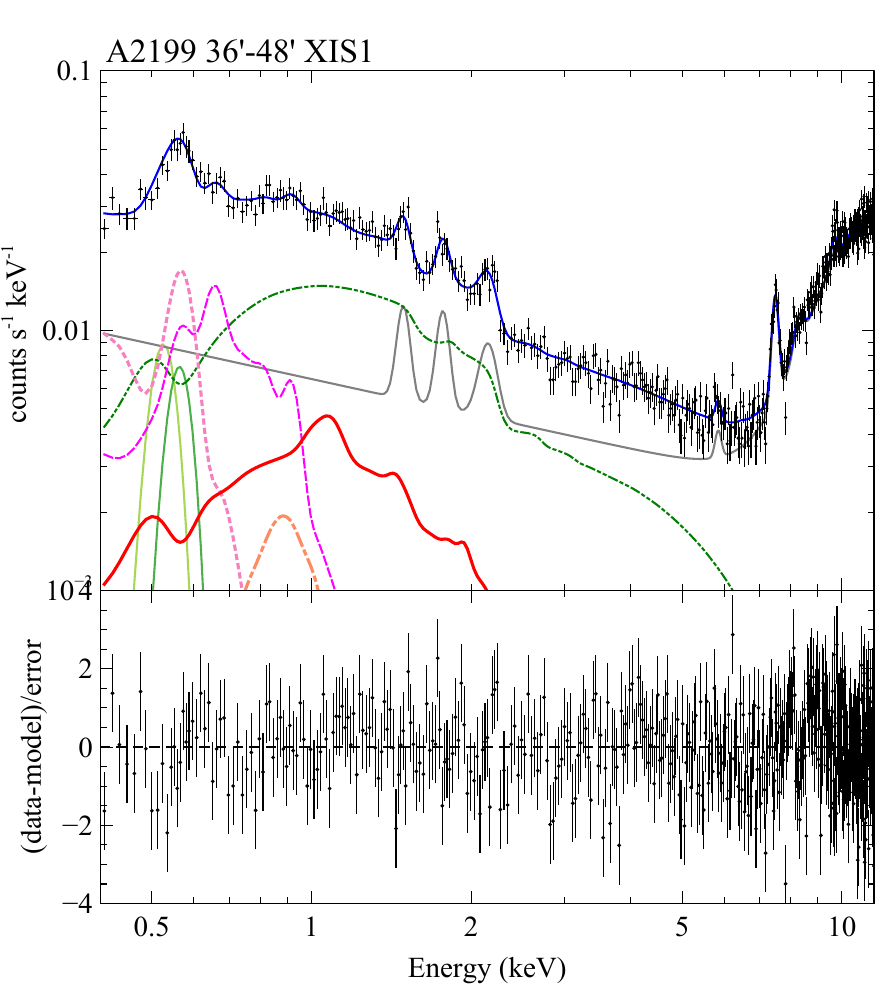}
	\includegraphics[width=0.23\textwidth,angle=0,clip]{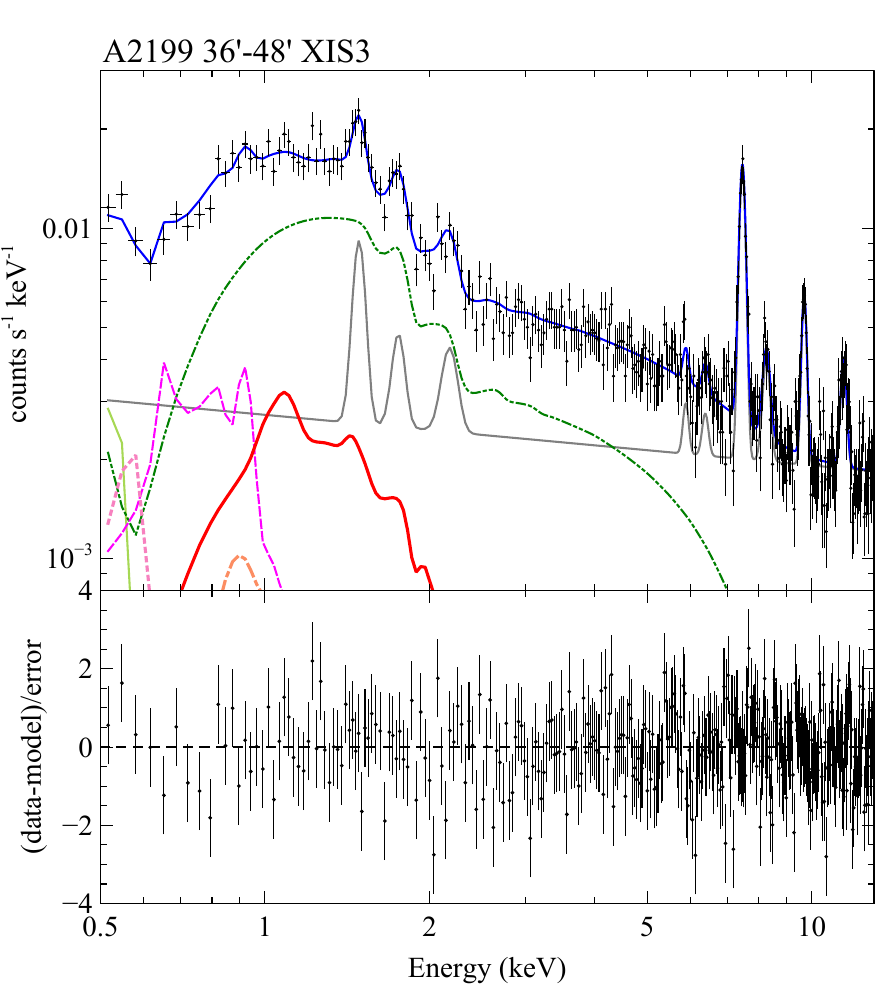}
		\caption{
			The representative XIS1 spectra of the Perseus cluster at 80$'$--100$'$ (1.7--2.2 Mpc) and A2199 at 36$'$--48$'$ (1.3--1.8 Mpc) 
			The contributions of the ICM, LXB, MWH, HG, CXB, O\,\emissiontype {I}, O\,\emissiontype {VII}, and NXB components are shown.
			The bottom panels show the residuals of the fit.
			{Alt text: Four subfigures, each consisting of two panels: the upper panels show the representative XIS1 and XIS3 spectra of the Perseus cluster and A2199 and contributions of the spectral components with different colors, and the lower panels display the residuals.  }
		}
		\label{fig:spec}
	\end{center}
\end{figure*}

We then fitted the spectra of each annular region by adding the ICM component to the background model.
The ICM emission was modeled as a single temperature plasma using the
$apec$ plasma code modified by Galactic absorption, $tbabs_{\rm Gal} \times apec_{\rm ICM}$.
The temperature, abundance, and normalization for each annular region were treated as free parameters,
and the redshift was fixed to the value given in table \ref{tb:sample}. 
The normalization of the CXB component was restricted within the ranges given in Appendix \ref{sec:lockman}, and the normalization of the HG emission was restricted within the 1 $\sigma$ statistical uncertainty derived from the background region of the corresponding cluster or group. 
For the Perseus cluster, we adopted the value derived from the stacked spectra in the 110$'$--130$'$ region.
For each arm of the Virgo cluster, we adopted the HG normalization derived for that arm, while for the stacked spectrum of the four arms, we adopted the result of the northern arm, which is close to the average of the four.
For the Coma cluster, we adopted the result beyond 110$'$ in the eastern arm, excluding the sub-cluster sector.
The normalizations of the Gaussians, LHB, and MWH were left free.
This spectral model provided acceptable fits, with a C-statistic/d.o.f close to unity.
Figure \ref{fig:spec} presents representative stacked XIS1 and XIS3 spectra from the Perseus cluster at the 80$'$--100$'$ annulus (corresponding to 1.7--2.2 Mpc)
and A2199 at 36$'$--48$'$ annulus (1.3--1.8 Mpc).
Our model reproduces these spectra well, although minor discrepancies appear in the Fe-L band of the XIS1 spectrum of the Perseus cluster, whereas the XIS3 spectrum is reasonably well fitted.
Hereafter, we refer to this model as "Model-HG08", and the derived emission measure ($\int n_{\rm e}n_{\rm H} ds$, where $n_{\rm e}$, $n_{\rm H}$, and $s$ are the electron, hydrogen densities, and the distance along the line of sight, respectively), temperature and metal abundance as EM$_{\rm HG08}$,  $kT_{\rm HG08}$, and Fe$_{\rm HG08}$, respectively, since the metal abundances are mainly derived from the fitting of Fe lines.
Because measuring metal abundances in the cluster outskirts is often uncertain,  we also refitted the spectra by fixing the metal abundances to 0.3 solar.  
We refer to this version as "Model-Z03", with the corresponding emission measures and temperatures denoted as EM$_{\rm Z03}$ and $kT_{\rm Z03}$, respectively.

\section{Results}
\label{sec:results}

\subsection{Emission measure profiles}

\begin{figure*}[tbhp]
	\begin{center}
	\includegraphics[width=0.9\textwidth]{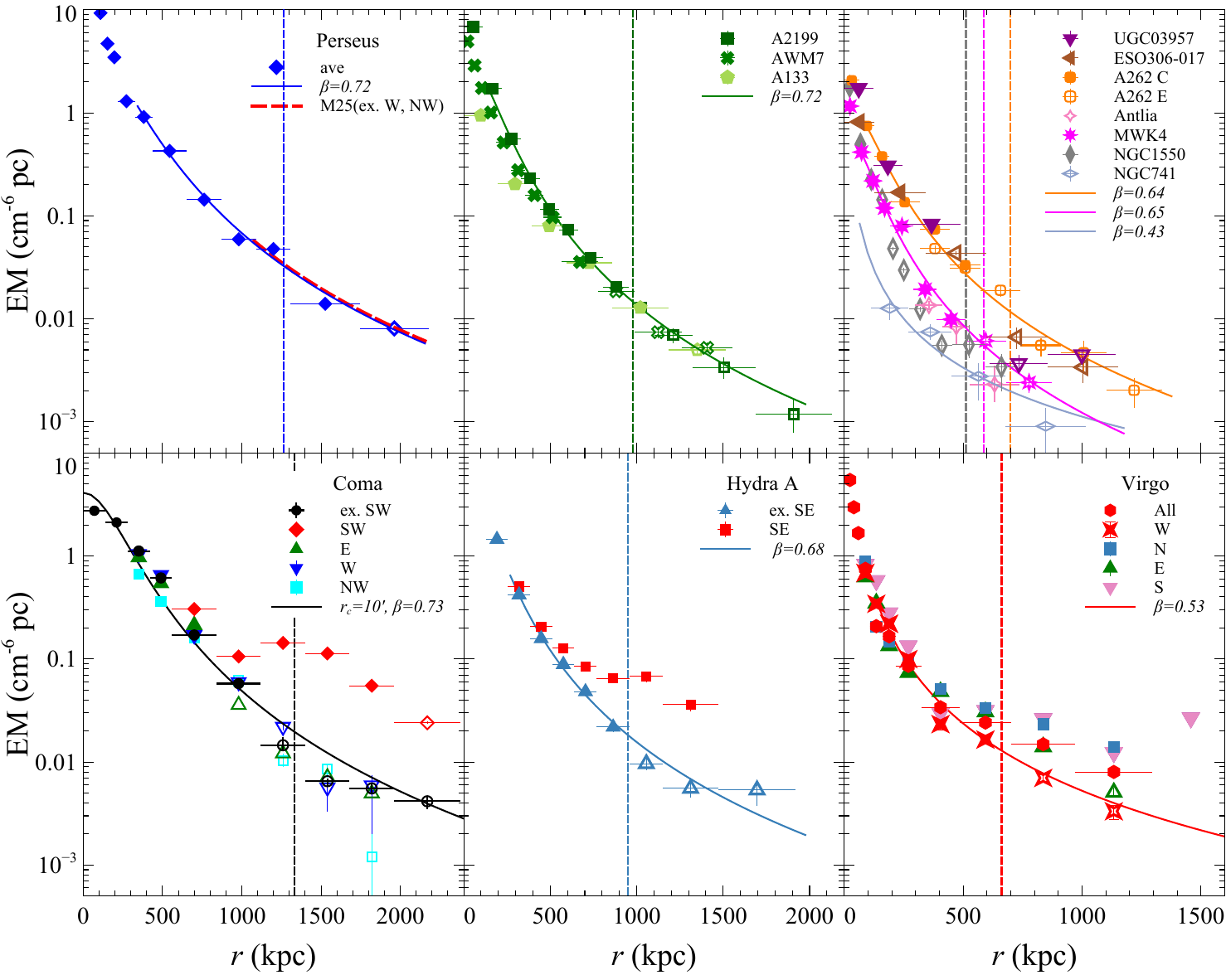}
		\caption{
		Radial profiles of the emission measure for the ICM, EM$_{\rm HG08}$ (filled symbols) and EM$_{\rm Z03}$ (open symbols), with representative $\beta$-model profiles. The dashed line in the top left panel shows the best-fit emission measure profile of the Perseus cluster excluding the W and NW arms from \citet{Matsushita2025}. The vertical lines indicate $r_{500}$ of representative clusters.
{Alt text: Six-panel figures showing radial profiles of ICM emission measure. The vertical axis is the emission measure in units of cm$^{-6}$ pc, and the horizontal axis is the radius in units of kpc. 
Different systems are plotted with colored symbols,  and the representative $\beta$-model curves are overlaid. } 		
		}
		\label{fig:em}
	\end{center}
\end{figure*}

Figure \ref{fig:em} shows the radial profiles of EM$_{\rm HG08}$ in the inner regions and EM$_{\rm Z03}$ in the outer regions for each cluster.
The relaxed systems exhibit smooth profiles that gradually decrease with radius.
Beyond several hundred kpcs from the center, Abell 2199, AWM7, and A133 all exhibit similar emission measure profiles. 
The profiles for Antlia, MKW4, and NGC 1550 are also similar.

In figure \ref{fig:em}, we also plot representative $\beta$-models described by
\begin{equation}
 {\rm EM}(r)={\rm EM}_c\left(1+\frac{r^2}{{r_c}^2}\right)^{-3\beta+1/2},
\label{eq:em}
 \end{equation}
 where ${\rm EM}(r)$ and ${\rm EM_c}$ are the emission measure and its central value, respectively,
 $r$ is the projected distance from the cluster center, and $r_c$ is the core radius.
As plotted in figure \ref{fig:em}, outside the cool cores regions, the emission-measure profiles of the relaxed clusters are well described by $\beta$-models, with best-fit values of $\beta\sim$0.6--0.7 for clusters and $\sim$0.4--0.65 for the groups.

The three "merging clusters," the Coma, Hydra A, and Virgo clusters, exhibit significantly azimuthal
variations in their radial profiles.
In figure \ref{fig:em}, we also show the best-fit $\beta$ model for the Coma cluster,  obtained from eROSITA observations \citep{Churazov2021}, with $r_c=10'$ and $\beta=0.73$, based on the data excluding the southwest sector. Even after
excluding the southwest sector, our Suzaku profiles for the Coma cluster deviate from this model, probably due to the asymmetry in the ICM and limited azimuthal coverage of Suzaku observations.
Excluding the bright sectors, the emission-measure profiles for Hydra A and the western arm of the Virgo cluster are consistent with $\beta$-models.



\subsection{Projected temperature profiles}
\begin{figure*}[tbhp]
	\begin{center}
	\includegraphics[width=0.85\textwidth]{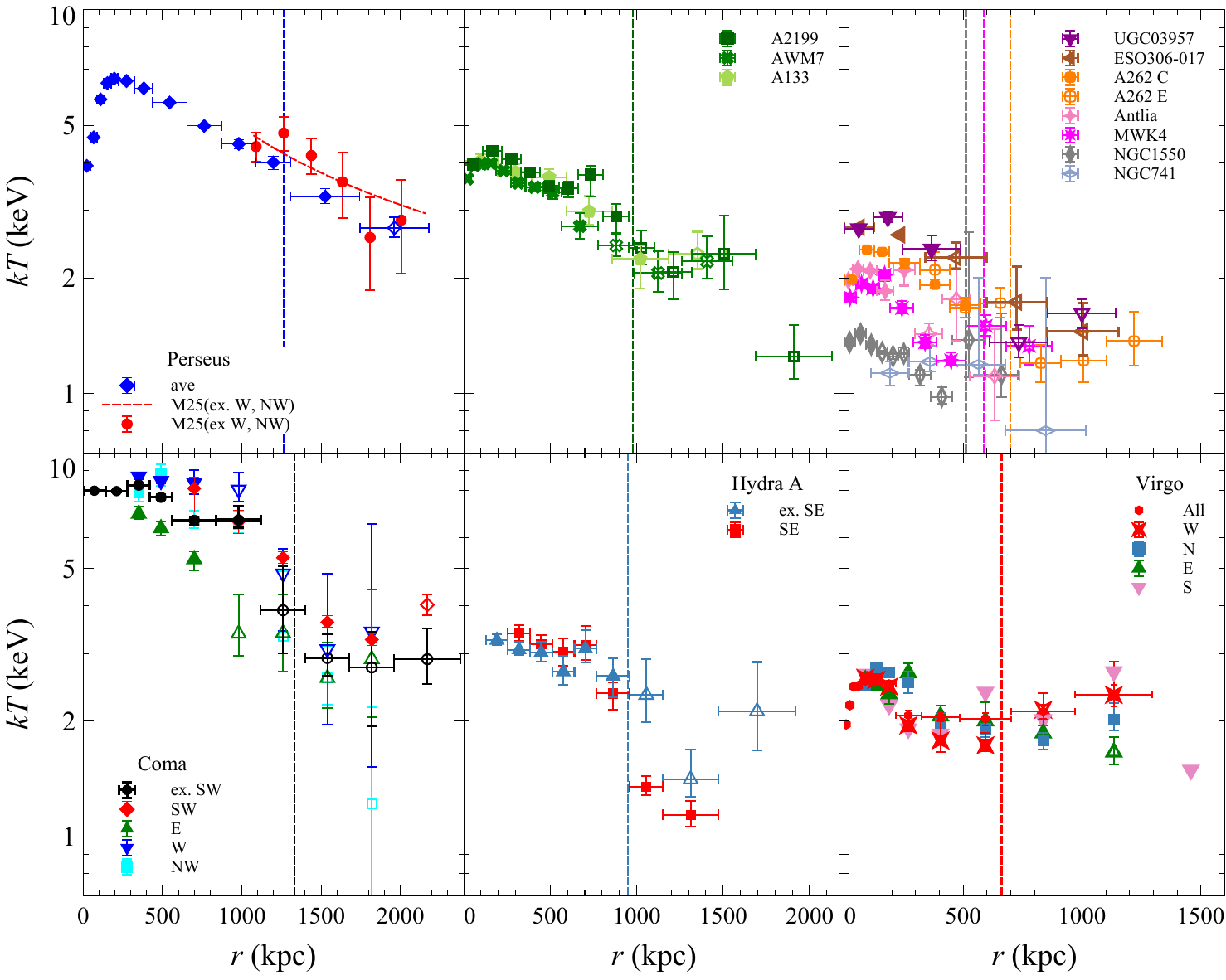}
		\caption{
		Radial profiles of the ICM temperature: $kT_{\rm HG08}$ (filled symbols) and $kT_{\rm Z03}$ (open symbols) . The dashed line and filled circles in the top left panel show the best-fit power-law relation and weighted averages of temperatures with similar radial distances of the Perseus cluster, excluding the W and NW arms, obtained by \citet{Matsushita2025}. The vertical lines indicate $r_{500}$ of representative clusters.	
		{Alt text: Six-panel figures showing radial profiles of ICM temperature. The vertical axis is the temperature in units of keV, and the horizontal axis is the radius in units of kpc. 
Different systems are plotted with colored symbols. } 	}
		\label{fig:kT}
	\end{center}
\end{figure*}

Figure \ref{fig:kT} shows the projected ICM temperature profiles, with $kT_{\rm HG08}$ in the inner regions and $kT_{\rm Z03}$ in the outer regions.
In relaxed clusters, the temperatures decrease smoothly beyond the cool core.
For example, in the Perseus cluster, the temperature drops from 6.6 keV at $\sim$200 kpc to $\sim$ 2.7 keV at $\sim$2000 kpc.
Other relaxed systems including Abell 2199, AWM7, Abell 133, UGC03957, and ESO306-107
exhibit similar temperature gradients, with values in the outermost regions reaching approximately one-half to two-thirds of their peak temperatures.

In contrast, the merging clusters show more irregular temperature profiles.
The Coma cluster shows temperature jumps at $\sim$1000 kpc in the W and NW arms, probably caused by shocks, 
 while the E arm has lower temperatures, as found by \citet{Simionescu2015} and \citet{Uchida2016}.
Hydra A also shows a temperature discontinuity in the southeast sector as reported by
\citet{DeGrandi2016}.
The Virgo cluster is distinguished by exhibiting relatively flat temperature profiles in all directions.

Hereafter, we excluded subcluster regions in the Coma and Hydra A clusters (the southwest sector in Coma and the southeast sector in Hydra A).
For the Virgo cluster, we used both the western (W) arm and the azimuthally averaged profile (hereafter, ``Virgo All'').
For A262, only the central regions and the eastern arm (E) were used, as the NE3 observation was heavily contaminated by SWCX emission.

\subsection{Systematic uncertainties in emission measure and temperature measurements}
\label{sec:sys}

\begin{figure*}[tbhp]
	\begin{center}
	\includegraphics[width=0.95\textwidth]{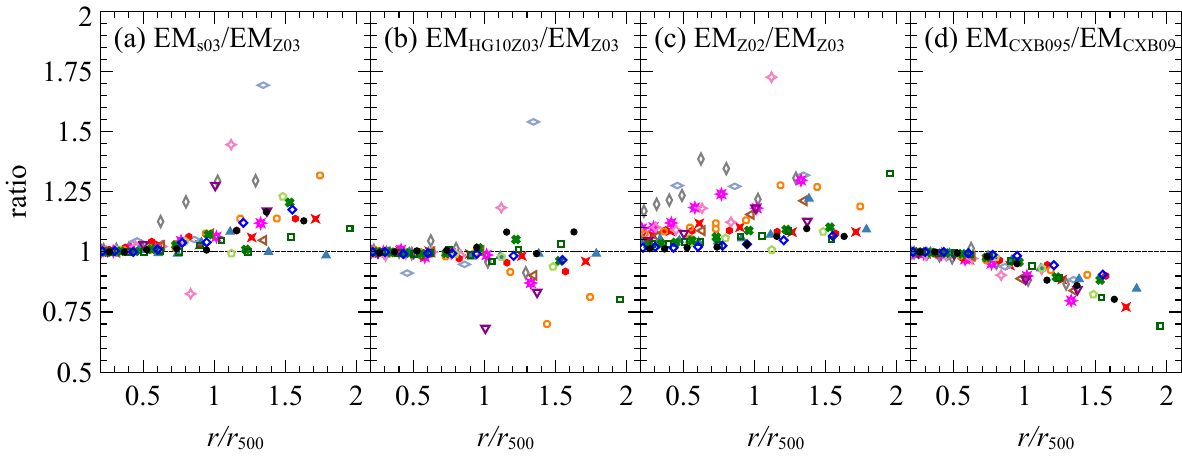}
		\caption{The ratio of the best-fit emission measures for (a) Model-s03 to Model-Z03 (b) Mode-Z03 to Model-HG10Z03 (c) Model-Z03/Model-Z02 (d) Model-CXB095/Model-CXB09.
		The meanings of the colors and symbols are the same as in figure 6.
		{Alt text: Four-panel figures showing the systematic uncertainties in the emission measure
		plotted against the radius in units of kpc. 
Different systems are plotted with colored symbols. } 
}
		\label{fig:emsys}
	\end{center}
\end{figure*}

\begin{figure*}[tbhp]
	\begin{center}
	\includegraphics[width=0.95\textwidth]{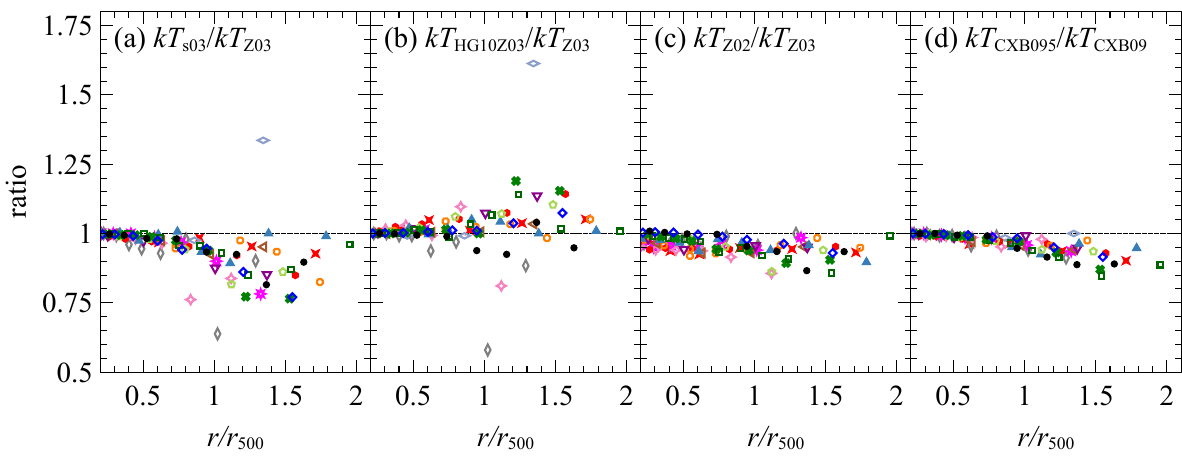}
		\caption{
		The ratio of the best-fit temperatures for (a) Model-s03 to Model-Z03 (b) Mode-Z03 to Model-HG10Z03 (c) Model-Z03/Model-Z02 (d) Model-CXB095/Model-CXB09.
		The meanings of the colors and symbols are the same as in figure 6.
			{Alt text: Four-panel figures showing the systematic uncertainties in the temperature
		plotted against the radius in units of kpc. 
Different systems are plotted with colored symbols. } 
		}
		\label{fig:kTsys}
	\end{center}
\end{figure*}

We examined how varying assumptions in our background modeling and other systematics affect emission measures and temperatures, in order to assess systematic uncertainties.

\subsubsection{Brightness of the HG component:}
At Galactic longitudes $75^\circ<l<285^\circ$ and latitudes $|b|>15^\circ$, the typical emission measure of the HG component is a few $\times 10^{-4} {\rm cm^{-6} pc}$, with a median value of $3\times10^{-4} {\rm cm^{-6} pc}$ when a 1 solar abundance is assumed \citep{Sugiyama2023}.
If a 0.3 solar abundance is adopted, as for the ICM, this corresponds to $\sim10^{-3} {\rm cm^{-6} pc}$, indicating a non-negligible contribution at cluster outskirts.
Similarly, the {HaloSat} survey reported that the emission measure of the HG component is typically $\sim10^{-3} {\rm cm^{-6} pc}$ over 85\% of the sky with $|b|>30^\circ$, with some regions, particularly within $90^\circ$ of the Galactic center, reaching $\sim10^{-2} {\rm cm^{-6} pc}$ when a 0.3 solar abundance is adopted \citep{Bluem2022}.
Such contamination can bias ICM temperature and density measurements in these regions.

To evaluate the impact of the HG brightness, we tested fits that exclude the HG component (Model-s03, assuming an ICM abundance of 0.3 solar).
To isolate the effect of the HG component, we compared the derived emission measures and abundances with those from Model-Z03.
Beyond $r_{500}$, the emission measures can be up to 70\% higher at most, and the temperatures typically 20--30\% lower than the values derived with Model-Z03.
Even around $\sim r_{500}$, temperature differences of up to 20\% arise when the HG component is omitted.
These effects are shown in Figures \ref{fig:emsys} and \ref{fig:kTsys}.

Although we included the HG component in our spectral model, spatial variations are expected if the HG component originates from stellar feedback processes in our Galaxy.
In particular, substantial azimuthal variations in the brightness of HG were reported beyond the virial radius of the Perseus cluster by \citet{Matsushita2025}.
Nevertheless, our analysis quantifies the systematic uncertainties in emission-measure and temperature measurements attributable to the HG contribution, even in the presence of such spatial variations.

 \subsubsection{Temperature of the HG component:}
Although we fixed the HG temperature at 0.8 keV in our baseline model, measurements in anti-Galactic center regions show that the HG temperature actually ranges from 0.6 to 1.2 keV, with a median of 0.8 keV \citep{Sugiyama2023}.
To assess the impact of this uncertainty, we refitted the spectra assuming a higher HG temperature of 1.0 keV and an ICM abundance of 0.3 solar (Model-HG10Z03).
As a result, the derived emission measures and temperatures differ by about 10--20\% beyond $r_{500}$ compared to those obtained using the baseline Model-Z03.
The exception is NGC741 at 1.35 $r_{500}$. In this case, $kT_{\rm HG10Z03}$ is 1.9 keV, much higher than the 1.2 keV obtained at 0.86 $r_{500}$.
If we instead assume 1.2 keV for $kT_{\rm HG10Z03}$, the resulting emission measure becomes nearly identical to that from Model-Z03.

\subsubsection{Fixed abundances:}
Because ICM abundances are difficult to measure at radii beyond $\sim r_{500}$, we also refitted spectra of the cluster regions by fixing the metal abundance at 0.2 solar (Model-Z02).
At low temperatures ($kT<1.5$ keV), the emission measures change by several tens of percent, as the strength of the Fe-L lines under the assumed fixed abundance primarily determines them.
However, for hotter systems, the impact on the emission measure is minor, whereas the temperatures vary by $\sim$10--20\%.
The exception is the Antlia data at $1.1$ $r_{500}$, which has a large statistical uncertainty, and the two values are consistent within $2\sigma$.

\subsubsection{CXB level:}
The CXB level carries systematic uncertainties of several percent in the spectra accumulated over the full FOV (Appendix \ref{sec:lockman}).
To evaluate the effect of the CXB level, we refitted the spectra by fixing the CXB normalization at $9.0\times10^{-4}$ and $9.5\times10^{-4} {\rm photons~ keV^{-1} cm^{-2} s^{-1}}/(400\pi~ {\rm arcmin^2})$ at 1 keV (Model-CXB09 and Model-CXB095).
A 7\% shift in the CXB normalization, corresponding to the expected cosmic variance within a Suzaku XIS FOV for the threshold flux of excluding the point sources of $F_{\rm th}=10^{-13} \rm{erg~s^{-1} cm^{-2}}$, 
 changes the emission measure by $\sim$10\% at $r_{500}$ and $\sim$20\% at $1.5\,r_{500}$, depending on the plasma temperature, and the temperatures vary by a similar amount.


\subsubsection{Point spread function and stray-light contamination:} 
\begin{figure*}[tbhp]
	\begin{center}
	\includegraphics[width=0.9\textwidth]{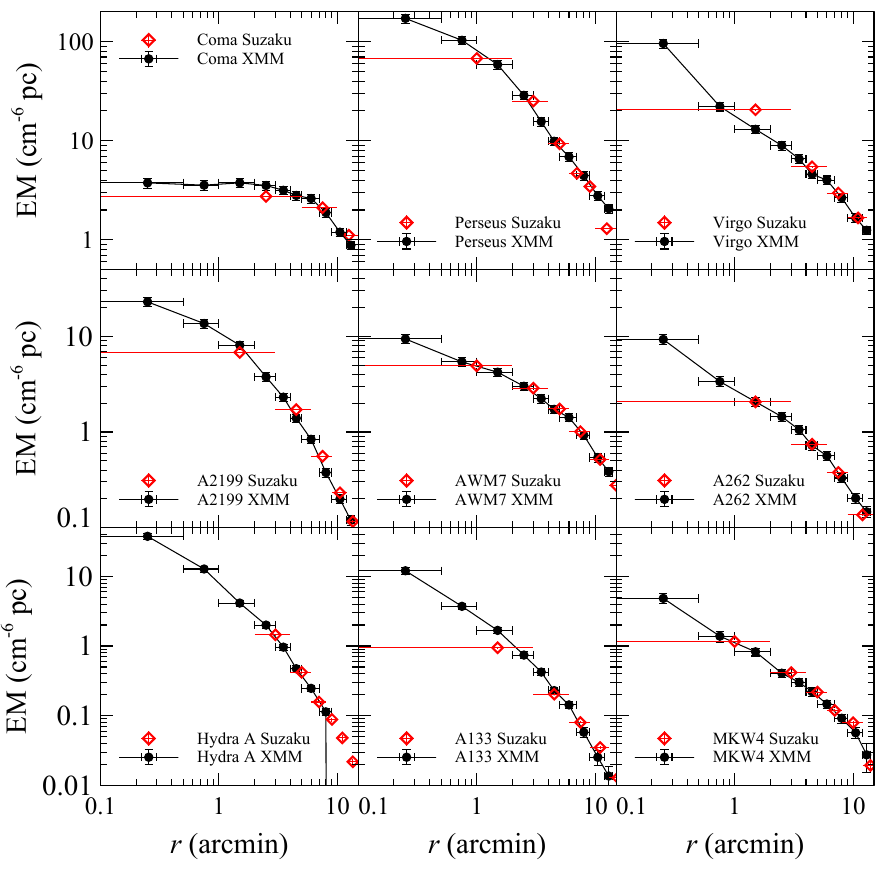}
		\caption{
	Comparison of Emission measure profiles: EM$_{\rm HG08}$ with Suzaku  (open diamonds) and with XMM (filled circles;\cite{Snowden2008}).
	{Alt text: Nine-panel figures showing radial profiles of ICM emission measure with Suzaku and XMM. The vertical axis is the emission measure in units of cm$^{-6}$ pc, and the horizontal axis is the radius in units of kpc. 
 } 		
	}
		\label{fig:EMxmm}
	\end{center}
\end{figure*}

The Suzaku X-ray telescope has a relatively large half-power diameter (HPD) of $2'$ \citep{SuzakuXRT}, and we adopted the flat ARFs rather than $\beta$-model ARFs (which assume a $\beta$-model gas distribution).
To assess the impact of the HPD and the adoption of flat ARFs, we compared our results with those from {XMM-Newton}.
The {XMM-Newton} cluster catalog by \citet{Snowden2008} includes nine clusters from our sample.
We converted the surface-brightness profiles in \citet{Snowden2008} into emission-measure profiles using their published temperatures and metal abundances, and compared them with our {Suzaku} results (Figure \ref{fig:EMxmm}).
Except for the innermost bins (radii of a few arcminutes), the {XMM-Newton} derived emission measures agree well with those from {Suzaku}.
This consistency likely arises because {Suzaku}’s PSF-induced mixing is dominated by photons from adjacent annuli, and the flat ARFs, which assume uniform surface brightness within $20'$, effectively account for this contamination.

Beyond $30'$ from a bright X-ray peak, stray light arising from non-standard reflection paths can contaminate the outskirts out to $\sim100'$ \citep{Mori2005, Takei2012}.
\citet{Urban2014} showed that, on the outskirts of the Perseus cluster, the regions shielded from stray light by the XRT geometry (“shaded” regions) and the regions exposed to stray light yield consistent temperatures and emission measures.
Since the core of Perseus is significantly brighter than that of the other clusters in our sample (see Section \ref{sec:emd}), stray light contamination is expected to be even less significant for the rest of the sample.
The Virgo cluster extends to several hundred arcminutes and is free of stray light.
In contrast, Hydra A and A133 are more distant systems with extents of $\sim30'$, where PSF effects dominate over stray light.

\subsubsection{The effect on the stacking for the Perseus cluster}

\citet{Matsushita2025} analyzed the same Suzaku data of the Perseus cluster beyond $\sim$1 Mpc from the center of the cluster.
They extracted spectra over the FOV of the XIS detectors without stacking spectra from different observations.  
They also accounted for the spatial variation of the HG component, using the "phabs" model for photoelectric absorption and the solar abundance table of \citet{Lodders2003}.
In their analysis, they focused on six azimuthal arms, excluding the W and NW directions that extend toward large-scale filaments of the Universe. In contrast, this study utilizes all eight available arms. 
They found that beyond $r_{500}$, the scatter in the emission measure among the six arms was relatively small, while the W and NW arms exhibited emission measures about 1.8 times higher.
Their best-fit power-law relation for the emission measures beyond 1070 kpc, excluding the W and NW arms, agrees well with the best-fit $\beta$-model for our data (figure \ref{fig:em}).
As shown in figure \ref{fig:kT}, our temperature profile is 10--15 \% lower than that derived by \citet{Matsushita2025} for the relaxed arms of Perseus.

\bigskip

 Based on these tests, we adopt systematic uncertainties of 10\% and $10^{-3} {\rm cm^{-6} pc}$  for the emission measure, and 10\% beyond $>r_{500}$ and 20\% beyond $>r_{200}$ for the temperature in our subsequent discussion.

\subsection{Electron density}
\label{sec:ne}

To derive electron density profiles for clusters with cool cores, listed in the X-ray cluster catalog by \citet{Snowden2008}, 
we simultaneously fitted emission measure profiles with Suzaku and XMM, using a sum of two $\beta$ models.
The innermost Suzaku data point (within 1$'$) was excluded to minimize the systematic uncertainties associated with the PSF of the Suzaku XRT and the use of "flat" ARFs.
The model is described as,
\begin{equation}
{\rm EM}={\rm EM}_{i}\left(1+\frac{r^2}{r_{ci}^2}\right)^{-3\beta+1/2}+{\rm EM}_{o}\left(1+\frac{r^2}{r_{co}^2}\right)^{-3\beta+1/2}
\end{equation}
where ${\rm EM}_{i}$ and ${\rm EM}_{o}$ are the normalizations of the inner and outer components,
and $r_{ci}$ and $r_{co}$ are their respective core radii. 
We assumed a common $\beta$ for the two components.
For clusters and groups not included in \citet{Snowden2008} and for the Coma cluster, we adopted a single $\beta$ model (equation \ref{eq:em}) to fit the data.
As shown in figures \ref{fig:EMfit} in the Appendix \ref{sec:beta}, these models reproduce the observed emission measure profiles of individual clusters well.
The residuals are relatively small: for example, in the Perseus cluster, deviations are at most a few tens of percent, corresponding to $\sim$10 percent in electron density.
Some discrepancies between the model and the data may arise from the remaining substructures or gas clumping, even after bright extended structures were excluded as point sources based on the X-ray images.
These $\beta$-model fits help suppress the influence of unresolved gas clumping, while deprojection methods,
commonly used in Suzaku analysis, may amplify small deviations from spherical symmetry.

For the NGC 741 group, we fixed $r_c$ to the best-fit value of 0.09$'$ (2.1 kpc),
based on the XMM observation by \citet{NGC741},
since Suzaku data points only cover radii beyond 5$'$ (110 kpc).
We obtained $\beta=0.40\pm0.06$ from our fit, which is in agreement with the XMM results of $\beta=0.428\pm 0.005$ derived within $\sim$100 kpc.

The best-fit emission measure profiles were converted into electron density ($n_{\rm e}$) profiles.
For clusters and groups not listed in \citet{Snowden2008}, and for Coma, we used the single $\beta$ model.
For the others, the sum of two $\beta$ models was used to derive the electron density profiles.

\subsection{Pressure profiles and $r_{500}$}
\label{sec:r500}

We calculate the electron pressure profiles, $P_{\rm e}=n_{\rm e} T$, using the best-fit electron density profiles and
projected temperature data.
 Within $\sim$ 0.5 $r_{500}$, where the temperature gradient is relatively small, 
the difference between the projected and deprojected temperature profiles remains within several percent.
Beyond $r_{500}$, this difference increases to 10--20 percent, comparable to the systematic uncertainties in the temperature measurements.

The pressure profile of galaxy clusters is expected to follow a generalized Navarro-Frenk-White (gNFW) model \citep{NagaiNFW}:

\begin{equation}
P_{\rm e} (R)= P_{500} f_P \frac{P_0}{(c_{500}x)^{\gamma_p}[1+(c_{500}x)^{\alpha_p}]^{\beta_p-\gamma_p}/\alpha_p}\label{eq:gnfw}
\end{equation}

where $R$ is the three-dimentional distance from the center and $x=R/r_{500}$.   $P_0$ and  $c_{500}$ are normalization and concentration parameters, respectively.
The parameters $\gamma_p$, $\alpha_p$, and $\beta_p$ describe the slopes in the core, intermediate, and outer regions, respectively.
The best-fit values from the SZ measurements with Planck \citep{Planck2013} are:
$    P_0 = 6.41, 
    \alpha_p = 1.33,
    \beta_p = 4.13, 
    \gamma_p = 0.31,
  c_{500} = 1.81$.
  
  The characteristic pressure $P_{\rm 500}$ is expected to scale
with the total mass of the cluster in the standard self-similar model,
based solely on gravitational heating \citep{Arnaud2010}:

\begin{equation}
P_{500} = 1.65 \times 10^{-3} h(z)^{8/3}\left( \frac{M_{500}}{3\times 10^{14} M_\odot} \right)^{2/3}  \mathrm{keV cm^{-3}}
\end{equation}

where $M_{500}$ is the mass enclosed within $r_{500}$.
We also introduce a correction factor, $f_P$,  to account for deviations from the self-similar evolution:
as given by \citet{Arnaud2010}:

\begin{equation}
f_P = \left( \frac{M_{500}}{3 \times 10^{14} M_\odot} \right)^{0.12}.
\end{equation}

We applied the Planck pressure profile \citep{Planck2013} to the observed pressure data beyond 0.25 $r_{500}$, allowing only $M_{500}$ (denoted as $M_{500p}$)
to vary, while fixing the other parameters to the Planck best-fit values.
The resulting $R_{500p}$ values are listed in table \ref{tb:mass}, and 
 the best-fit pressure profiles are shown in figure \ref{fig:Pfit} in Appendix \ref{sec:pfit}.
For example, the pressure profile of the  Perseus cluster is well reproduced by the Planck profile, yielding $M_{500p}=(5.9\pm0.1)\times 10^{14}M_\odot$  and $\chi^2/d.o.f=3.0/6$.
More than half of the clusters are well described by this Planck profile; however, some systems, especially lower-mass ones, show pressure excesses beyond $R_{500p}$ ($R_{500}$ corresponding to $M_{500p}$.
The poorest system in our sample, NGC 741, exhibits a noticeably shallower pressure slope.
We also fitted the profiles by allowing both $M_{500}$ and $\beta_p$ to vary, thus better accounting for such deviations.
Figure \ref{fig:Pfit} also shows the best-fit profiles.
Most clusters yield $\beta_p$ values consistent with the Planck value of 4.13.  For example,
this new fit gives $\beta_p=4.25\pm 0.21$ for the Perseus cluster,  consistent with the fixed-$\beta_p$ fit.
In contrast, some lower-mass systems tend to have smaller $\beta_p$ values, indicating flatter pressure slopes in their outskirts.
Even with this model,  significant pressure excesses remain beyond $R_{500p}$ of the Virgo cluster (stacked spectra of the four arms),  Virgo W arm, and MKW4.

\begin{table}
	\caption{$M_{500}$ and $M_{2500}$ of the sample clusters and groups. 
$M_{500}$ is derived either from the pressure profile using the Planck model or from hydrostatic equilibrium fits, 
with special sectors excluded as noted in the footnotes.}
	\label{tb:mass}
	\begin{center}
		\begin{tabular}{llllllllll}
			\hline
target & $M_{\rm 500p}^*$ & 	$M_{\rm 500HE}^\dagger$ & $M_{\rm 2500HE}^\dagger$ \\
   & ($10^{13} M_\odot$)   & ($10^{13} M_\odot$)   & ($10^{13} M_\odot$)\\\hline
  Coma$^\ddagger$ & 68.6$\pm$1.4 & $68.2^{+7.0}_{-6.4}$ & $33.3^{+2.8}_{-2.7}$\\ 
   Perseus & 58.7$\pm$1.3 & $48.1^{+4.5}_{-4.2}$ & $19.5^{+1.6}_{-1.6}$\\ 
  A2199 & 27.3$\pm$0.6 & $24.8^{+2.0}_{-1.9}$ & $10.9^{+0.9}_{-0.8}$\\ 
   A133 & 22.7$\pm$0.7 & $19.5^{+3.0}_{-2.8}$ & $7.2^{+0.9}_{-0.9}$\\ 
  AWM7 & 22.5$\pm$0.4 & $19.7^{+1.9}_{-1.8}$ & $10.1^{+0.9}_{-0.9}$\\ 
  HydraA$^\S$ & 25.6$\pm$0.7 & $18.8^{+2.3}_{-2.2}$ & $6.3^{+0.8}_{-0.7}$\\ 
   UGC03957 & 11.4$\pm$0.6 & $11.9^{+2.1}_{-2.0}$ & $5.2^{+1.0}_{-0.8}$\\ 
  ESO306-017 & 12.1$\pm$0.4 & $12.3^{+2.3}_{-2.1}$ & $4.9^{+0.8}_{-0.7}$\\ 
  A262E$^|$ & 9.9$\pm$0.3 & $8.9^{+0.9}_{-0.8}$ & $3.5^{+0.3}_{-0.3}$\\ 
   Virgo All & 10.7$\pm$0.3 & $9.2^{+1.1}_{-1.0}$ & $3.6^{+0.4}_{-0.4}$\\ 
  Virgo W$^\#$ & 8.3$\pm$0.2 & $8.9^{+1.1}_{-1.1}$ & $4.3^{+0.6}_{-0.5}$\\ 
  Antlia & 5.2$\pm$0.2 & $7.7^{+1.7}_{-1.5}$ & $3.0^{+0.5}_{-0.4}$\\ 
  MKW4 & 5.9$\pm$0.1 & $6.1^{+0.7}_{-0.6}$ & $3.5^{+0.3}_{-0.4}$\\ 
  NGC1550 & 3.9$\pm$0.1 & $4.5^{+0.4}_{-0.4}$ & $2.4^{+0.2}_{-0.2}$\\ 
  NGC741 & 2.1$\pm$0.1 & $2.5^{+0.9}_{-0.7}$ & $0.5^{+0.1}_{-0.1}$\\ \hline
  \end{tabular}\\
 \end{center}
 $^*$ $M_{500}$  derived from the pressure profiles using the Planck pressure model\\
 $^\dagger$ $M_{500}$ and $M_{2500}$ derived from the hydrostatic mass\\
 $^\ddagger$ Excluding the southwest subsector\\
 $^\S$ Excluding the southeast subsector\\
 $^|$ Excluding the three observations toward northeast\\
 $^\#$ Virgo west arm\\
\end{table}


\subsection{Hydrostatic mass}
\label{sec:he}

\begin{figure}[tb]
	\begin{center}
	\includegraphics[width=0.45\textwidth]{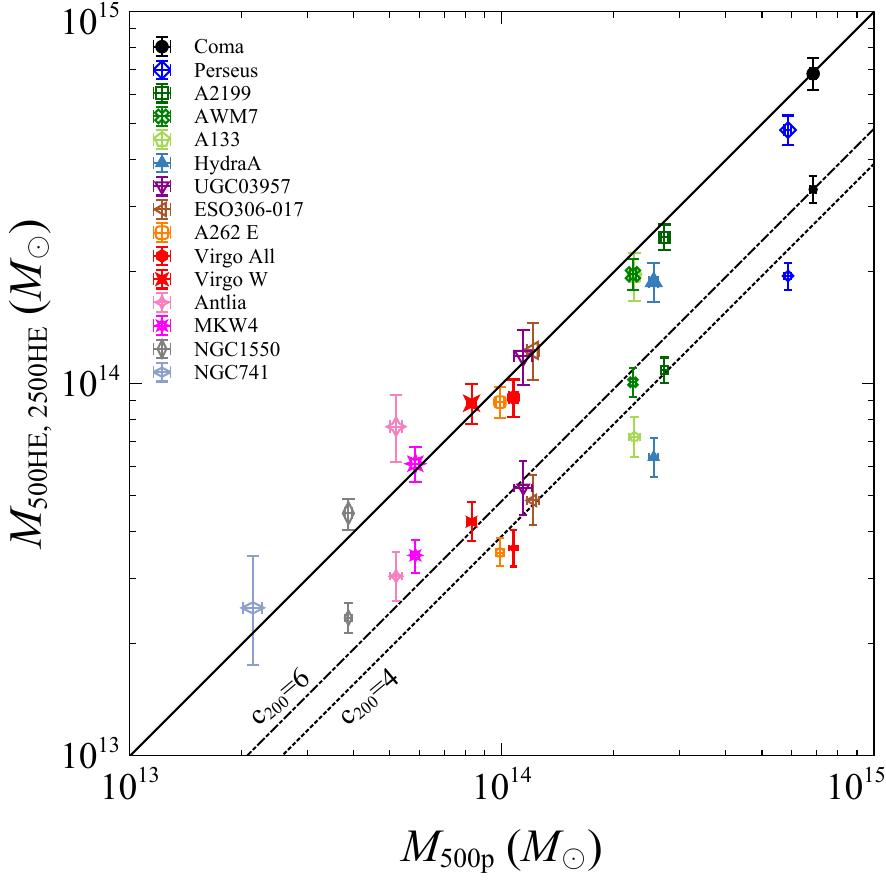}
		\caption{
			$M_{\rm 500HE}$ (larger symbols) and $M_{\rm 2500HE}$ (smaller sysmbols) are plotted against $M_{\rm 500p}$. 
			The solid line corresponds to equal mass, and dotted and dot-dashed lines correspond to $c_{200}=4$ and 6, respectively.
		 {Alt text: The scatter plot comparing the integrated hydrostatic masses within $r_{2500}$
		 and $r_{500}$ with $M_{500}$ derived from the pressure profiles, in solar mass units.}	
			}
		\label{fig:HE}
	\end{center}
\end{figure}

Assuming spherical symmetry and hydrostatic equilibrium,
the hydrostatic mass (hereafter $M_{\rm H.E.} (<R)$) within a three-dimensional radius $R$ is given by
\begin{equation}
M_{\rm H.E.}(<R) =\frac{kT(r)r }{\mu m_p G} \left( \frac{d {\rm ln} \rho_g(r) }{d {\rm ln} r} +  \frac{d {\rm ln} kT(r) }{d {\rm ln} r} \right)\\
\end{equation}
where $\rho_g$ is the mass density of the ICM, $G$, $k$, $\mu$, $m_p$, and $\mu$ are the gravitational constant, Boltzmann constant, mean molecular weight and proton mass, respectively. 
We adopt $\mu=0.62$ as the mean molecular weight.

Since the gNFW model reproduces the observed pressure profile reasonably well beyond 0.25 $r_{500}$, 
we used the best-fit pressure and electron density profiles to derive $M_{2500}$ and $M_{500}$ (hereafter $M_{\rm 2500 HE}$ and
$M_{\rm 500HE}$, respectively), assuming hydrostatic equilibrium. This method avoids
directly differentiating the temperature and gas density profiles.
For Perseus, A2199, AWM7, Hydra A, UGC03957, MKW4, and NGC1550, we used the gNFW pressure profile with a fixed $\beta_p$.  For the other systems, we used the best-fit gNWF pressure profiles, with both $M_{500p}$ and $\beta_p$ allowed to vary.
Table \ref{tb:mass} lists the resulting values for $M_{\rm 2500HE}$ and $M_{\rm 500HE}$ and
figure \ref{fig:HE} compares $M_{500p}$ with $M_{\rm 500HE}$. 
The two $M_{500}$ values agree well for lower mass systems, while $M_{500p}$ is slightly higher than $M_{\rm 500HE}$.
The ratios of $M_{\rm 2500HE}$ to $M_{500p}$ are close to those of $c_{200}\sim 4$ for massive clusters and $c_{200}\sim 6$ for lower systems, with some scatter.
These results are consistent with the average for the X-COP cluster sample, which reports $c_{200}=3.69^{+0.39}_{-0.36}$  for $M_{200}=8.6\times 10^{14}M_\odot$ \citep{Eckert2022}

A Subaru weak-lensing analysis for the Perseus cluster, assuming a single NFW halo,  gives $M_{200} = (6.82 \pm 1.76) \times 10^{14} M_\odot$ and $c_{200} = 3.83^{+0.03}_{-0.01}$ \citep{Perseus2024}, assuming the mass-concentration relation from Ishiyama et al. (2021). Adopting $c_{200} = 4$, this corresponds to $M_{500} = (4.7 \pm 1.2) \times 10^{14} M_\odot$, which is consistent with our results: $M_{500p} = (5.9 \pm 0.1) \times 10^{14} M_\odot$ and $M_{500\rm HE} = (4.8 \pm 0.5) \times 10^{14} M_\odot$ of the Perseus cluster.
For the Coma cluster, \citet{Okabecoma} reported a weak lensing $M_{500}$ of (4.4--7) $\times 10^{14} M_\odot$, which is also consistent with our estimate of $6.9 \times 10^{14} M_\odot$, despite the strong azimuthal dependence of the cluster in X-ray emission.

We do not present $M_{200}$ values because our data points beyond $r_{500}$ are limited, 
making it difficult to constrain the pressure and density slopes in the larger radii.
\citet{Matsushita2025} reported a possible steepening of the electron density profile beyond $r_{500}$
in the Perseus cluster with Suzaku, but our analysis is not sensitive enough to detect such slope changes.
Systematic uncertainties at the background level introduce additional errors, 
particularly affecting the determination of the outer profile slopes.
Furthermore, nonthermal pressure support may bias hydrostatic mass estimates in the cluster outskirts (e.g., \cite{Eckert2022}).
Although Hitomi and XRISM observations indicate negligible contributions from nonthermal pressure near the core regions
for clusters such as Perseus \citep{HitomiPerseus2018, XRISMPerseus}, A2029 \citep{XRISMA2029}, and Centaurus \citep{XRISMCentaurus}, its role at larger radii remains uncertain.

\subsection{Comparison with previous Suzaku studies}
Our electron density and temperature profiles are generally consistent with previous measurements out to $r_{500}$ \citep{Urban2014, Simionescu2013,UGC, Antlia,  Simionescu2017, A2199Suzaku, MKW4, A133Suzaku}. Beyond $r_{500}$, we sometimes obtain slightly lower electron densities and slightly higher temperatures, differences that are plausibly attributable to the adopted background treatment. In particular, outermost bins with apparent temperatures near $kT \sim 1$~keV in earlier work can be difficult to distinguish from emission associated with the soft X-ray emitting foreground components, which may bias soft-band temperatures low and densities high.

\section{Discussion}

We analyzed \textit{Suzaku} data for 14 galaxy clusters and groups, covering a mass range of $M_{500p}$ from $3\times10^{13}\,M_\odot$ to $7\times10^{14}\,M_\odot$, with coverage to the virial radius. 
Our sample includes lower-mass systems than the \textit{Planck} and XCOP cluster samples, allowing us to explore ICM properties on a broader mass scale.
Based on the presence of bright subclusters, we classify Coma, Hydra A, and Virgo clusters as ``merging'' systems
and the others as "relaxed" systems.
Hereafter, we adopt $M_{500}\equiv M_{500p}$, i.e., the mass values obtained from fitting the pressure profiles with only $M_{500p}$ allowed to vary.
To calculate the scale radii,  we adopted an NFW mass profile with $c_{200} =4$, which yields $r_{2500}=0.43~ r_{500}, r_{1000}=0.71 ~ r_{500}$, and $r_{200}= 1.53~r_{500}$.
Even if $c_{200}=6$, the resulting ratios of $r_\Delta$ change only marginally.

In this section, we exclude sub-cluster regions in the Coma and Hydra A clusters (the southwest sector in Coma and the southeast sector in Hydra A). 
For the Virgo cluster, we use both the western (W) arm and the azimuthally averaged profile (hereafter, ``Virgo All''). 
For A262, only the central regions and the eastern (E) arm are used, as the NE3 observation is heavily contaminated by SWCX emission.

\subsection{Emission measure profiles}
\label{sec:emd}

\begin{figure*}[tbhp]
	\begin{center}
	\includegraphics[width=0.45\textwidth]{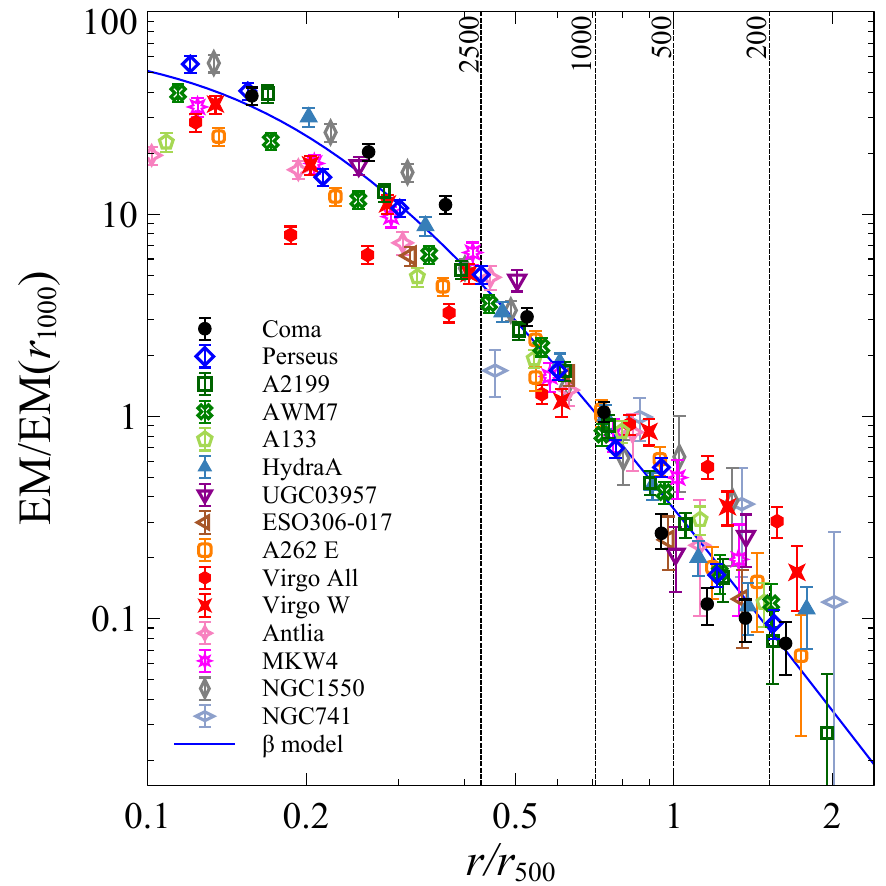}
		\includegraphics[width=0.45\textwidth]{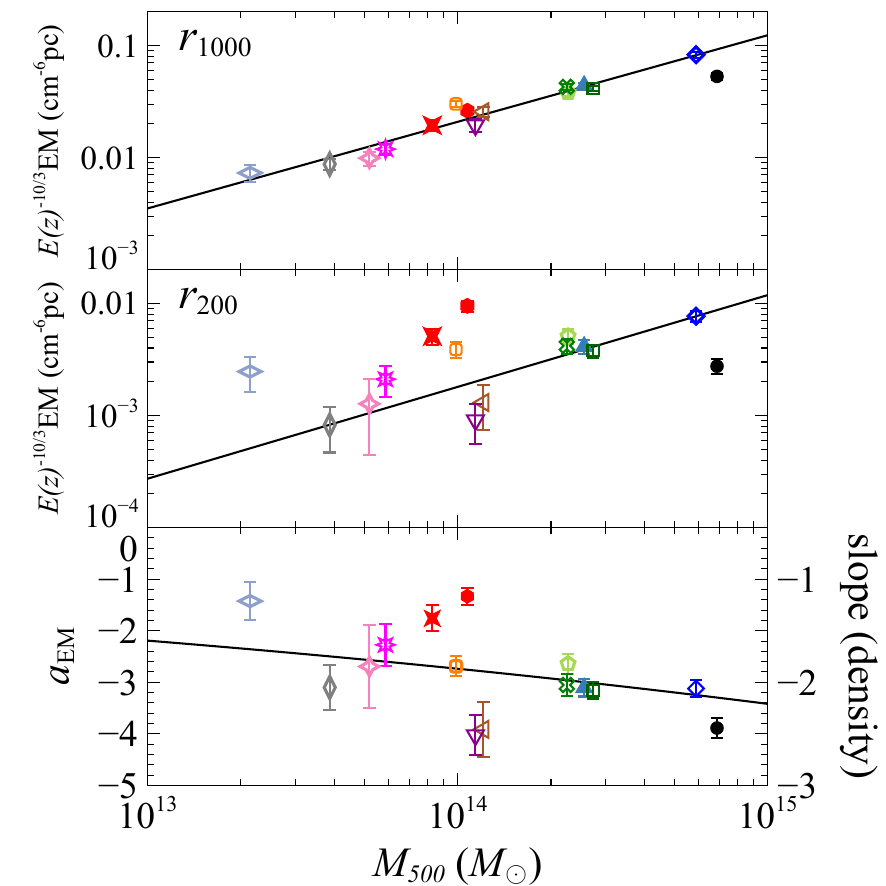}
		\caption{
		(left) Radial profiles of the emission measure, scaled by $r_{500}$ and EM$_{1000}$. The solid line shows  a representative $\beta$-model profile $r_c=0.21 r_{500}$ and $\beta=0.73$. (right) 
		 $E(z)^{-10/3} $EM($r_{1000}$), $E(z)^{-10/3} $EM($r_{200}$), and the radial emission measure slope beyond $0.4 r_{500}$ are plotted against $M_{500}$. 
		 {Alt text: (left)  A plot of scaled radial profiles of the emission measure with different colors and symbols representing different systems. (right) Three-panel figure showing emission measure
		 at $r_{1000}$ and $r_{200}$, as well as the radial slope, plotted against $M_{500}$ in units of solar mass. }}
		\label{fig:ems}
	\end{center}
\end{figure*}

As shown in Figure \ref{fig:em}, outside the core regions,  the emission measure profiles are well represented by a $\beta$-model,  indicating that they follow a power-law distribution at $r\gg r_c$.
We therefore fitted  the emission measure profiles beyond $r>0.4~r_{500}$ for each cluster using the following power-law form:

\begin{equation}
{\rm EM}(r)={\rm EM (r_{\rm \Delta})} \left(r/r_{\rm \Delta}\right)^{-a_{\rm EM}}
\end{equation}

where EM($r_{\rm \Delta}$) is the emission measure at $r_{\rm \Delta}$ and $a_{\rm EM}$ is the power-law slope.
In this analysis, we used EM$_{\rm HG08}$ for the inner regions and EM$_{\rm Z03}$ for the outer regions, and included systematic uncertainties of 10\% and $10^{-3} \rm{cm^{-6} pc}$ in each data point (sec \ref{sec:sys}).
The emission measure profiles, normalized by EM$(r_{1000})$ and scaled in radius by $r_{500}$, are
shown in the left panel of the figure \ref{fig:ems}.
Excluding the two merging clusters (Virgo and Coma),  and the lowest-mass system (the NGC 741 group), 
the relaxed clusters exhibit similar emission measure profiles with small scatter 
beyond 0.4 $r_{500}$, extending out to $\sim$ 1.5--2 $r_{\rm 500}$. 
For comparison, figure \ref{fig:ems} also shows a representative $\beta$-model with $ r_c=0.21~ r_{500}$ and $\beta=0.73$,
corresponding to an emission measure slope of 3.4 and a density slope of 2.2.
This $\beta$-model best fits the Perseus cluster, and the derived values of $r_c$ and $\beta$ are consistent with 
the best-fit $\beta$ model using $r>10'$ of the all arms of the Perseus data reported by \citet{Urban2014}

The resultant values of $E(z)^{-10/3}$EM ($r_{\rm 1000}$), $E(z)^{-10/3}$EM ($r_{\rm 200}$), and radial slope $a_{\rm EM}$ for each system are plotted against $M_{\rm 500}$ in the right panel of figure \ref{fig:ems}.
Again, excluding the NGC 741 group and the two merging clusters, these quantities exhibit clear correlations with 
$M_{\rm 500}$, and can be described by a power-law relation of the form, 

\begin{equation}
E(z)^{-10/3} {\rm EM}(r_\Delta)=A \left(M_{500}/10^{14}M_\odot\right)^{\alpha}
\end{equation}
The deviations from this best-fit relation among the relaxed clusters are relatively small, particularly at $r_{1000}$.
The best-fit normalization $A$ and slope $\alpha$ of the best-fit power-law relation for the 11 relaxed systems are listed in Table \ref{tb:EM}.
At $r_{\rm 1000}$ and $_{\rm 200}$, the emission measure scales with the system mass with
a slope of $\sim$0.8.
The radial slope of the emission measure beyond 0.4 $r_{\rm 500}$ is well described by 

\begin{equation}
\alpha_{\rm EM}=A \left(M_{500}/10^{14}M_\odot\right)^{\alpha}
\end{equation}
As listed in table \ref{tb:EM}, for the 11 relaxed systems, we obtain
 $A=-2.74 \pm0.10$, which corresponds to a density slope of $1.87\pm0.05$, with a only weak mass dependence ($\alpha=0.10\pm0.03$).
For massive systems such as Perseus, the corresponding density slope is $\sim$2.

\begin{table*}
	\caption{Scaling relations of emission measure for the 11 relaxed systems, where
	parameter$=A(M_{500}/10^{14}M_\odot)^\alpha$
}
	\label{tb:EM}
	\begin{center}
		\begin{tabular}{llllllllll}
			\hline
 parameter &  A   &      $\alpha$  & $\chi^2$/d.o.f\\\hline
$E(z)^{-10/3}$EM($r_{1000}$) (cm$^{-6}$pc)     & 0.021$\pm$0.001 & 0.78$\pm$0.03 & 45/9\\
$E(z)^{-10/3}$EM($r_{200}$) (cm$^{-6}$pc)  & 0.0018$\pm$0.0002 & 0.82$\pm$0.09 & 35/9\\
$a_{\rm EM}$   & -2.74$\pm$0.10 & 0.10$\pm$0.03 & 28/9\\\hline
\end{tabular}
\end{center}
\end{table*}

\subsection{Electron density profiles}
\begin{figure*}[tbhp]
	\begin{center}
	\includegraphics[width=0.45\textwidth]{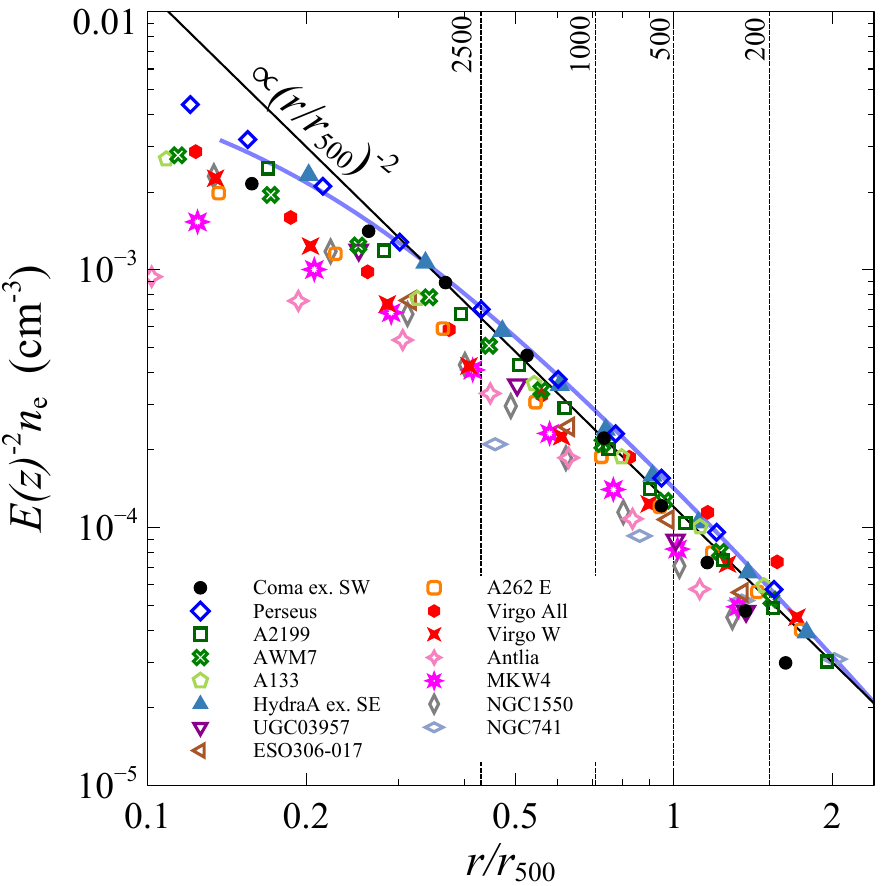}
	\includegraphics[width=0.45\textwidth]{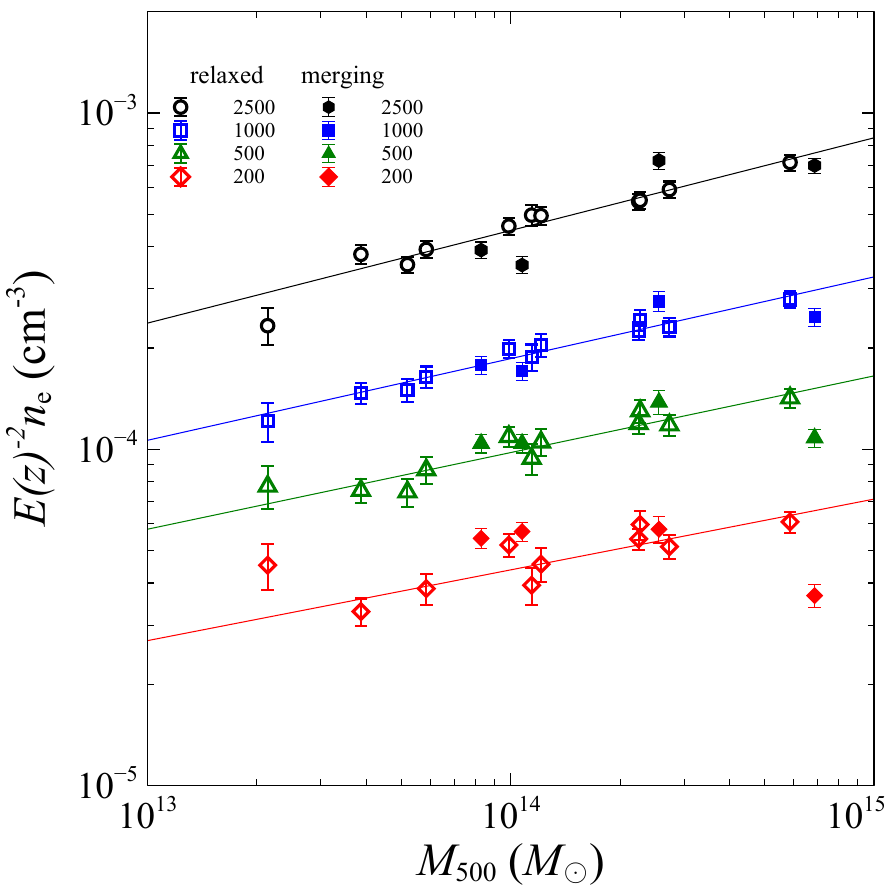}
			\caption{
		(left) Radial profiles of the electron density. The blue line shows the Vikhlinin profile that best fits the Perseus cluster. The solid line shows the power-law profile of $r^{-2}$	(right)  Electron densities at $r_{2500}$ (circles), $r_{1000}$ (squares), $r_{500}$ (triangles), and $r_{200}$ (diamonds), plotted against $M_{500}$. The solid lines show the best-fit power-law relations. Open and filled symbols correspond to the relaxed systems and the merging systems (Coma, Hydra-A, Virgo All, and Virgo W).
		 {Alt text: (left)  Scaled radial electron density profiles for different systems.  (right) Scatter plot showing the electron densities in units of cm$^{-3}$ at  $r_{2500}$, $r_{1000}$, $r_{500}$, and $r_{200}$, plotted against $M_{500}$ in units of solar mass. }
		}
		\label{fig:nes}
	\end{center}
\end{figure*}

\begin{table*}
	\caption{Scaling relations of electron density for the relaxed clusters, expressed as $E(z)^{-2}n_{\rm e}=A (M_{500}/10^{14}M_\odot)^\alpha$
}
	\label{tb:ne}
	\begin{center}
		\begin{tabular}{lllllllll}
			\hline
$\Delta$  & N$^*$ & A   (cm$^{-3}$) &      $\alpha$  & $\chi^2$/d.o.f\\\hline	
2500        & 11&   	$(4.45\pm 0.09)\times 10^{-4} $ & 0.28$\pm0.02$ & 9.6/9 \\		
1000        & 11 &    	$(1.86\pm 0.04)\times 10^{-4} $ & 0.24$\pm0.02$ & 3.3/9 \\	
500        &  11 &   	$(9.8\pm 0.3)\times 10^{-5} $ & 0.23$\pm0.03$ & 7.6/9 \\		
200        &  10$^\dagger$ & 	$(4.4\pm 0.2)\times 10^{-5} $ & 0.19$\pm0.03$ &15.3/8 \\		
 		\hline
\end{tabular}\\
\end{center}
$ ^*$:  The number of relaxed systems used to derive the relation.\\
$^\dagger$: Except for the Antlia cluster.\\
\end{table*}

Figure \ref{fig:nes} shows the radial profiles of the electron density, where the radius is normalized by  $r_{500}$. 
While the profiles appear similar in shape,  
smaller systems exhibit systematically lower electron densities at a given scale radius.
 Beyond $\sim  0.3  r_{500}$, the density slopes are consistent with $n_{\rm e}\propto (R/r_{500})^{-2}$.

\citet{Vikhlinin2006} proposed the following function for the ICM density profile, which we adopt for regions outside the core:
\begin{equation}
    n_p n_e = n_0^2 \,\frac{(R/r_c)^{-\alpha_n}}{\left[1 + (R/r_c)^2 \right]^{3\beta_n - \alpha_n/2}}
    \,\frac{1}{\left[1 + (R/r_s)^{\gamma_n} \right]^{\epsilon_n/\gamma_n}} .
\end{equation}
Here $n_p$ and $n_e$ are the proton and electron number densities, respectively; $n_0$ is the central electron density; $r_c$ is the core radius; $\beta_n$ and $\alpha_n$ describe the inner slopes; and $r_s$ and $\epsilon_n$ parameterize a steepening at larger radii. We fixed $\gamma_n=3$ and fitted this model to the electron-density profile of the Perseus cluster.
The fit reproduces the observed profile of the Perseus cluster well (Figure \ref{fig:nes}), with best-fit parameters
$\alpha_n=0.58$, $\beta_n=0.62$, $r_c=0.19\,r_{500}$, $r_s=1.0\,r_{500}$, and $\epsilon_n=1.06$.
Both the XCOP sample \citep{XCOP2019} and the \textit{Chandra} sample \citep{Morandi2015} report a steepening of the density slope at large radii; the agreement of the Perseus profile with the Vikhlinin form is consistent with such steepening. However, because the number of radial bins at $r>r_{500}$ is limited, we do not pursue this topic further in this context.

Based on the classical self-similar scaling relation of clusters of galaxies,  the
gas densities at a given scaled radius are expected to be independent of the cluster mass.
To examine this, 
the right panel in figure \ref{fig:nes} plots $E(z)^{-2}n_{\rm e}$ at $r_{2500}$, $r_{1000}$, $r_{500}$, and $r_{200}$. 
At each $r_\Delta$, the relaxed clusters show tight correlations between the electron density
and $M_{500}$.
We fitted the relation between  $E(z)^{-2}n_{\rm e}$  and $M_{500}$ using a power-law:
\begin{equation}
E(z)^{-2}n_{\rm e}=A \left(M_{500}/10^{14}M_\odot \right)^\alpha
\end{equation}
with the best-fit parameters $A$ and $\alpha$ listed in Table \ref{tb:ne}.
This power-law provides an excellent fit at all radii,  with $\chi^2/d.o.f$ values close to unity, except for the NGC 741 at $r_{200}$.
Given the relatively small sample size and measured scatter, 
we do not attempt to derive the intrinsic scatter in these relations.
At $r_{2500}$, the relaxed clusters follow  $n_{\rm e}\propto M_{500}^{0.28\pm 0.02}$, 
deviating from the expected self-similar scaling, as previously reported by Chandra and XMM observations \citep{Vikhlinin2009, Sun2009}.
The mass slope $\alpha$ gradually decreases with radius: $\sim$ 0.24 at $r_{1000}$ and  $\sim$ 0.23 at $r_{500}$.
Even at $r_{200}$, the correlation persists, 
 with a slightly flatter slope of $n_{\rm e} \propto M_{500}^{0.19 \pm 0.03}$,
indicating that the gas density still retains a weak dependence on the mass of the system in the outskirts.
In contrast, the merging clusters exhibit significant deviations.
At $r_{200}$, for example, Coma has a lower $n_{\rm e}$ than expected from the mass trend.
 
\citet{Sun2012} compiled the electron densities at scaled radii out to $r_{500}$ for a large number of samples.
 Their results are consistent with ours at $r_{2500}$.
 However, their sample does not show mass dependence at $r_{500}$,
  although our $n_{\rm e}$ values remain within their observed scatter.
At $r_{500}$ and $r_{200}$, the Perseus cluster ($M_{500}=6\times10^{14} M_\odot$) has electron densities of
$n_{\rm e}\sim1.5\times10^{-4} \mathrm{cm^{-3}}$ and $6\times10^{-5} \mathrm{cm^{-3}}$, respectively.
These values are close to the XCOP median densities of $1.4\times10^{-4} \mathrm{cm^{-3}}$ and $5\times10^{-5} \mathrm{cm^{-3}}$ for clusters with $M_{500}$ in the range $3$--$9\times10^{14} M_\odot$ \citep{XCOP2019}.
They are also close to the average values from a large \textit{Chandra} sample of massive clusters:
$1.3\times10^{-4} \mathrm{cm^{-3}}$ at $r_{500}$ and $5\times10^{-5} \mathrm{cm^{-3}}$ at $r_{200}$ \citep{Morandi2015}.

In our analysis, we excluded bright subclusters, not only the prominent ones in merging systems but also those identified in A133 and other clusters.
Moreover, our use of best-fit single- or double-$\beta$ models mitigates the impact of substructures.
With Suzaku, small clump candidates cannot be excised as effectively as with XMM-Newton or Chandra; however, their contribution to gas-density overestimation appears modest.
For example, \citet{Walker2022} analyzed surface-brightness fluctuations in the western sector of Perseus from $20'$ to $80'$ with XMM-Newton and found clumping factors up to 1.08 beyond $r_{500}$.
Similarly, \citet{Virgoclumps2021} reported relatively mild clumping in the northern arm of Virgo using \textit{XMM-Newton}.
In A133, most of the small clump candidates identified with Chandra are background sources, implying only a minor effect on gas clumping \citep{A133Suzaku}.

In summary, the electron density profiles of relaxed clusters show a weak but persistent dependence on system mass beyond $r_{2500}$, while merging clusters exhibit some deviations, highlighting the impact of dynamical state on the ICM structure.

\subsection{Gas fraction}

\begin{table*}
\caption{Scaling relations of the gas fraction for the relaxed clusters, expressed as $f_{\rm gas}=A (M_{500}/10^{14}M_\odot)^\alpha$
}
	\label{tb:fgas}
	\begin{center}
		\begin{tabular}{llllllllll}
			\hline
$\Delta$  & N$^*$ & A   &      $\alpha$  & $\chi^2$/d.o.f\\\hline
2500 &      11  &	0.064$\pm$0.001 & 0.34$\pm$0.02& 55.8/9\\
1000 &      11 &  	0.080$\pm$0.002 & 0.29$\pm$0.02 & 13.9/9\\
500        &  11 &   	0.092$\pm$0.002 & 0.27$\pm$0.02 & 5.5/9\\
200        &  10$^\dagger$ & 	0.111$\pm$0.003 & 0.23$\pm$0.03 & 4.8/8 \\	
 		\hline
\end{tabular}
\end{center}
$ ^*$:  The number of relaxed systems used to derive the relation.\\
$^\dagger$: Except for the Antlia cluster\\
\end{table*}

\begin{figure*}[htpb]
	\begin{center}
	\includegraphics[width=0.45\textwidth]{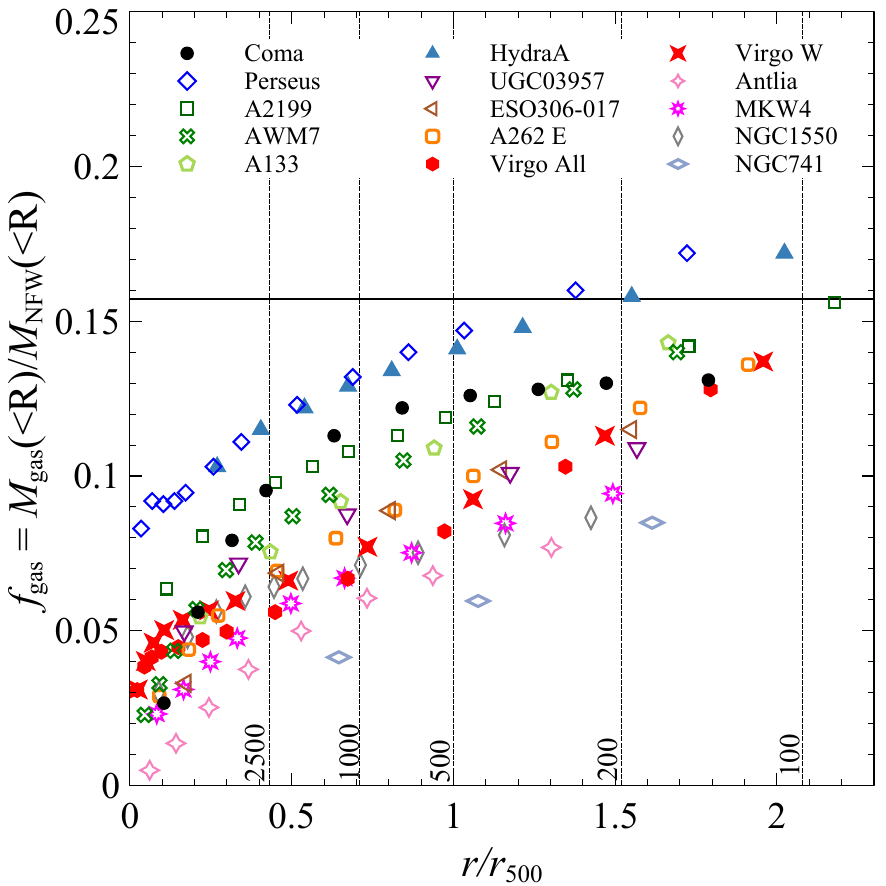}
	\includegraphics[width=0.45\textwidth]{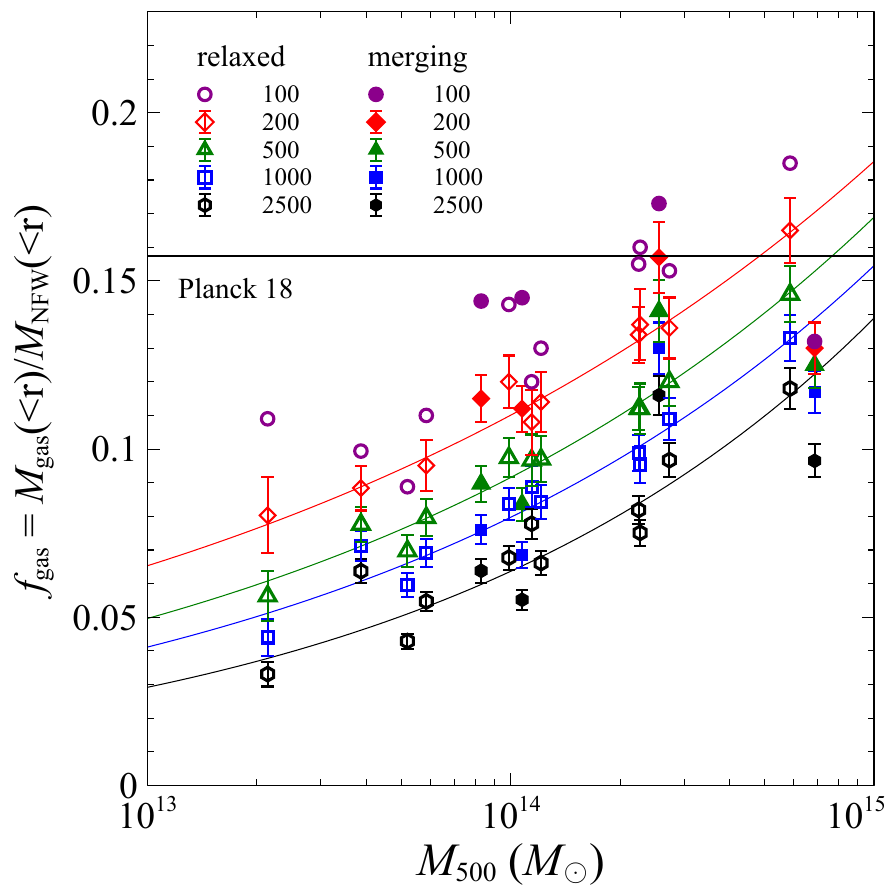}
		\caption{
	(left)		Radial profile of integrated gas fraction, the ratio of the integrated gas mass to the NFW mass with $c_{200}$=4. (right) The integrated  gas fraction at $r_{2500}, r_{1000}, r_{500}$, and $r_{200}$, plotted against $M_{500}$.	 Those of $r_{100}$ are extrapolated values.
	The horizontal solid lines represent the cosmic baryon fraction as determined by \citet{Planck18}.
	 {Alt text: (left)  Integrated gas fraction for different systems.  (right) Scatter plot showing the gas fraction at  $r_{2500}$, $r_{1000}$, $r_{500}$, and $r_{200}$, plotted against $M_{500}$ in units of solar mass.
	 The extrapolated values at $r_{100}$ are also shown.  }
	}
		\label{fig:fgas}
	\end{center}
\end{figure*}

Galaxy clusters form through the accumulation of dark matter and baryons, so their baryon fraction is expected to approach the cosmic mean. 
In massive clusters, most baryons reside in the ICM, with stars contributing a smaller portion. 
Therefore, if clusters behave as "closed boxes," their baryon fraction can serve as a useful cosmological probe.
However, X-ray observations have shown that the gas fraction varies with radius and system mass.
At $r_{500}$, massive clusters typically have gas fractions of  $\sim 0.1$--0.15, 
while poorer systems ($M_{500}\sim 10^{13}M_\odot$) show values as low as  $\sim$ 0.05 (e.g \cite{Sun2012}).
Some Suzaku studies have even reported that the gas-to-hydrostatic mass ratios exceed the cosmic mean beyond $r_{500}$  (\cite{Kawaharada2010}, \cite{Ichikawa2013}).  \citet{Simionescu2011} proposed
that gas clumpings at cluster outskirts cause an overestimation of the gas density and, hence, an overestimation of the gas fraction.
However, such excesses are not seen when compared to weak lensing masses (\cite{Kawaharada2010}, \cite{Ichikawa2013}).
By combining X-ray data with XMM and SZ measurements from Planck, and excluding possible clump candidates
using a much better PSF of XMM, 
the X-COP project measured the gas fraction out to $r_{200}$ close to the cosmic mean \citep{Eckert2019}.

We define the gas fraction, $f_{\rm gas}$,  as the ratio of the integrated gas mass within $R$, $M_{\rm gas}(<R)$,
to the total mass assuming an NFW profile, $M_{\rm NFW}(<R)$, assuming $c_{\rm 200}=4$ (section \ref{sec:he}).
This is because weak-lensing studies have shown that cluster mass distributions are well described by the NFW model (e.g. \cite{Okabe20102}).
In addition, it is difficult to determine the hydrostatic mass beyond $r_{500}$ due to our limited data points for evaluating pressure gradients.
Thus, we adopted the NFW mass profile to derive $f_{\rm gas}$.
We calculated the integrated gas mass from the electron density ($n_{\rm e}$) profiles.
We include a 5\%  systematic uncertainty in the gas mass to reflect the 10\%  systematic uncertainty in the emission measure.
The radial profiles of $f_{\rm gas}$ are plotted in the left panel of figure \ref{fig:fgas}.
Throughout this study, we adopt the cosmic baryon fraction from the Planck 2018 results \citep{Planck18}.
 $f_{\rm gas}$ increases with radius, reaching the cosmic mean near $r_{\rm 200}$ for the Perseus and Hydra A clusters.
For the Abell 2199 cluster,  $f_{\rm gas}$ reaches the cosmic mean around $2\times r_{200}$ or $\sim r_{\rm 100}$.
In contrast, relaxed lower-mass systems and the Coma cluster exhibit
$f_{\rm gas}$ values below the cosmic mean across the entire radial range.

The right panel of figure \ref{fig:fgas} plots $f_{\rm gas}$ at $r_{2500}$, $r_{1000}$, $r_{500}$, and $r_{200}$, 
as a fanction $M_{500}$.
At each overdensity radius $r_\Delta$,  $f_{\rm gas}$  positively correlates with system mass and shows small scatter.
These relations are well fitted by:
\begin{equation}
f_{\rm gas}=\frac{ M_{\rm gas}(<R)}{M_{\rm NFW}(<R)}=A (M_{500}/10^{14} M_\odot)^\alpha
\end{equation}
with best-fit values listed in Table \ref{tb:fgas}.
The normalization, $A$,  representing the gas fraction at $M_{500}=10^{14}M_\odot$,  increases with radius from 0.06 at $r_{\rm 2500}$ to 0.11 at $r_{\rm 200}$.
The slope $\alpha$ becomes shallower at larger radii,  decreasing from 0.34 at $r_{2500}$ to 0.23 at $r_{200}$.
 Even at $r_{200}$, poor clusters with  $M_{500}\sim 10^{14}M_\odot$ show gas fractions significantly below the cosmic mean.
Our $f_{\rm gas}$ values at $r_{\rm 500}$ lie within the scatter of previous studies (e.g \cite{Sun2009}, \cite{Sun2012}, \cite{Vikhlinin2009},\cite{Gonz2013}, \cite{Akino2022}). 
For example, Perseus has $f_{\rm gas} = 0.146\pm0.008$ at $r_{500}$ and 0.165$\pm$0.010 at $r_{200}$, consistent within 10 percent with the X-COP clusters of similar masses \citep{Eckert2022} and
the Chandra sample \citep{Morandi2015}.

We also estimate $f_{\rm gas}$ at $r_{100}$ extrapolating the best-fit $\beta$-model profiles (figure \ref{fig:fgas}).
Since the density profile is unlikely to become flatter at larger radii, this assumption likely gives an upper limit on the gas fraction at $r_{100}$.
Under this assumption, clusters with $M_{500}\sim$ a few $\times 10^{14}M_\odot$ may reach the cosmic mean at $r_{100}$,
while lower-mass systems with $M_{500}\sim$ several $\times 10^{13} M_\odot$ still show $f_{\rm gas}$$\sim$0.1.

In addition to the ICM, a non-negligible fraction of baryons in clusters exists in the form of stars. The stellar mass fraction $f_{\rm star}$ also depends on the system mass.
\citet{Lagana2013},  using data from XMM-Newton, Chandra, and SDSS, reported that the stellar fraction decreases with system mass: $f_{\rm star} \sim 1\%$ for massive galaxy clusters with $M_{500} > $several $\times 10^{14} M_\odot$, and a few percent for groups with $M_{500} \sim 1$--$5 \times 10^{13} M_\odot$. 
\citet{Budzynski2014} revisited the stellar fraction in systems with $M_{500} > 5 \times 10^{13} M_\odot$ and found that, including intracluster light, the stellar mass fraction shows a weak negative dependence on system mass, yielding values of 1--1.3\%.
Consequently, the total baryon fraction, $f_{\rm baryon} = f_{\rm gas} + f_{\rm star}$, remains below the cosmic mean in low-mass systems. 
In massive clusters, however, the combined baryon fraction approaches the cosmic value near $r_{200}$. 
Our results show that $f_{\rm gas}$ increases with radius and mass, reaching $\sim$0.15 at $r_{500}$ for massive clusters. 

Cosmological simulations show that, in the absence of AGN feedback, the baryon fraction within the virial radius (i.e., $\sim r_{200}$) approaches the cosmic mean, and clusters behave nearly as ``closed-box'' systems. 
When AGN feedback is included, a clear mass dependence emerges: massive clusters retain baryon and gas fractions close to the cosmic mean, whereas lower-mass systems remain baryon-poor because feedback suppresses gas retention more efficiently in shallower potentials 
(e.g., \cite{OWLS2014,McCarthy2017,Illustris2023}).
With AGN feedback, these simulations predict gas fractions of $\sim$0.07--0.1 for $10^{14}\,M_\odot$ systems at $r_{200}$, rising to $\sim$0.13--0.15 for the most massive clusters; the exact depletion depends on the adopted AGN feedback strength. 
Our measured gas mass fractions follow the same trends and lie within these ranges, supporting the view that non-gravitational feedback, primarily AGN, regulates baryon retention out to the virial boundary.

\section{Conclusion}

We analyzed Suzaku observations of 14 nearby galaxy clusters and groups, extending to $\sim r_{200}$, and derived radial profiles of electron density, temperature, and Fe abundance across a wide mass range by carefully modeling the soft X-ray background, including the 0.8 keV Galactic component, as well as NXB and CXB levels. Beyond $0.4,r_{500}$, relaxed systems exhibit remarkably self-similar emission measure, density, and pressure profiles with a small scatter of $\sim$20\%, and power-law fits yield density slopes of $\sim$2. The electron density scales tightly with total mass, with slopes of 0.28 at $r_{2500}$ and 0.16 at $r_{200}$. The gas fraction increases with both radius and mass, reaching $\sim$0.15 near $r_{200}$ in massive clusters ($M_{500} \gtrsim 5 \times 10^{14},M_\odot$), while remaining as low as $\sim$0.05--0.1 in lower-mass systems.

 \appendix

\chapter{Observation logs}
\label{sec:obslog}
The observation logs of the clusters of galaxies and Lockman Hole data analyzed in this paper are shown in tables \ref{tb:suzakuobslog} and \ref{tb:LHobs}, respectively.

%
%
\begin{longtable}{ll}
  \caption{$Suzaku$ observation logs.}
  \label{tb:suzakuobslog}
      \hline

\endhead
\hline
\endfoot
  \multicolumn{2}{@{}l@{}}{\hbox to 0pt{\parbox{100mm}{\footnotesize
%
       }    \hss}}
\endlastfoot

\hline Target & Sequence Number (Exposure time in units of ks.)  \\ 
\hline
Coma & 801044010 (71.6)
801097010 (159.8)
802047010 (23.9)
802048010 (29.3)
802082010 (46.0)
802083010 (25.3)\\ &
802084010 (28.0)
803051010 (163.2)
805079010 (82.4)
806020010 (37.7)
806021010 (18.8)
806022010 (24.3)\\ &
806023010 (42.5)
806024010 (24.8)
806025010 (16.1)
806030010 (9.2)
806031010 (7.5) 
806032010 (4.7)\\ &
806032020 (9.2)
806033010 (18.2)
806034010 (12.1)
806035010 (10.1) 
806036010 (9.5)
806037010 (11.1) \\ &
806038010 (13.8)
806039010 (6.3)
806040010 (8.1) 
806040020 (16.3)
806041010 (18.2)
806042010 (16.3) \\ &
806043010 (13.6)
806044010 (9.0) 
806045010 (14.1) 
806046010 (11.9)
806047010 (6.6)
806048010 (13.8)\\ &
806049010 (10.5)
806050010 (14.7)
806051010 (12.1)
808018010 (26.4)
808019010 (20.3)
808020010 (16.2)\\ &
808021010 (23.7)
808022010 (18.4)
808090010 (11.9)
808091010 (14.4) \\\hline
Perseus &
800010010 (35.3)
801049010 (24.9)
801049020 (27.1)
801049030 (29.6)
801049040 (7.6)
804056010 (6.9) \\ &
804057010 (11.8)
804058010 (11.6)
804059010 (17.7)
804060010 (21.6)
804061010 (27.6)
804062010 (26.9)\\ &
804063010 (13.1)
804064010 (9.7)
804065010 (11.8)
804066010 (20.9)
804067010 (21.7)
804068010 (30.0)\\ &
804069010 (29.7)
805045010 (26.0)
805046010 (17.0)
805047010 (16.2)
805048010 (13.8)
805096010 (8.0)\\ &
805097010 (10.1)
805098010 (6.6)
805099010 (9.3)
805100010 (9.2)
805101010 (14.8)
805102010 (13.0)\\ &
805103010 (6.4)
805104010 (6.9)
805105010 (11.0)
805106010 (9.7)
805107010 (7.7)
805108010 (12.3)\\ &
805109010 (15.1)
805110010 (8.8)
805111010 (6.5)
805112010 (12.8)
805114010 (6.9)
805115010 (9.7)\\ &
805116010 (12.9)
806099010 (11.1)
806100010 (8.7)
806101010 (9.5)
806102010 (7.7)
806103010 (9.9)\\ &
806104010 (12.8)
806105010 (8.4)
806106010 (12.9)
806107010 (14.7)
806108010 (10.0)
806109010 (7.0)\\ &
806110010 (10.0)
806111010 (10.4)
806112010 (10.6)
806113010 (9.2)
806114010 (8.9)
806115010 (11.6)\\ &
806116010 (10.6)
806117010 (9.8)
806118010 (13.6)
806119010 (15.8)
806120010 (8.3)
806121010 (7.4)\\ &
806122010 (9.9)
806123010 (9.8)
806124010 (9.5)
806125010 (5.7)
806126010 (8.3)
806127010 (10.3)\\ &
806128010 (10.3)
806129010 (6.8)
806130010 (14.8)
806131010 (14.0)
806132010 (7.5)
806133010 (8.2)\\ &
806134010 (11.1)
806135010 (9.2)
806136010 (6.8)
806137010 (10.2)
806138010 (9.6)
806139010 (8.6)\\ &
806140010 (7.1)
806141010 (11.1)
806142010 (15.8)
806143010 (10.3)
806144010 (10.4)
806145010 (12.7)\\ &
806146010 (7.3)
807019010 (11.9)
807020010 (22.1)
807021010 (16.3)
807022010 (21.8)
807023010 (11.5)\\ \hline
Abell 2199 &
801056010 (18.9)
801057010 (23.0)
801058010 (18.9)
801059010 (18.3)
801060010 (23.1)
805042010 (52.4)\\ &
806147010 (19.4)
806148010 (13.0)
806149010 (20.3)
806150010 (12.8)
806151010 (19.7)
806152010 (23.0)\\ &
806153010 (22.4)
806154010 (24.9)
806155010 (25.1)
806156010 (28.0)
806157010 (19.6)
806158010 (20.7)\\ &
806159010 (25.8)
806160010 (24.9)
806161010 (26.0)
806162010 (20.5)
808050010 (39.5)
808051010 (37.3)\\ \hline
Abell 133 &
805019010 (38.8)
805020010 (41.9)
805021010 (44.2)
805022010 (41.2)
808081010 (47.4)
808082010 (42.5)\\&
808083010 (45.4)
808084010 (46.8)\\ \hline
AWM7 &
801035010 (15.7)
801036010 (33.7)
801037010 (34.4)
802044010 (74.4)
802045010 (27.5)
802045020 (76.8)\\&
806008010 (32.8)
806009010 (30.0)
806010010 (29.3)
808023010 (9.9)
808024010 (31.8)
808025010 (14.0)\\&
808026010 (28.2)
808027010 (20.2) \\ \hline
Hydra A &
805007010 (33.9)
805008010 (33.1)
807087010 (18.7)
807088010 (34.3)
807089010 (34.7)
807090010 (32.6)\\&
807091010 (14.5) \\ \hline
UGC 03957 &
801072010 (9.3)
806091010 (37.8)
806092010 (43.4)
806093010 (43.9)
806094010 (42.3)
\\ \hline
ESO 0306-017 &
805075010 (24.7)
805076010 (58.3)\\ \hline
Abell 262 &
802001010 (33.0)
802079010 (49.6)
802080010 (48.0)
804049010 (39.1)
808108010 (21.3)
808109010 (28.2)\\&
808110010 (36.1)
808111010 (29.6)
808112010 (22.5)
808113010 (28.7)
808114010 (34.7)
808115010 (30.6)\\ \hline
Virgo &
701037010 (17.0)
800017010 (109.3)
801038010 (87.3)
801039010 (45.1)
801064010 (101.2)
801065010 (184.4)\\&
803043010 (81.0)
803069010 (58.4)
803070010 (21.6)
805054010 (91.7)
806060010 (19.3)
806061010 (15.8)\\&
806062010 (20.3)
806063010 (22.8)
806064010 (34.0)
806065010 (33.0)
807094010 (9.2)
807095010 (11.7)\\&
807096010 (8.0)
807097010 (13.4)
807098010 (16.5)
807099010 (17.9)
807100010 (23.1)
807101010 (17.8)\\&
807102010 (26.0)
807103010 (24.7)
807104010 (21.0)
807105010 (25.8)
807106010 (8.6)
807107010 (7.7)\\&
807108010 (12.0)
807109010 (12.5)
807110010 (15.2)
807111010 (18.9)
807112010 (15.3)
807113010 (18.1)\\&
807114010 (18.6)
807115010 (12.2)
807116010 (12.0)
807117010 (12.7)
807118010 (12.5)
807119010 (11.3)\\&
807120010 (9.8)
807121010 (14.8)
807122010 (14.1)
807123010 (17.2)
807124010 (11.8)
807125010 (17.1)\\&
807126010 (14.7)
807127010 (10.3)
807128010 (15.4)
807129010 (15.4)
807130010 (20.0)\\ &
807131010 (16.8)
807132010 (17.6)
807133010 (18.7)
808045010 (84.5)
808116010 (8.6)
808117010 (10.1)\\ &
808118010 (9.8)
808119010 (9.0)
808120010 (11.8)
808121010 (10.5)
808122010 (15.4)
808123010 (15.4)\\&
808124010 (15.1)
808125010 (18.4)
808126010 (19.6)
808127010 (16.6)
808128010 (18.4)\\ \hline
Antlia &
802035010 (55.5)
807066010 (20.8)
807067010 (21.3)
807068010 (19.0)
807069010 (35.8)
807070010 (38.9)\\ &
807071010 (37.7)\\ \hline
MWK~4 &
805081010 (68.4)
805082010 (72.8)
808065010 (83.0)
808066010 (28.2)
808067010 (90.3)
809062010 (82.2)\\ \hline
NGC~1550 &
803017010 (72.5)
803018010 (35.0)
803046010 (54.0)
808060010 (47.6)
808061010 (25.3) \\ \hline
NGC~741 &
804052010 (10.8)
804052020 (10.9)
804052030 (7.4)
804052040 (11.0)
804053010 (11.3)
804054010 (8.3)\\ &
804054020 (11.3)
804054030 (9.4)
804054040 (8.8)
804055010 (6.7)
804055020 (10.2) \\\hline
\end{longtable}

\begin{table*}
	\caption{$Suzaku$ Lockman Hole observation logs.}
	\begin{center}
		\label{tb:LHobs}
		\begin{tabular}{llll}
			\hline
			Sequence &  Date-Obs.\footnotemark[$*$] & (R.A., decl.)\footnotemark[$\dagger$]  &  Exposure\footnotemark[$\ddagger$]  \\
			Number & & (J2000.0) & (ksec) \\ \hline
			101002010 & 2006-05-17T17:44:06 & $\timeform{10h51m44.78s},\timeform{+57D15'20.52"}$ & 71.7 \\
			102018010 & 2007-05-03T23:12:08 & $\timeform{10h51m42.17s},\timeform{+57D15'29.16"}$ & 85.4 \\
			103009010 & 2008-05-18T11:07:29 & $\timeform{10h51m44.86s},\timeform{+57D15'16.56"}$ & 69.4 \\
			104002010 & 2009-06-12T07:17:40 & $\timeform{10h51m45.05s},\timeform{+57D15'17.64"}$ & 64.5 \\
			105003010 & 2010-06-11T07:29:06 & $\timeform{10h51m45.17s},\timeform{+57D15'02.52"}$ & 70.1 \\
			107001010 & 2012-05-05T21:21:52 & $\timeform{10h51m40.73s},\timeform{+57D15'17.28"}$ & 31.8 \\
			108001010 & 2013-11-06T23:05:05 & $\timeform{10h51m46.66s},\timeform{+57D16'31.44"}$ & 34.4 \\
			109014010 & 2014-11-30T00:17:59 & $\timeform{10h51m45.65s},\timeform{+57D16'40.80"}$ & 34.2 \\
			\hline
			\multicolumn{4}{@{}l@{}}{\hbox to 0pt{\parbox{180mm}{\footnotesize
						\par\noindent
						\footnotemark[$*$] Start date of the observation written in the event FITS files as DATE-OBS keyword.      
						\par\noindent
						\footnotemark[$\dagger$] The nominal position of the observation written in the event FITS files as RA\_NOM and DEC\_NOM  keywords. 
						\par\noindent
						\footnotemark[$\ddagger$] After data screening. 
					}    \hss}}
		\end{tabular}
	\end{center}
\end{table*}

\chapter{Non X-ray Background}
\label{sec:nxb}

\begin{figure}[htpd]
	\begin{center}
		\includegraphics[width=0.45\textwidth,clip]{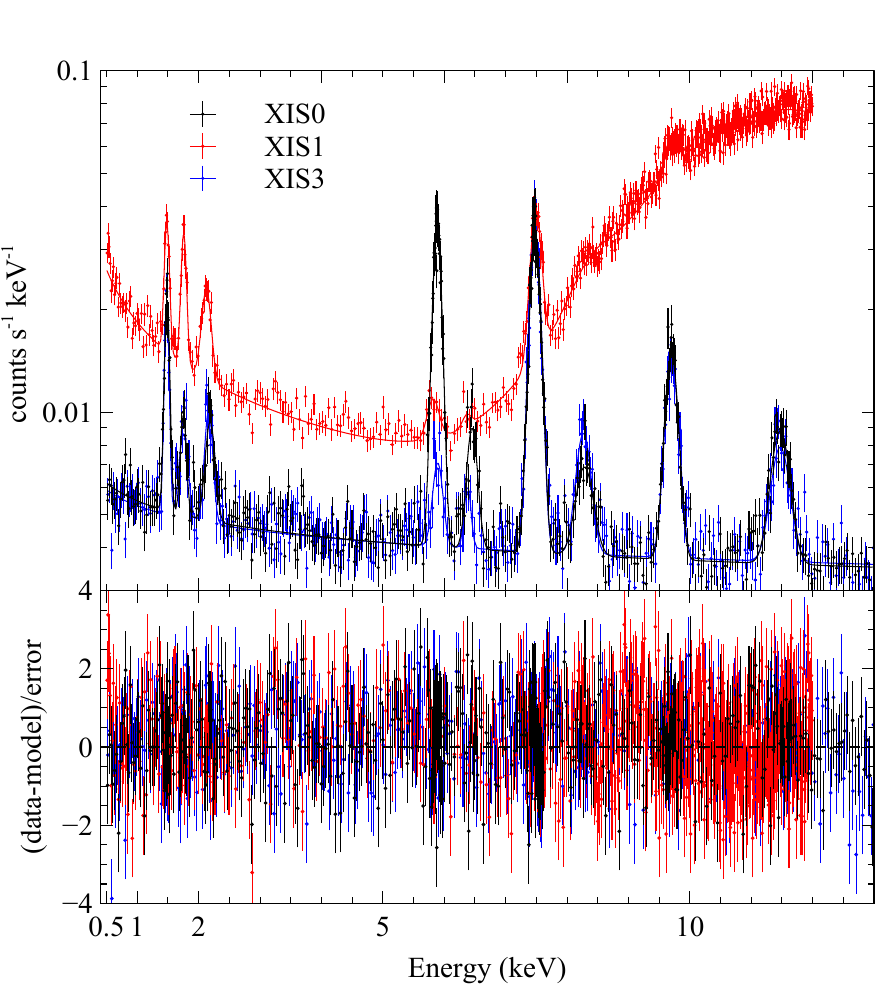}
		\caption{
		The NXB spectra of XIS0 (black),  XIS1 (red), and XIS3 (blue) spectra for the Lockman Hole data with obsid=101002010, fitted with a power-law and Gaussians using the diagonal responses. 	 The bottom panel shows the residuals of the fits.
		{Alt text: Two-panel plot showing the three spectra plotted with different colors and marks, and
		a logarithmic vertical axis and the residuals plotted against the X-ray energy in units of kiloelectron volt.}
		}
		\label{fig:specnxb}
	\end{center}
\end{figure}

To model the NXB components, we first fitted the NXB spectra of each XIS detector and individual observations using an RMF file that includes only diagonal components.
The NXB spectra were modeled with a power-law and nine Gaussians representing the following fluorescent lines: Al-K$\alpha$, Si-K$\alpha$, Au-M$\alpha$, Mn-K$\alpha$, Mn-K$\beta$, Ni-K$\alpha$, Ni-K$\beta$, Ni-K$\beta$,  Au-L$\alpha$, and Au-L$\beta$ (\cite{Tawa2008}).
Additionally,  for the XIS1 detector, we included a broad Gaussian component centered at 11.4 keV with $\sigma$=1.9 keV to reproduce the continuum above 7 keV. 
Figure \ref{fig:specnxb} shows representative X-ray spectra from a Lockman hole observation (obsid=101002010), illustrating that this model
 reasonably reproduces the observed NXB features.
The Mn-K$\alpha$ and K$\beta$ lines scattered from the calibration sources are stronger in the XIS0 detector than in the others.
In this observation, the best-fit line central energies of the Gaussian component modeling Mn-K$\beta$ are 6.45$\pm$0.01 keV for XIS0 and 6.38$\pm$0.02 keV for XIS3,
while the Mn-K$\alpha$ line appears at 5.89 keV,  consistent with the theoretical value.
This discrepancy in the Mn-K$\beta$ line energy may indicate contamination from the Fe-K$\alpha$ fluorescent line at 6.4 keV, especially in the NXB spectra of XIS3.

\begin{figure*}[htpd]
		\includegraphics[width=0.45\textwidth,clip]{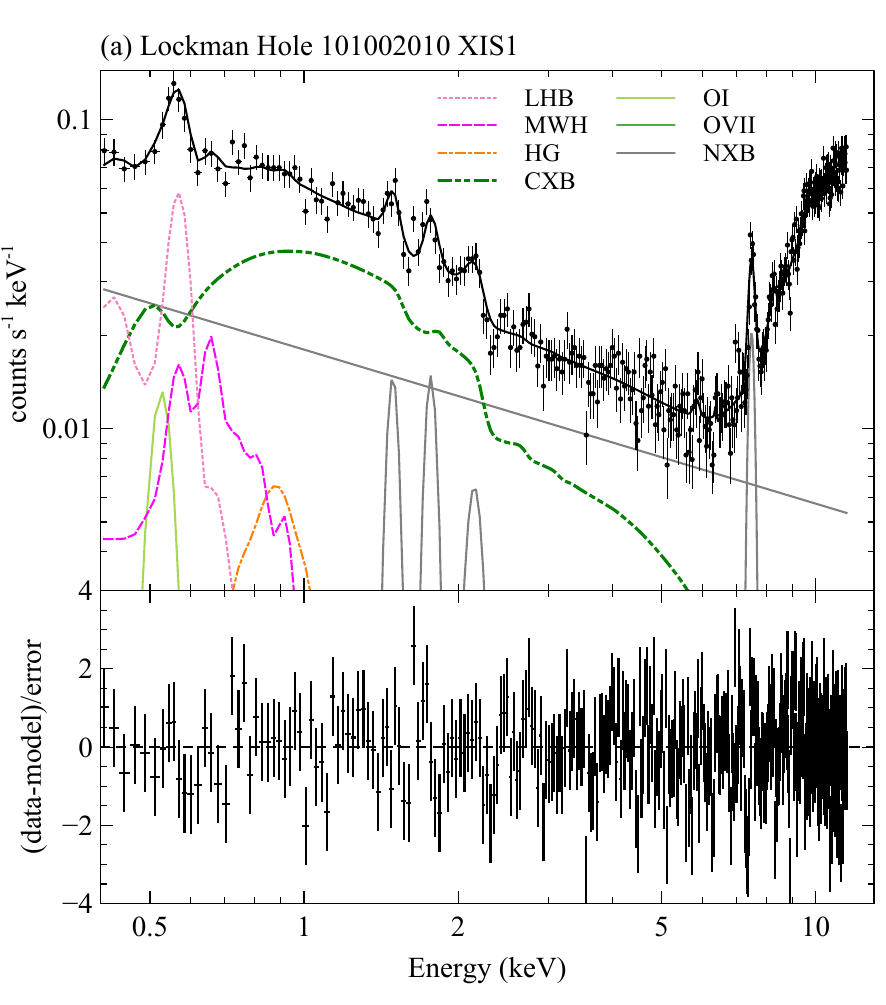}
		\includegraphics[width=0.45\textwidth,clip]{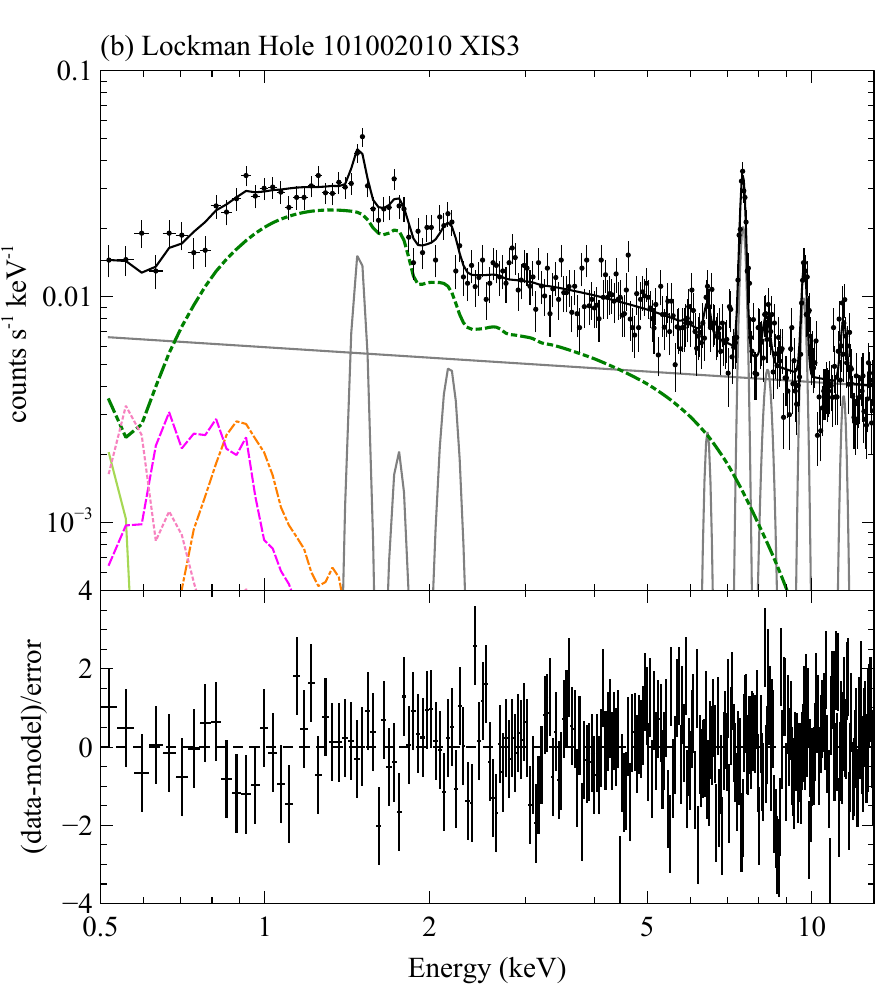}
		\caption{
		The XIS1 (a) and XIS3 (b) spectra of the Lockman Hole data with obsid=101002010. The thick solid, dashed, dotted, and dot-dashed lines show the contributions of the CXB, LHB, MWH, and HG components, respectively. The gray thin lines and Gaussians show the contributions of NXB.
		{Alt text: Two two-panel figures showing the spectra of the two kinds of detectors with a logarithmic vertical axis and the residuals plotted against the X-ray energy in units of kiloelectron volt. 
		Contributions from the different spectral components are indicated with different colors and line styles.}
		}
		\label{fig:speclockman}
\end{figure*}

\begin{figure}[htpd]
	\begin{center}
		\includegraphics[width=0.45\textwidth,clip]{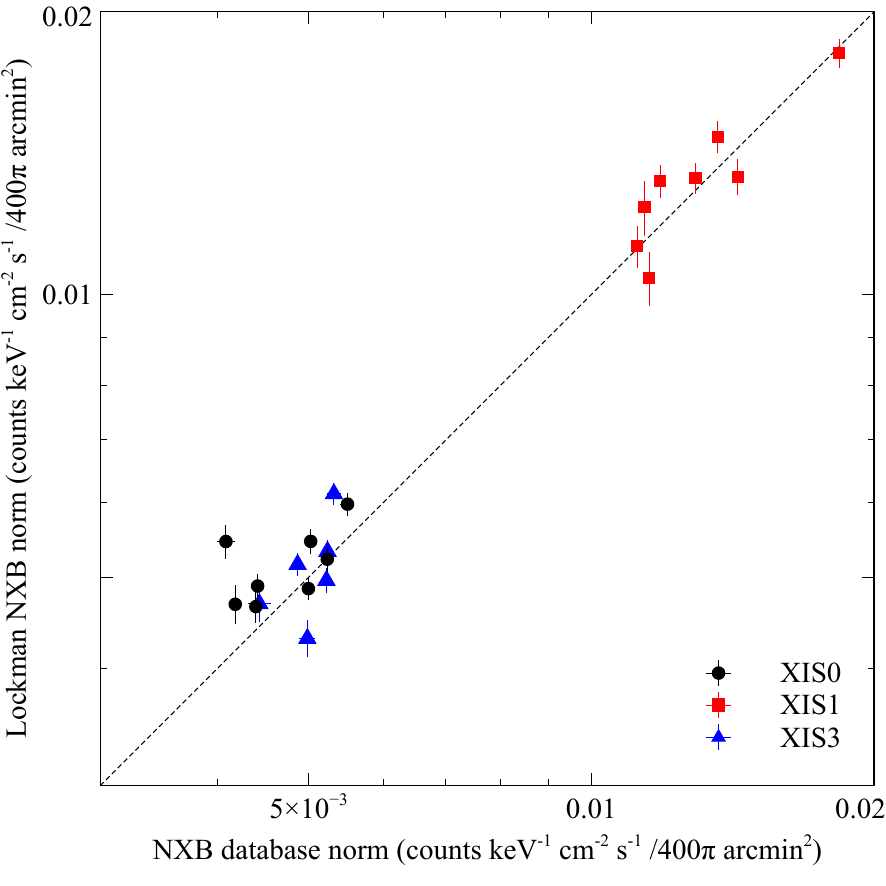}
		\caption{
		The normalizations for the power-law component of the NXB for the Lockman hole observations are plotted against those from the corresponding NXB spectra
		for XIS0 (circles), XIS1 (squares), and XIS3 (triangles)
		{Alt text: Scatter plot comparing the normalization of the non-X-ray background. Different colors and marks indicate different detectors.}
		}
		\label{fig:normnxb}
	\end{center}
\end{figure}

We first fitted the corresponding NXB spectra using the above model to construct NXB models for the corresponding sky spectra.
In these fits, the central energies of the Gaussian components were fixed to the average values derived from the Lockman Hole observations for each detector, 
while the power-law index and normalizations of the power-law and Gaussian components were allowed to vary.
When fitting the sky spectra, we fixed the power-law index to the best-fit value obtained from the corresponding NXB spectral fit.
The normalizations of the NXB  components were left free during the sky spectral fitting.
Figure \ref{fig:speclockman} shows representative Lockman Hole spectra (obsid=101002010), which are reasonably fitted.
Figure \ref{fig:normnxb} compares the normalization of the NXB power-law component derived from the Lockman hole observations and the corresponding NXB spectra.
While they generally correlate, there is some scatter.
This suggests that directly modeling the NXB is preferable to subtracting it, as it better accounts for variability from one observation to another.

\chapter{The level of the cosmic X-ray background}
\label{sec:lockman}

We analyzed eight Lockman Hole observations with nearly identical pointing positions to estimate the flux level of the CXB component and its dependence on the threshold flux used to exclude point sources. 
Point sources were detected following the procedure described in Section \ref{sec:obs}. 
For each observation, we excluded circular regions with a radius of 1.5\arcmin\ around point sources whose fluxes exceeded given threshold fluxes, i.e., 
$
F_{\rm th}=2\times10^{-14}, 3\times10^{-14}, 5\times10^{-14}$ { and } $10\times10^{-14} {\rm erg\,s}^{-1}\,{\rm cm}^{-2} 
$
in the 2.0--10.0 keV band. We then extracted spectra over the remaining FOV of the XIS detectors. 

We fitted the spectra from each observation using the same background model applied in the background regions (Section \ref{sec:fit}). In these fits, we fixed the temperatures of the MWH and HG components at 0.22 keV and 0.8 keV, respectively, while allowing the normalizations of all components to vary. 
As shown in Figure \ref{fig:speclockman}, all spectra are well reproduced by this model.

\begin{figure*}[htpd]
	\begin{center}
		\includegraphics[width=0.9\textwidth,clip]{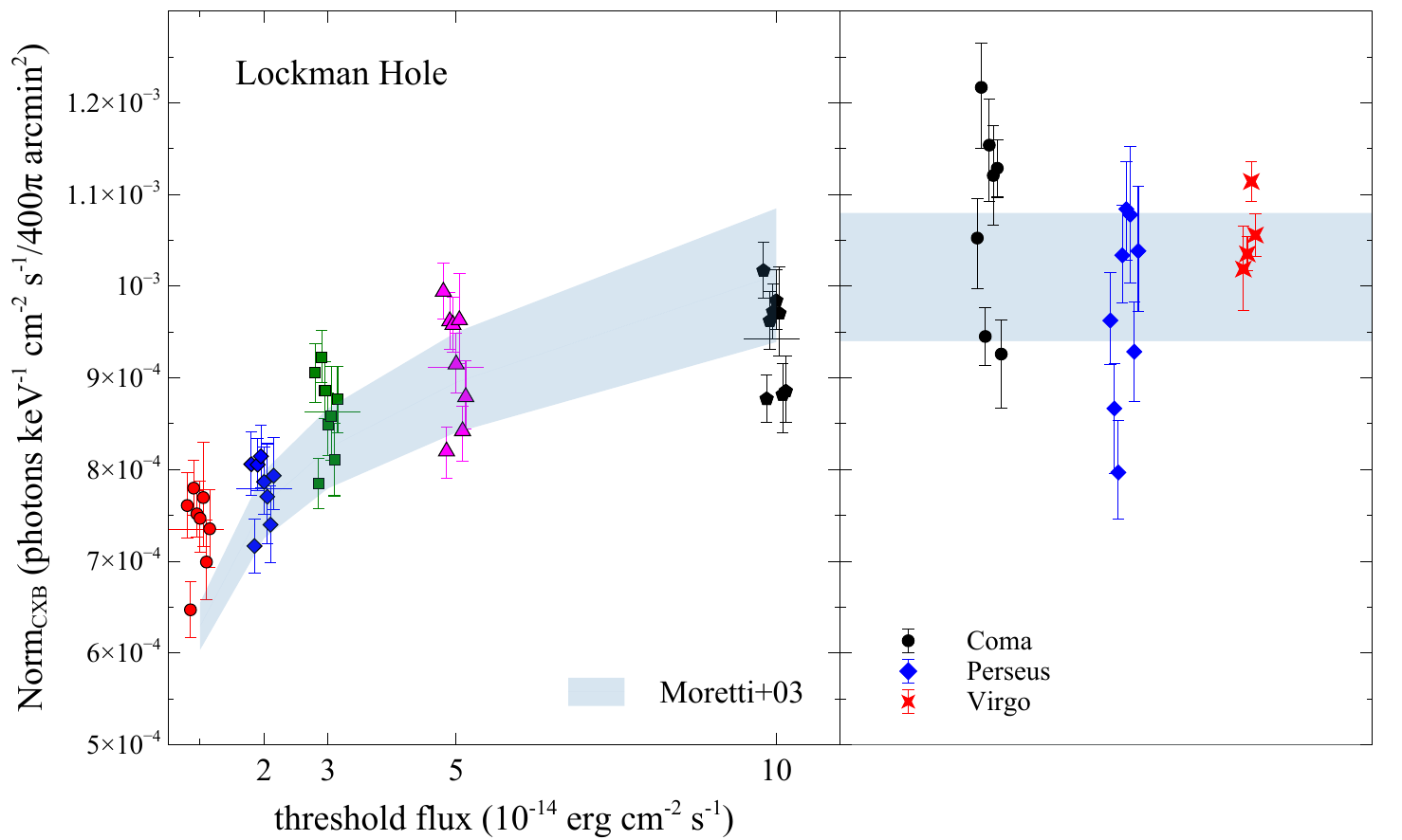}
		\caption{ (left)
		The normalization of the CXB power-law with Suzaku Lockman hole data plotted against threshold flux, excluding point sources. 
		The shaded area corresponds to the result of \citet{Moretti2003} with the cosmic variance. (right) The same for the outskirts of the Coma, Perseus, and Virgo clusters,
		whose threshold flux is $10^{-13}{\rm erg cm^{-2} s^{-1}}$.
		{Alt text: Two-panel figures with a vertical axis show the normalization of the cosmic X-ray background from the Lockman Hole and the Coma, Perseus, and Virgo clusters.}
		}
		\label{fig:cxbnorm2}
	\end{center}
\end{figure*}

Figure \ref{fig:cxbnorm2} shows the best-fit CXB normalizations correlate with $F_{\rm th}$. 
While the pointing positions are nearly identical,  the CXB normalization shows significant scatter, especially for larger $F_{\rm th}$.
Because most background point sources contributing to the CXB are distant active galactic nuclei (AGNs) that vary randomly, the CXB level may not be strictly constant.
\citet{Moretti2003}, using \textit{Chandra} data, estimated how the CXB level depends on $F_{\rm th}$. Their result is also plotted in Figure \ref{fig:cxbnorm2}. Additionally, we estimated the cosmic variance for the XIS FOV based on \citet{Moretti2003} and \citet{Bautz2009}, which is included in the figure. The observed Lockman Hole CXB level and its scatter are consistent with the values derived by \citet{Moretti2003}.
This scatter likely reflects the intrinsic variation of distant, unresolved AGNs that dominate the CXB. 

\begin{figure*}[htpd]
	\begin{center}
		\includegraphics[width=1\textwidth,clip]{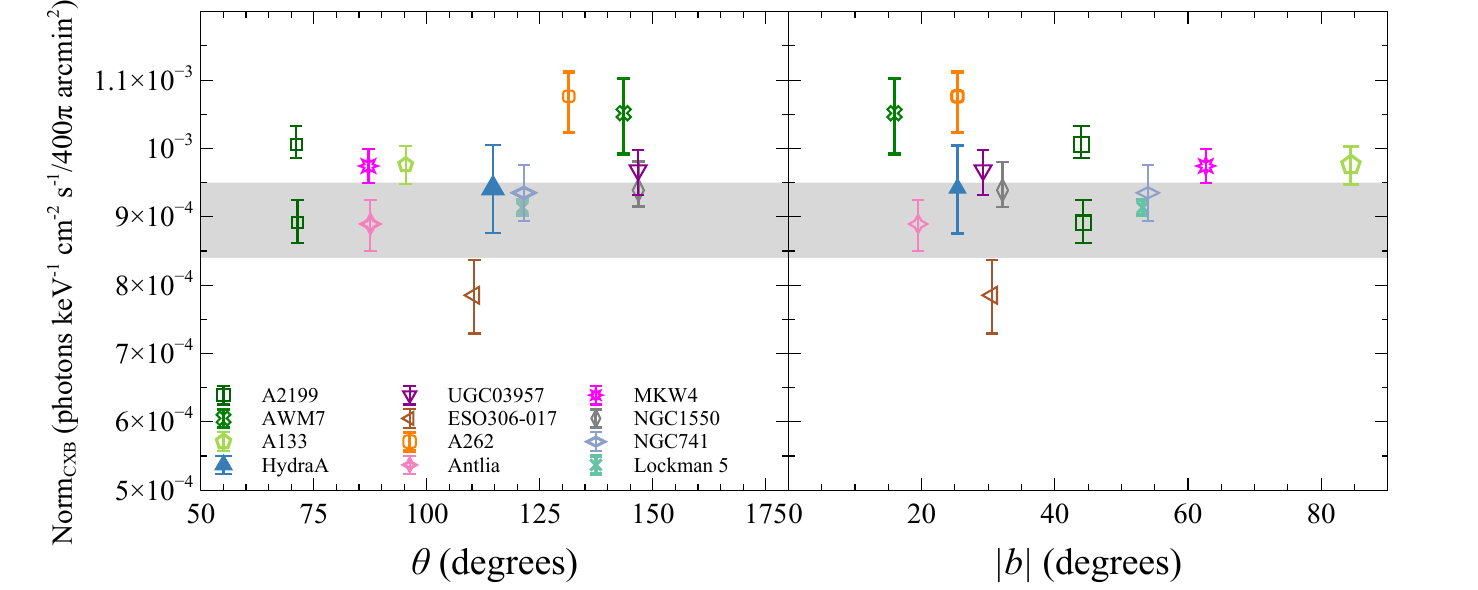}
		\caption{
		The normalizations of the power-law component for CXB for the outskirts of our sample, and the average of the Lockman Hole data,
		whose threshold flux is $5\times 10^{-14}{\rm erg cm^{-2} s^{-1}}$, plotted against the angle from the Galactic center ($\theta$; left panel) and the absolute value of the Galactic latitude ($|b|$; right panel).
		 The shaded area represents the results by \citet{Moretti2003} with cosmic variance included.
		{Alt text: Two-panel figures plotting the normalization of the cosmic X-ray background for the
		background regions plotted against the Galactic coordinates.}
		}
		\label{fig:cxbnorm3}
	\end{center}
\end{figure*}

The right panel of Figure \ref{fig:cxbnorm2} shows the CXB level measured in the outskirts of the Coma, Perseus, and Virgo clusters, where $F_{\rm th} = 10\times10^{-14}$ erg\,s$^{-1}$\,cm$^{-2}$. 
Here, the spectra were accumulated over the full XIS FOV beyond 100$'$ for the Coma cluster.
We used stacked spectra for the eight arms (110$'$--130$'$) for Perseus and for the four arms (180$'$--240$'$ for W and 240$'$--300$'$ for the others) for Virgo.
These results also agree with the relation by \citet{Moretti2003}.
Figure \ref{fig:cxbnorm3} shows the CXB level measured in the background regions of other clusters for $F_{\rm th} = 5\times10^{-14}$ erg\,s$^{-1}$\,cm$^{-2}$, plotted against the angle from the Galactic center, $\theta$ and absolute value of the Galactic latitude, $b$. 
The CXB level shows no systematic dependence on sky position.
The figure also includes the CXB levels from the weighted-average Lockman Hole observations at 
$
F_{\rm th} = 5\times10^{-14}{\rm erg\,s}^{-1}\,{\rm cm}^{-2}.
$
For A2199, we plot the CXB levels from the two regions: the two pointings at 90$'$ (3.4 $r_{500}$) offset and the stacking spectra from the 46$'$--58$'$ annular region (1.7--2.2 $r_{500}$).
The latter shows a $\sim$ 10\% higher CXB level than the former. 
Given that the expected cosmic variance for these two is less than 5 \%, this difference may suggest the presence of faint residual ICM emission in the annular region.
Similar slight excesses are also observed in A133, UGC03957, ESO306-017, and MWK4,  where the background regions were taken from $\sim r_{200}$--$\sim r_{100}$ regions,
and some residual ICM emission may remain, raising the apparent CXB level.
Thus, we can conclude that observed CXB levels are consistent with \citet{Moretti2003} for $F_{\rm th} = 5\times10^{-14}$ erg\,s$^{-1}$\,cm$^{-2}$. 


Beyond $r_{500}$, the typical spectral accumulation area is larger than half of the XIS FOV, and the systematic uncertainties among detectors are around 10\%. 
Noting that the Lockman Hole data slightly prefer a lower CXB level for $F_{\rm th} = 10\times10^{-14}$ erg\,s$^{-1}$\,cm$^{-2}$, 
taken into these factors into account,  we adopt the following ranges for the CXB level in the analyses:
$(0.88-1.08)\times 10^{-3} {\rm photons ~keV^{-1} cm^{-2} s^{-1}/400\pi arcmin^2}$ and 
$(0.84-0.97)\times 10^{-3} {\rm photons ~keV^{-1} cm^{-2} s^{-1}/400\pi arcmin^2}$
for $F_{\rm th} = 10\times10^{-14}$ and $5\times10^{-14}$ erg\,s$^{-1}$\,cm$^{-2}$, respectively.

\chapter{The soft X-ray background}
\label{sec:soft}

\begin{figure*}[htpd]
				\includegraphics[width=1\textwidth,clip]{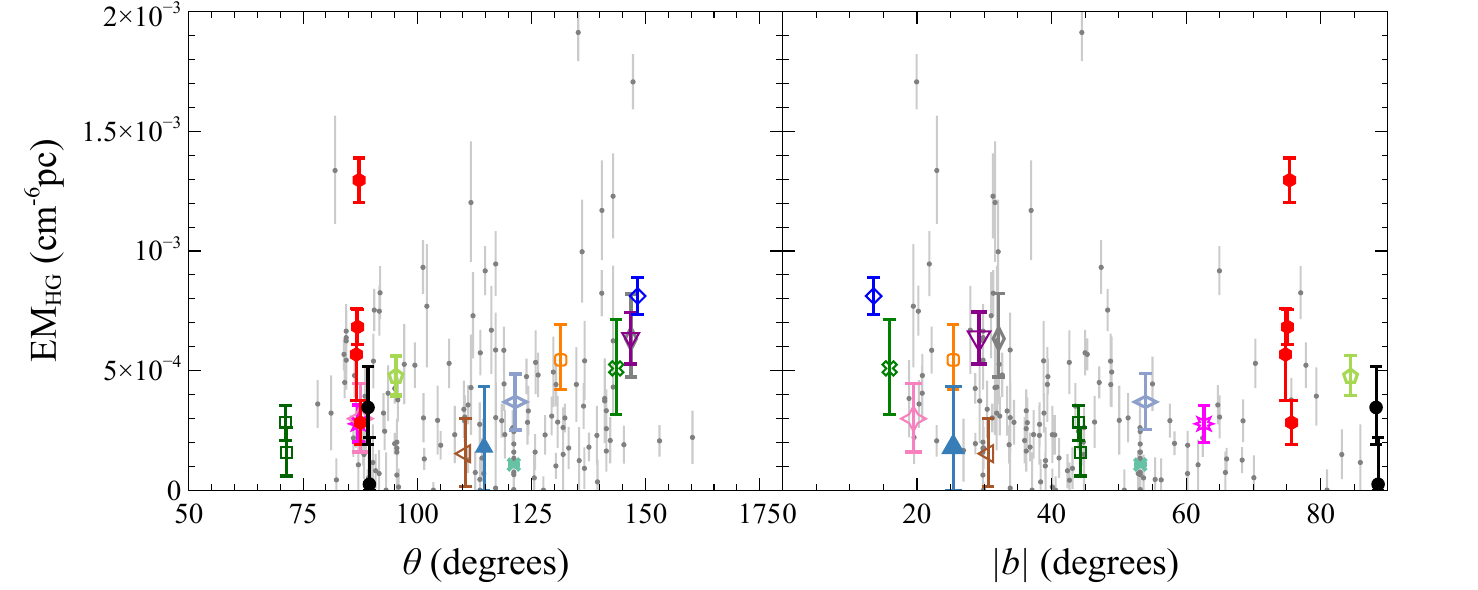}
				\includegraphics[width=1\textwidth,clip]{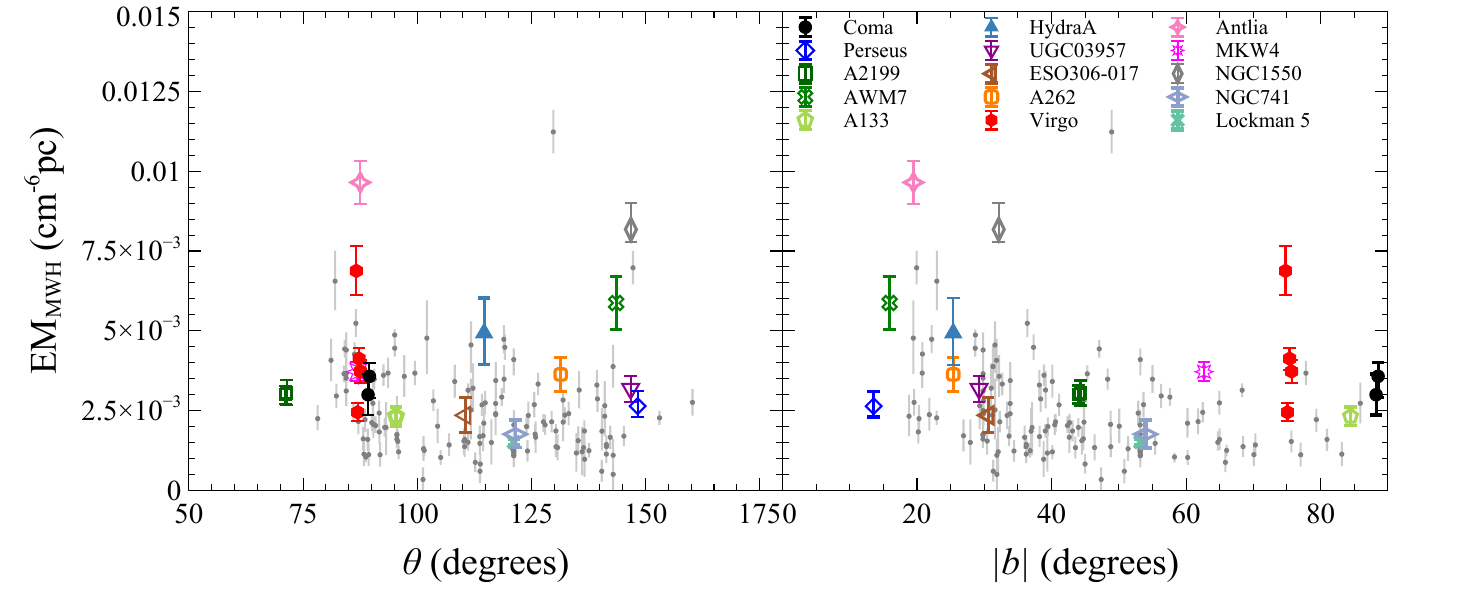}
		\caption{
		Emission measure of the HG component (top panels) and the MWH component (bottom panels) plotted against the angle from the Galactic center ($\theta$; left panels) and the absolute value of the Galactic latitude ($|b|$; right panels).
		{Alt text: Two two-panel figures plotting the emission measures of the HG and MWH components for the
		background regions plotted against the Galactic coordinates.}
		}
		\label{fig:emnorm}
\end{figure*}

\citet{Ueda2022} reported a strong correlation between the MWH emission measure and the sunspot number, suggesting that solar-activity-related emission, mainly O\,\emissiontype {VII} He$\alpha$, can contaminate the MWH. They also showed that observations taken around solar minimum and in the regions, $105^\circ < l< 255^\circ$ and $|b|>35^\circ$,
  have a fairly uniform MWH temperature of $\sim$ 0.22 keV. This plasma may fill the halo in near-hydrostatic equilibrium at the Milky Way’s virial temperature.
\citet{Yoshino2009} found excess emissions at 0.7--1 keV in spectra of some regions without clusters and bright X-ray sources observed with Suzaku, which could be explained by either a higher 
 Ne abundance (Ne/O$\sim$ 3 solar) in the hot gas or an additional higher temperature (0.5--0.9 keV) component.   Further studies with Suzaku data  \citep{Nakashima2018,Gupta2021,Gupta2023,Ueda2022,Sugiyama2023} similarly identified a 0.6--1.0 keV emission component.

If the hot ISM in the Galaxy is spherically distributed in the Galactic halo, its emission measures would depend on the angle from the Galactic center,  $\theta$. In contrast, if higher emission measures increase at lower $|b|$,  the ISM concentrates on the Galactic disk.  Both spherical and disk-like morphology components likely exist for the HG and Halo components (\cite{Ueda2022}, \cite{Sugiyama2023}).
The origin of these components is out of the scope of this paper, but the past supernovae may have heated the surrounding ISM. In some cases, they create superbubbles like Orion-Eridanus superbubbles at $l\sim 200^\circ$ and $b\sim-30^\circ$ extending over several tens of degrees in the sky. Such superbubbles may contribute to the disk-like components.
With the HaloSAT survey, \citet{Bluem2022} reported that both 0.2 keV and 0.8 keV components decrease with $\theta$, indicating that 
there are spherical components, even for the HG component. As a result, even around $b\sim 90^\circ$, 
around the Coma cluster and A133 cluster, EM$_{\rm HG}$ can be relatively bright.
Moreover, the eROSITA all-sky map shows additional bright features, such as the eROSITA bubbles \citep{Predehl2020} and the Orion-Eridanus superbubble  \citep{Burrows1993, Brown1995, Snowden1995}.

In figure \ref{fig:emnorm}, we plot emission measures of the HG component (EM$_{\rm HG}$) and the MWH component (EM$_{\rm MWH}$) from the cluster background regions against $\theta$ and $|b|$ 
and compared them with Suzaku observations of non-cluster fields  (\cite{Ueda2022}, \cite{Sugiyama2023}), under fixed temperatures  $kT_{\rm halo}=0.22$ keV and $kT_{\rm HG}=0.80$ keV . Here, the effect of the difference in the solar abundance was corrected.  
We employed the stacked spectra of the E and SW arms beyond 110$'$ for the Coma cluster
and  the stacked 110$'$--130$'$ spectra for the Perseus cluster.
The resulting EM$_{\rm HG}$ and EM$_{\rm halo}$ values in the cluster background regions lie within the range of those derived for the non-cluster sample, indicating that the  HG component significantly contributes to the emission in the cluster outskirts.

The Virgo and MKW4 clusters lie close to the edge of the {eROSITA} bubbles/North Polar Spur. According to \citet{Gupta2023},  the X-ray emission from the {eROSITA} bubbles can be well described by the sum of two CIE (collisional ionization equilibrium) components at 0.2 keV and 0.8 keV.  The higher EM$_{\rm HG}$ at the south outskirts of the Virgo cluster likely reflects that eROSITA bubbles are located in that direction.

The Antlia cluster lies outside the brightest {eROSITA} bubbles but is located in the middle of an extended X-ray enhancement, spanning 20$^\circ$--30$^\circ$ across. This enhancement may be associated with a supernova remnant, as suggested by \citet{Antlia2021}. Although the emission measure of the hot gas (EM$_{\rm HG}$) in Antlia is comparable to those of regions at comparable Galactic latitude and longitude, the halo emission measure (EM$_{\rm halo}$) is notably high.

NGC 1550 is situated within the Orion--Eridanus superbubble \citep{Burrows1993, Brown1995, Snowden1995}. The X-ray emission of the superbubble can be successfully modeled by the sum of 0.2 keV and 0.8 keV CIE components (\cite{Fuller2023}, Fukushima et al. submitted). The relatively high fluxes of EM$_{\rm HG}$ and EM$_{\rm halo}$ seen toward NGC 1550 are likely attributable to the superbubble.


\chapter{The SWCX emissions}
\label{sec:SWCX}

\begin{figure*}[tbhp]
	\begin{center}
		\includegraphics[width=0.9\textwidth,angle=0,clip]{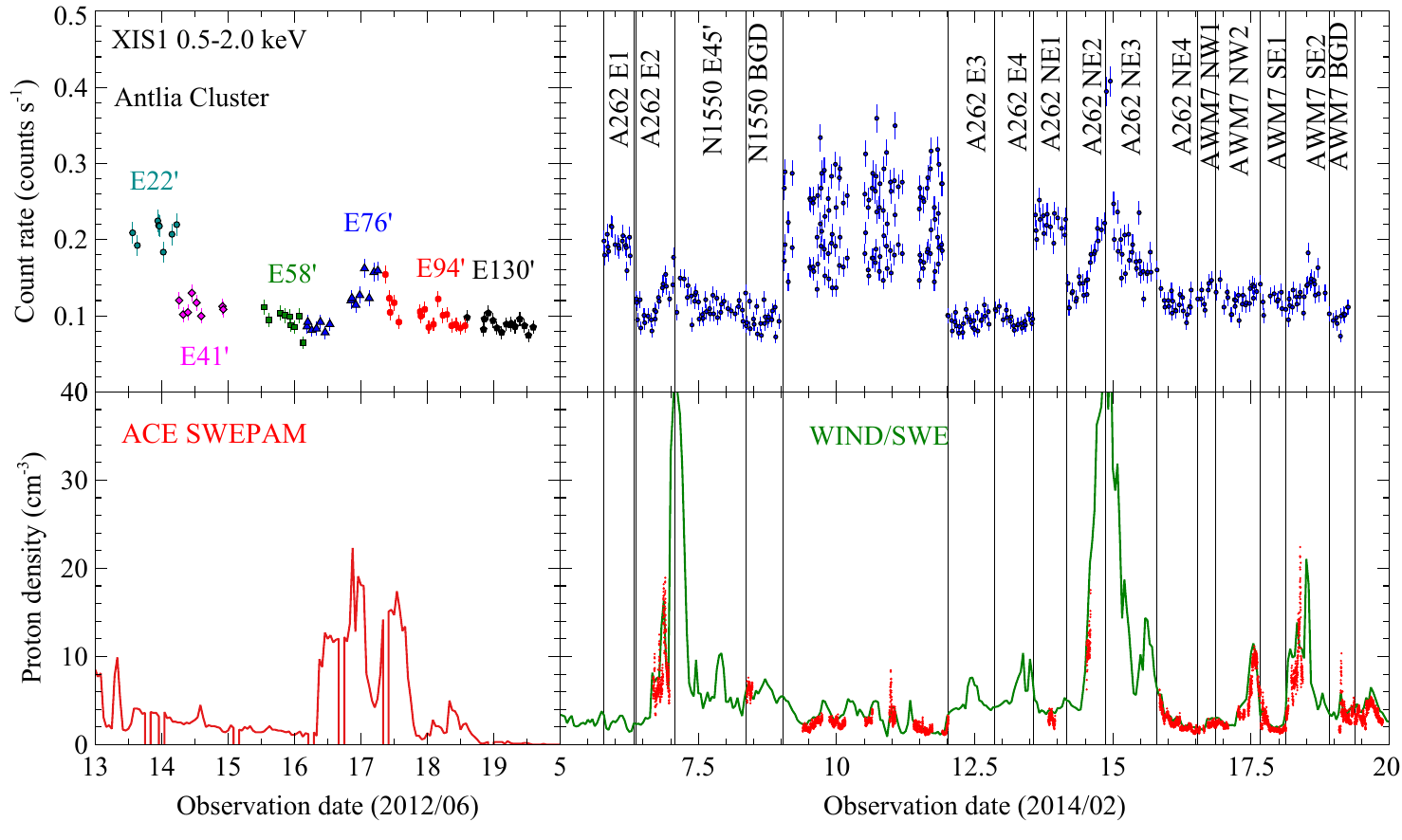}
		\end{center}
		\caption{(Upper panels) The 0.5-2.0 keV light curves of XIS1 during the Antlia offset observations and during  February 2014, when some parts of A262, NGC 1550, and AWM7 were observed.
		(Bottom panels) Solar proton density measured by ACE/SWEPAM (red) and WIND/SWE (green)
		{Alt text: Four-panel figure showing the light curves of the XIS detectors in units of counts s$^{-1}$ and the proton density in units of cm$^{-3}$ plotted against the observation dates. }}
		\label{fig:lc}
\end{figure*}

\begin{figure}[tbhp]
	\begin{center}
		\includegraphics[width=0.45\textwidth,angle=0,clip]{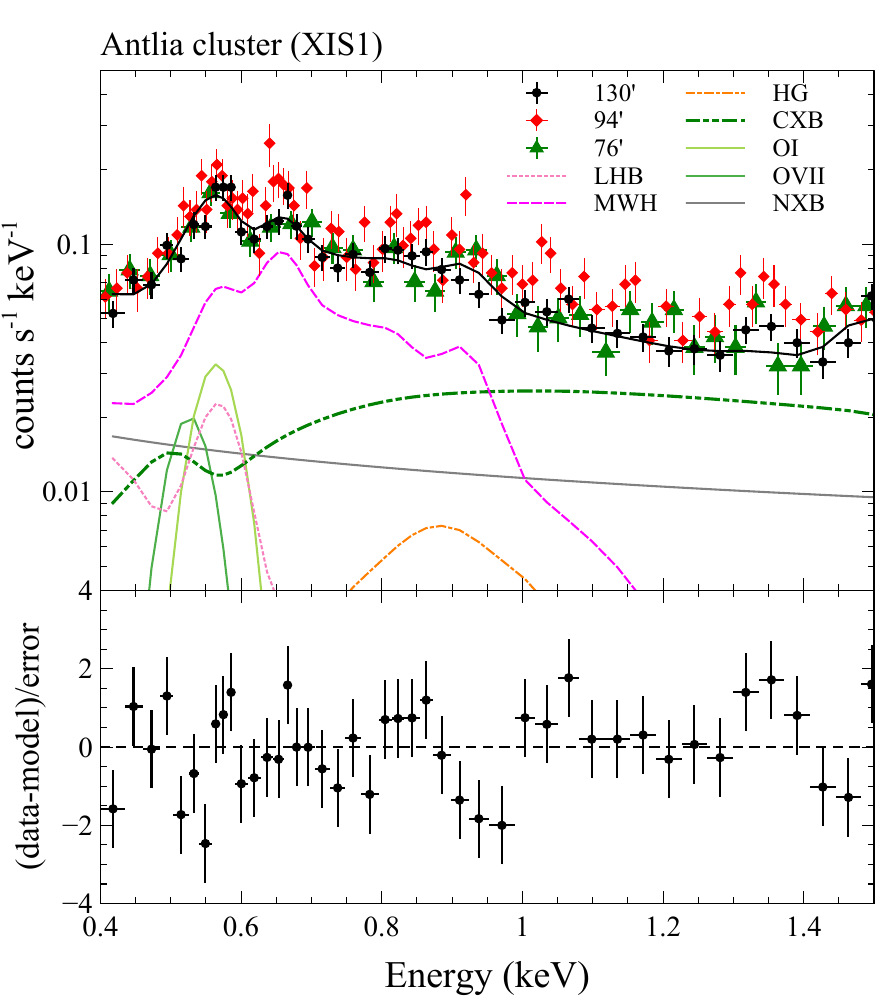}
		\end{center}
		\caption{The XIS1 spectra of the Antlia cluster for the pointings of 76$'$ (64$'$--82$'$; green triangles), 94$'$ (82$'$--100$'$; red diamonds), and 130$'$ (BGD: black circles) toward east from the cluster center. For the 76$'$ and 94$'$ pointing, the events were filtered using the 0.5--2.0 keV light curve.  The dashed, dotted, dot-dashed, and solid thick lines represent the contributions from the LHB, MWH, HG, and CXB components, respectively. 
		The bottom panel shows the residuals for the 130$'$ region.
		{Alt text: Two-panel figure showing the XIS1 spectra comparing different reginons and residuals.}}
		\label{fig:antliaspec}
\end{figure}

During the offset observations of the Antlia cluster, A262, the NGC 1550 group, and AWM7,
 several enhancements were observed in the 0.5-2.0 keV light curves of the XIS1 detector (Figure \ref{fig:lc}).
These enhancements occurred during the 76$'$ and 94$'$ offset observations of Antlia, 
A262/E2 and NGC1550/E45$'$, as well as from A262/NE2 to A262/NE3, AWM7SE2.
In these cases, corresponding increases in the solar proton density were also detected.

Figure \ref{fig:antliaspec} shows the XIS spectra of the Antlia cluster, extracted over the FOV of the observations for 76$'$ (1.55 $r_{500}$), 94$'$ (1.92 $r_{500}$), and 130$'$ (2.65 $ r_{500}$)
pointings toward the east from the cluster center.
Here, the events were filtered using the 0.5--2.0 keV light curve of the XIS1 detector.
Our background model reproduces the XIS1 spectrum of the 130$'$ pointing well, except for a hint of residual structure at the Mg\,\emissiontype {XI}  He-$\alpha$ line.
No significant difference is seen between the 76$'$ and 130$'$ spectra, suggesting that 
 the ICM contribution at 76$'$ is negligible compared to the luminous MWH component, possibly from the Antlia SNR.
The enhancements in the light curves indicate that the SWCX emissions contaminated the last part of the 76$'$ and the first part of the 94$'$ observations.
Although the strengths of the O\,\emissiontype {VII} He$\alpha$ line at 0.56 keV are comparable in the two spectra, 
the 94$'$ spectrum exhibits apparent excesses at several line-like features, such as O\,\emissiontype {VII} He$\alpha$ (0.56 keV), Ne\,\emissiontype {X}  Ly$\alpha$ (1.0 keV),  and Mg\,\emissiontype {XI} He$\alpha$ (1.34 keV), even after applying the 0.5-2.0 keV light curve filtering.
These features are often seen from geocoronal SWCX emissions \citep{Fujimoto2007}.
Although the 130$'$ pointing was taken after 94$'$ observation and its light curve appeared stable,
a faint hint of Mg\,\emissiontype {XI} He$\alpha$ emission remains.
These results indicate that light curve filtering alone is insufficient to remove SWCX contamination following significant solar proton flare events.

\begin{figure}[tbhp]
	\begin{center}
		\includegraphics[width=0.45\textwidth,angle=0,clip]{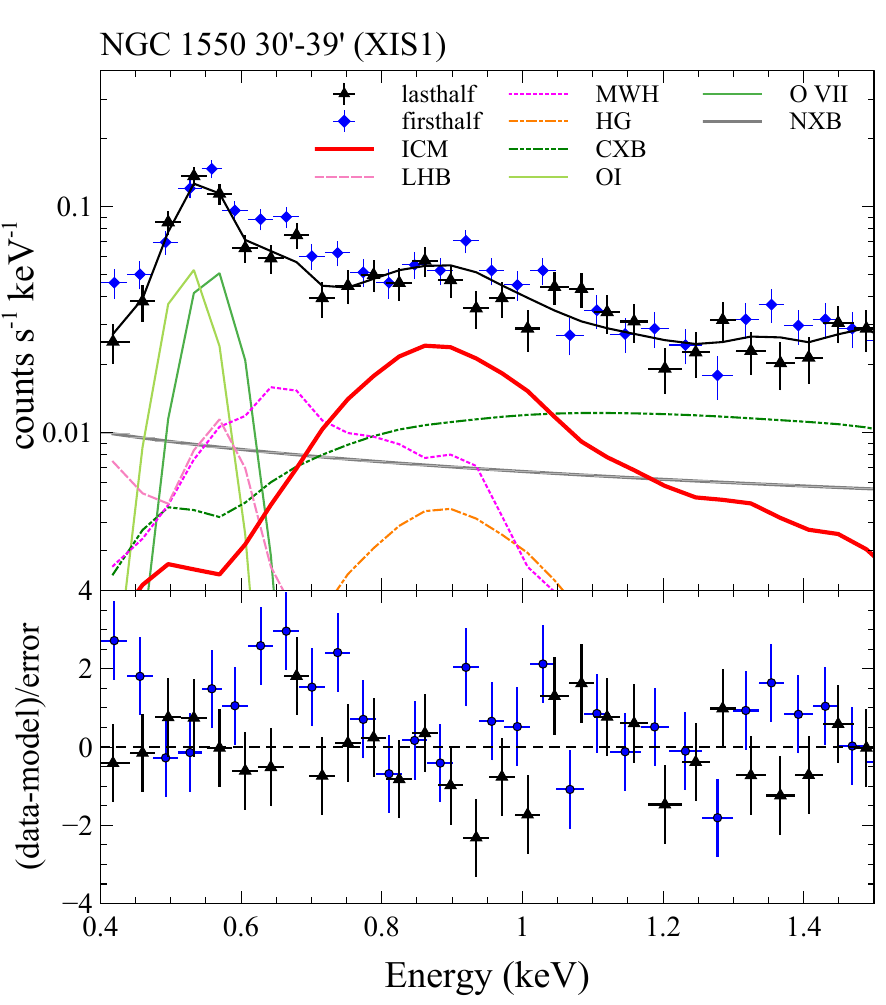}
		\end{center}
		\caption{The XIS1 spectra of the NGC 1550 group for the annular region of 30$'$--39$'$ (0.89--1.16 $r_{500}$), for the earlier interval (first 27 ks; blue diamonds) and the later interval (last 21 ks; black circles).
		The model contributions fitted to the latter half period are also shown: ICM (thick solid line), LHB (long dashed), MHW (dotted), HG (dot-dashed),
		CXB (dot-dot-dashed), OI (thin solid), and NXB (thick gray) to the latter half period are also shown. The bottom panel shows the residuals for the best-fit model of the latter half. {Alt text: Two-panel figure showing the XIS1 spectra comparing different reginons and residuals.}}
		\label{fig:NGC155045}
\end{figure}

SWCX emissions also contaminated the first half of the NGC 1550 45$'$ offset observation.
We divided this observation into two time intervals, the first 27 ks and the latter 20 ks, and extracted spectra
from the two annular regions: 30$'$--39$'$ and 39$'$--48$'$.
The XIS1 spectra for 30$'$--39$'$ are shown in figure \ref{fig:NGC155045}. 
In the first half, clear enhancements are observed in several lines compared to the latter half,
 including  CVI Ly $\gamma$ (0.46 keV),  O\,\emissiontype {VII} He $\alpha$, O\,\emissiontype {VII}I Ly$\alpha$,  Mg \,\emissiontype {XI}  He$\alpha$
and a feature around 0.9 keV (possibly Fe or Ne?).
Since the Antlia cluster observation may have also been contaminated by similar line emission after the flare in the light curve, we fitted the latter half spectra of the NGC 1550 45$'$ offset by adding these lines to Model-Z03.
Here, the normalization of the MWH component was fixed to the value obtained from the NGC 1550 background.
We note that the two spectra are consistent in the 0.8--0.9 keV band, where the ICM component dominates.

During A262 E2, NE2, NE3 observations, no time intervals with stable light curves were identified due to strong flaring events.
Therefore, we did not apply light curve filtering to these data.
To account for potential SWCX contamination, we added seven Gaussian components to Model-Z03,
with their central energies fixed to the values reported for the SWCX spectrum by \citet{Fujimoto2007}.
Using this model fit, we obtained consistent results for the east arm in the overlapping regions.
However,  the SWCX contamination in the NE3 observation was too strong to be adequately removed.
As a result, we excluded the NE arm from the spectral analysis and used only the east arm.
We adopt only 21$'$--27$'$ among the three annular regions from the E2 observation, since the other two regions overlap the
E1 and E3 pointings, without flaring events.

For the AWM7 cluster, the SE2 (61$'$ southeast) observation, which was affected by a proton enhancement,
exhibits increased counts at the O\,\emissiontype {VII} He$\alpha$ line
compared to the NW2 (61$'$ north west) observation, which shows a flat light curve.
However, the two spectra around 1 keV are quite similar.
 Therefore,  we stacked spectra within annular regions
and allowed normalization of the Gaussian for the O\,\emissiontype {VII} He line to vary, as we did for the other spectra.

\chapter{$\beta$-model fits of the emission measure profiles}
\label{sec:beta}

\begin{figure*}[tbhp]
	\begin{center}
		\includegraphics[width=1\textwidth,angle=0,clip]{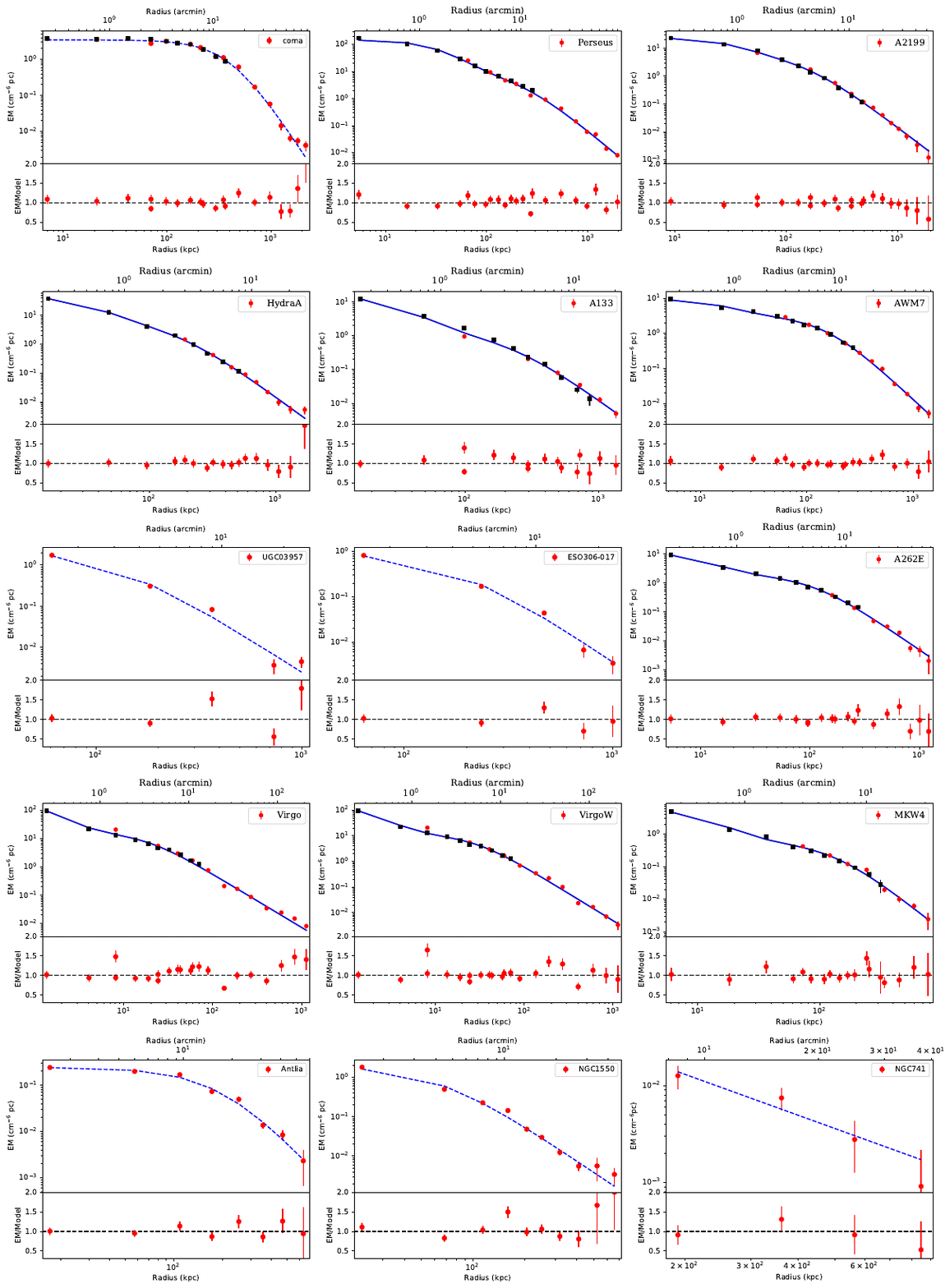}
\end{center}
		\caption{
Radial profiles of the emission measure (our results and the XMM results by Snowden et al. 2008), fitted with a double $\beta$-model (blue solid lines) and a single $\beta$-model (red dashed lines).
The bottom panels show the data-to-model ratio.
{Alt text: Fifteen figures, each consisting of two subpanels showing the radial profiles of the emission measure with the best-fit models. The lower subpanels display the residuals.}
}
		\label{fig:EMfit}
\end{figure*}

The emission measure profiles with the best-fit single or two $\beta$ models (section \ref{sec:ne}) are plotted in figure \ref{fig:EMfit}.

\chapter{Pressure profiles}
\label{sec:pfit}

\begin{figure*}[tbhp]
	\begin{center}
		\includegraphics[width=1\textwidth,angle=0,clip]{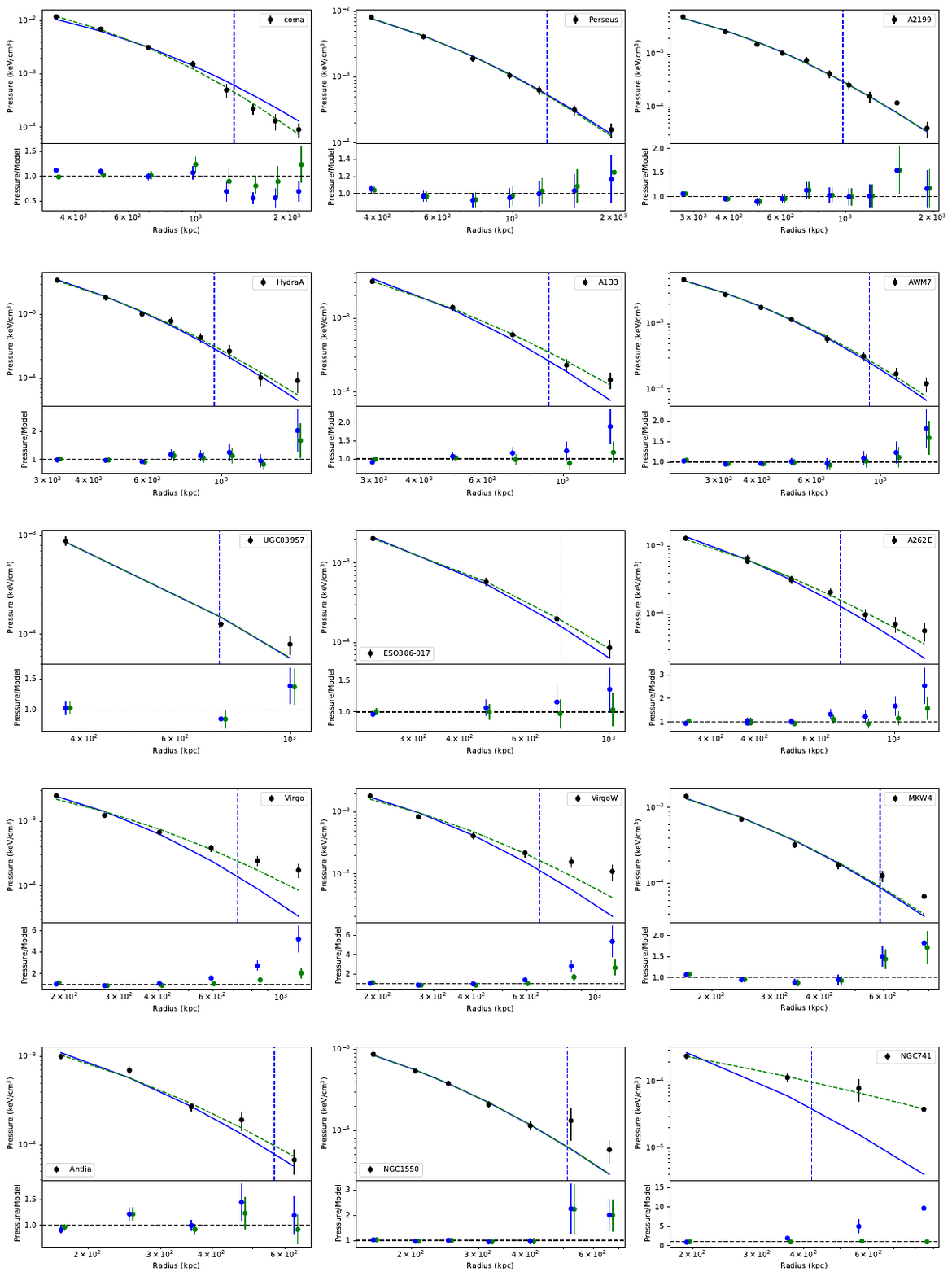}
			\end{center}
			\caption{
Radial profiles of the pressure beyond 0.2 $r_{500}$ fitted with the best-fit Planck pressure profile (solid lines).
The bottom panels show the data-to-model ratio.
{Alt text: Fifteen figures, each consisting of two subpanels showing the radial profiles of the pressure with the best-fit models. The lower subpanels display the residuals.}}
		\label{fig:Pfit}
\end{figure*}

The electron pressure profiles with the gNFW model (section \ref{sec:r500}) are plotted in figure \ref{fig:Pfit}.

\bibliographystyle{apj}
\bibliography{clusters_apj}  

\begin{thebibliography}{}
\expandafter\ifx\csname natexlab\endcsname\relax\def\natexlab#1{#1}\fi

\bibitem[{{Akino} {et~al.}(2022){Akino}, {Eckert}, {Okabe}, {Sereno}, {Umetsu},
  {Oguri}, {Gastaldello}, {Chiu}, {Ettori}, {Evrard}, {Farahi}, {Maughan},
  {Pierre}, {Ricci}, {Valtchanov}, {McCarthy}, {McGee}, {Miyazaki},
  {Nishizawa}, \& {Tanaka}}]{Akino2022}
{Akino}, D., {Eckert}, D., {Okabe}, N., {et~al.} 2022, \pasj, 74, 175

\bibitem[{{Arnaud} {et~al.}(2010){Arnaud}, {Pratt}, {Piffaretti},
  {B{\"o}hringer}, {Croston}, \& {Pointecouteau}}]{Arnaud2010}
{Arnaud}, M., {Pratt}, G.~W., {Piffaretti}, R., {et~al.} 2010, \aap, 517, A92

\bibitem[{{Ayromlou} {et~al.}(2023){Ayromlou}, {Nelson}, \&
  {Pillepich}}]{Illustris2023}
{Ayromlou}, M., {Nelson}, D., \& {Pillepich}, A. 2023, \mnras, 524, 5391

\bibitem[{{Bautz} {et~al.}(2009){Bautz}, {Miller}, {Sanders}, {Arnaud},
  {Mushotzky}, {Porter}, {Hayashida}, {Henry}, {Hughes}, {Kawaharada},
  {Makashima}, {Sato}, \& {Tamura}}]{Bautz2009}
{Bautz}, M.~W., {Miller}, E.~D., {Sanders}, J.~S., {et~al.} 2009, \pasj, 61,
  1117

\bibitem[{{Bluem} {et~al.}(2022){Bluem}, {Kaaret}, {Kuntz}, {Jahoda},
  {Koutroumpa}, {Hodges-Kluck}, {Fuller}, {LaRocca}, \& {Zajczyk}}]{Bluem2022}
{Bluem}, J., {Kaaret}, P., {Kuntz}, K.~D., {et~al.} 2022, \apj, 936, 72

\bibitem[{{Brown} {et~al.}(1995){Brown}, {Hartmann}, \& {Burton}}]{Brown1995}
{Brown}, A.~G.~A., {Hartmann}, D., \& {Burton}, W.~B. 1995, \aap, 300, 903

\bibitem[{{Budzynski} {et~al.}(2014){Budzynski}, {Koposov}, {McCarthy}, \&
  {Belokurov}}]{Budzynski2014}
{Budzynski}, J.~M., {Koposov}, S.~E., {McCarthy}, I.~G., \& {Belokurov}, V.
  2014, \mnras, 437, 1362

\bibitem[{{Burrows} {et~al.}(1993){Burrows}, {Singh}, {Nousek}, {Garmire}, \&
  {Good}}]{Burrows1993}
{Burrows}, D.~N., {Singh}, K.~P., {Nousek}, J.~A., {Garmire}, G.~P., \& {Good},
  J. 1993, \apj, 406, 97

\bibitem[{{Cash}(1979)}]{Cash}
{Cash}, W. 1979, \apj, 228, 939

\bibitem[{{Churazov} {et~al.}(2021){Churazov}, {Khabibullin}, {Lyskova},
  {Sunyaev}, \& {Bykov}}]{Churazov2021}
{Churazov}, E., {Khabibullin}, I., {Lyskova}, N., {Sunyaev}, R., \& {Bykov},
  A.~M. 2021, \aap, 651, A41

\bibitem[{{De Grandi} {et~al.}(2016){De Grandi}, {Eckert}, {Molendi},
  {Girardi}, {Roediger}, {Gaspari}, {Gastaldello}, {Ghizzardi}, {Nonino}, \&
  {Rossetti}}]{DeGrandi2016}
{De Grandi}, S., {Eckert}, D., {Molendi}, S., {et~al.} 2016, \aap, 592, A154

\bibitem[{{Eckert} {et~al.}(2022){Eckert}, {Ettori}, {Pointecouteau}, {van der
  Burg}, \& {Loubser}}]{Eckert2022}
{Eckert}, D., {Ettori}, S., {Pointecouteau}, E., {van der Burg}, R.~F.~J., \&
  {Loubser}, S.~I. 2022, \aap, 662, A123

\bibitem[{{Eckert} {et~al.}(2019){Eckert}, {Ghirardini}, {Ettori}, {Rasia},
  {Biffi}, {Pointecouteau}, {Rossetti}, {Molendi}, {Vazza}, {Gastaldello},
  {Gaspari}, {De Grandi}, {Ghizzardi}, {Bourdin}, {Tchernin}, \&
  {Roncarelli}}]{Eckert2019}
{Eckert}, D., {Ghirardini}, V., {Ettori}, S., {et~al.} 2019, \aap, 621, A40

\bibitem[{{Fesen} {et~al.}(2021){Fesen}, {Drechsler}, {Weil}, {Strottner},
  {Raymond}, {Rupert}, {Milisavljevic}, {Subrayan}, {di Cicco}, {Walker},
  {Mittelman}, \& {Ludgate}}]{Antlia2021}
{Fesen}, R.~A., {Drechsler}, M., {Weil}, K.~E., {et~al.} 2021, \apj, 920, 90

\bibitem[{{Foster} {et~al.}(2012){Foster}, {Ji}, {Smith}, \&
  {Brickhouse}}]{Foster2012}
{Foster}, A.~R., {Ji}, L., {Smith}, R.~K., \& {Brickhouse}, N.~S. 2012, \apj,
  756, 128

\bibitem[{{Fujimoto} {et~al.}(2007){Fujimoto}, {Mitsuda}, {Mccammon}, {Takei},
  {Bauer}, {Ishisaki}, {Porter}, {Yamaguchi}, {Hayashida}, \&
  {Yamasaki}}]{Fujimoto2007}
{Fujimoto}, R., {Mitsuda}, K., {Mccammon}, D., {et~al.} 2007, \pasj, 59, 133

\bibitem[{{Fuller} {et~al.}(2023){Fuller}, {Kaaret}, {Bluem}, {Kuntz},
  {Hodges-Kluck}, \& {Jahoda}}]{Fuller2023}
{Fuller}, C.~A., {Kaaret}, P., {Bluem}, J., {et~al.} 2023, \apj, 943, 61

\bibitem[{{Ghirardini} {et~al.}(2019){Ghirardini}, {Eckert}, {Ettori},
  {Pointecouteau}, {Molendi}, {Gaspari}, {Rossetti}, {De Grandi}, {Roncarelli},
  {Bourdin}, {Mazzotta}, {Rasia}, \& {Vazza}}]{XCOP2019}
{Ghirardini}, V., {Eckert}, D., {Ettori}, S., {et~al.} 2019, \aap, 621, A41

\bibitem[{{Gonzalez} {et~al.}(2013){Gonzalez}, {Sivanandam}, {Zabludoff}, \&
  {Zaritsky}}]{Gonz2013}
{Gonzalez}, A.~H., {Sivanandam}, S., {Zabludoff}, A.~I., \& {Zaritsky}, D.
  2013, \apj, 778, 14

\bibitem[{{Gupta} {et~al.}(2021){Gupta}, {Kingsbury}, {Mathur}, {Das},
  {Galeazzi}, {Krongold}, \& {Nicastro}}]{Gupta2021}
{Gupta}, A., {Kingsbury}, J., {Mathur}, S., {et~al.} 2021, \apj, 909, 164

\bibitem[{{Gupta} {et~al.}(2023){Gupta}, {Mathur}, {Kingsbury}, {Das}, \&
  {Krongold}}]{Gupta2023}
{Gupta}, A., {Mathur}, S., {Kingsbury}, J., {Das}, S., \& {Krongold}, Y. 2023,
  Nature Astronomy, 7, 799

\bibitem[{{Henley} \& {Shelton}(2013)}]{Henley2013}
{Henley}, D.~B., \& {Shelton}, R.~L. 2013, \apj, 773, 92

\bibitem[{{Hitomi Collaboration} {et~al.}(2018){Hitomi Collaboration},
  {Aharonian}, {Akamatsu}, {Akimoto}, {Allen}, {Angelini}, {Audard}, {Awaki},
  {Axelsson}, {Bamba}, {Bautz}, {Blandford}, {Brenneman}, {Brown}, {Bulbul},
  {Cackett}, {Canning}, {Chernyakova}, {Chiao}, {Coppi}, {Costantini}, {de
  Plaa}, {de Vries}, {den Herder}, {Done}, {Dotani}, {Ebisawa}, {Eckart},
  {Enoto}, {Ezoe}, {Fabian}, {Ferrigno}, {Foster}, {Fujimoto}, {Fukazawa},
  {Furuzawa}, {Galeazzi}, {Gallo}, {Gandhi}, {Giustini}, {Goldwurm}, {Gu},
  {Guainazzi}, {Haba}, {Hagino}, {Hamaguchi}, {Harrus}, {Hatsukade}, {Hayashi},
  {Hayashi}, {Hayashi}, {Hayashida}, {Hiraga}, {Hornschemeier}, {Hoshino},
  {Hughes}, {Ichinohe}, {Iizuka}, {Inoue}, {Inoue}, {Inoue}, {Ishida},
  {Ishikawa}, {Ishisaki}, {Iwai}, {Kaastra}, {Kallman}, {Kamae}, {Kataoka},
  {Katsuda}, {Kawai}, {Kelley}, {Kilbourne}, {Kitaguchi}, {Kitamoto},
  {Kitayama}, {Kohmura}, {Kokubun}, {Koyama}, {Koyama}, {Kretschmar}, {Krimm},
  {Kubota}, {Kunieda}, {Laurent}, {Lee}, {Leutenegger}, {Limousin},
  {Loewenstein}, {Long}, {Lumb}, {Madejski}, {Maeda}, {Maier}, {Makishima},
  {Markevitch}, {Matsumoto}, {Matsushita}, {McCammon}, {McNamara}, {Mehdipour},
  {Miller}, {Miller}, {Mineshige}, {Mitsuda}, {Mitsuishi}, {Miyazawa},
  {Mizuno}, {Mori}, {Mori}, {Mukai}, {Murakami}, {Mushotzky}, {Nakagawa},
  {Nakajima}, {Nakamori}, {Nakashima}, {Nakazawa}, {Nobukawa}, {Nobukawa},
  {Noda}, {Odaka}, {Ohashi}, {Ohno}, {Okajima}, {Ota}, {Ozaki}, {Paerels},
  {Paltani}, {Petre}, {Pinto}, {Porter}, {Pottschmidt}, {Reynolds},
  {Safi-Harb}, {Saito}, {Sakai}, {Sasaki}, {Sato}, {Sato}, {Sato}, {Sawada},
  {Schartel}, {Serlemtsos}, {Seta}, {Shidatsu}, {Simionescu}, {Smith}, {Soong},
  {Stawarz}, {Sugawara}, {Sugita}, {Szymkowiak}, {Tajima}, {Takahashi},
  {Takahashi}, {Takeda}, {Takei}, {Tamagawa}, {Tamura}, {Tanaka}, {Tanaka},
  {Tanaka}, {Tanaka}, {Tashiro}, {Tawara}, {Terada}, {Terashima}, {Tombesi},
  {Tomida}, {Tsuboi}, {Tsujimoto}, {Tsunemi}, {Tsuru}, {Uchida}, {Uchiyama},
  {Uchiyama}, {Ueda}, {Ueda}, {Uno}, {Urry}, {Ursino}, {Wang}, {Watanabe},
  {Werner}, {Wilkins}, {Williams}, {Yamada}, {Yamaguchi}, {Yamaoka},
  {Yamasaki}, {Yamauchi}, {Yamauchi}, {Yaqoob}, {Yatsu}, {Yonetoku},
  {Zhuravleva}, \& {Zoghbi}}]{HitomiPerseus2018}
{Hitomi Collaboration}, {Aharonian}, F., {Akamatsu}, H., {et~al.} 2018, \pasj,
  70, 9

\bibitem[{{HyeongHan} {et~al.}(2024){HyeongHan}, {Jee}, {Lee}, {ZuHone},
  {Zhuravleva}, {Kang}, \& {Hwang}}]{Perseus2024}
{HyeongHan}, K., {Jee}, M.~J., {Lee}, W., {et~al.} 2024, arXiv e-prints,
  arXiv:2405.00115

\bibitem[{{Ichikawa} {et~al.}(2013){Ichikawa}, {Matsushita}, {Okabe}, {Sato},
  {Zhang}, {Finoguenov}, {Fujita}, {Fukazawa}, {Kawaharada}, {Nakazawa},
  {Ohashi}, {Ota}, {Takizawa}, {Tamura}, \& {Umetsu}}]{Ichikawa2013}
{Ichikawa}, K., {Matsushita}, K., {Okabe}, N., {et~al.} 2013, \apj, 766, 90

\bibitem[{{Ishisaki} {et~al.}(2007){Ishisaki}, {Maeda}, {Fujimoto}, {Ozaki},
  {Ebisawa}, {Takahashi}, {Ueda}, {Ogasaka}, {Ptak}, {Mukai}, {Hamaguchi},
  {Hirayama}, {Kotani}, {Kubo}, {Shibata}, {Ebara}, {Furuzawa}, {Iizuka},
  {Inoue}, {Mori}, {Okada}, {Yokoyama}, {Matsumoto}, {Nakajima}, {Yamaguchi},
  {Anabuki}, {Tawa}, {Nagai}, {Katsuda}, {Hayashida}, {Bamba}, {Miller},
  {Sato}, \& {Yamasaki}}]{Ishisaki2007}
{Ishisaki}, Y., {Maeda}, Y., {Fujimoto}, R., {et~al.} 2007, \pasj, 59, S113

\bibitem[{{Jetha} {et~al.}(2008){Jetha}, {Hardcastle}, {Babul}, {O'Sullivan},
  {Ponman}, {Raychaudhury}, \& {Vrtilek}}]{NGC741}
{Jetha}, N.~N., {Hardcastle}, M.~J., {Babul}, A., {et~al.} 2008, \mnras, 384,
  1344

\bibitem[{{Kalberla} {et~al.}(2005){Kalberla}, {Burton}, {Hartmann}, {Arnal},
  {Bajaja}, {Morras}, \& {P{\"o}ppel}}]{Kalberla2005}
{Kalberla}, P.~M.~W., {Burton}, W.~B., {Hartmann}, D., {et~al.} 2005, \aap,
  440, 775

\bibitem[{{Kawaharada} {et~al.}(2010){Kawaharada}, {Okabe}, {Umetsu},
  {Takizawa}, {Matsushita}, {Fukazawa}, {Hamana}, {Miyazaki}, {Nakazawa}, \&
  {Ohashi}}]{Kawaharada2010}
{Kawaharada}, M., {Okabe}, N., {Umetsu}, K., {et~al.} 2010, \apj, 714, 423

\bibitem[{{Lagan{\'a}} {et~al.}(2013){Lagan{\'a}}, {Martinet}, {Durret}, {Lima
  Neto}, {Maughan}, \& {Zhang}}]{Lagana2013}
{Lagan{\'a}}, T.~F., {Martinet}, N., {Durret}, F., {et~al.} 2013, \aap, 555,
  A66

\bibitem[{{Lodders}(2003)}]{Lodders2003}
{Lodders}, K. 2003, \apj, 591, 1220

\bibitem[{{Lodders} {et~al.}(2009){Lodders}, {Palme}, \& {Gail}}]{Lodders2009}
{Lodders}, K., {Palme}, H., \& {Gail}, H.~P. 2009, Landolt B{\"o}rnstein, 4B,
  712

\bibitem[{{Matsushita} {et~al.}(2025){Matsushita}, {Sugiyama}, {Ueda}, {Okabe},
  {Fukushima}, {Kobayashi}, {Yamasaki}, \& {Sato}}]{Matsushita2025}
{Matsushita}, K., {Sugiyama}, H., {Ueda}, M., {et~al.} 2025, arXiv e-prints,
  arXiv:2509.12624

\bibitem[{{McCarthy} {et~al.}(2017){McCarthy}, {Schaye}, {Bird}, \& {Le
  Brun}}]{McCarthy2017}
{McCarthy}, I.~G., {Schaye}, J., {Bird}, S., \& {Le Brun}, A. M.~C. 2017,
  \mnras, 465, 2936

\bibitem[{{Mirakhor} \& {Walker}(2020)}]{A2199Suzaku}
{Mirakhor}, M.~S., \& {Walker}, S.~A. 2020, \mnras, 497, 3943

\bibitem[{{Mirakhor} \& {Walker}(2021)}]{Virgoclumps2021}
---. 2021, \mnras, 506, 139

\bibitem[{{Morandi} {et~al.}(2015){Morandi}, {Sun}, {Forman}, \&
  {Jones}}]{Morandi2015}
{Morandi}, A., {Sun}, M., {Forman}, W., \& {Jones}, C. 2015, \mnras, 450, 2261

\bibitem[{{Moretti} {et~al.}(2003){Moretti}, {Campana}, {Lazzati}, \&
  {Tagliaferri}}]{Moretti2003}
{Moretti}, A., {Campana}, S., {Lazzati}, D., \& {Tagliaferri}, G. 2003, \apj,
  588, 696

\bibitem[{{Mori} {et~al.}(2005){Mori}, {Iizuka}, {Shibata}, {Haba}, {Hayakawa},
  {Hayashi}, {Inoue}, {Inoue}, {Ishida}, {Itoh}, {Itoh}, {Kunieda}, {Maeda},
  {Misaki}, {Naitou}, {Okada}, {Shimizu}, \& {Yokoyama}}]{Mori2005}
{Mori}, H., {Iizuka}, R., {Shibata}, R., {et~al.} 2005, \pasj, 57, 245

\bibitem[{{Nagai} {et~al.}(2007){Nagai}, {Kravtsov}, \& {Vikhlinin}}]{NagaiNFW}
{Nagai}, D., {Kravtsov}, A.~V., \& {Vikhlinin}, A. 2007, \apj, 668, 1

\bibitem[{{Nakashima} {et~al.}(2018){Nakashima}, {Inoue}, {Yamasaki}, {Sofue},
  {Kataoka}, \& {Sakai}}]{Nakashima2018}
{Nakashima}, S., {Inoue}, Y., {Yamasaki}, N., {et~al.} 2018, \apj, 862, 34

\bibitem[{{Okabe} {et~al.}(2014){Okabe}, {Futamase}, {Kajisawa}, \&
  {Kuroshima}}]{Okabecoma}
{Okabe}, N., {Futamase}, T., {Kajisawa}, M., \& {Kuroshima}, R. 2014, \apj,
  784, 90

\bibitem[{{Okabe} {et~al.}(2010){Okabe}, {Takada}, {Umetsu}, {Futamase}, \&
  {Smith}}]{Okabe20102}
{Okabe}, N., {Takada}, M., {Umetsu}, K., {Futamase}, T., \& {Smith}, G.~P.
  2010, \pasj, 62, 811

\bibitem[{{Planck Collaboration} {et~al.}(2013){Planck Collaboration}, {Ade},
  {Aghanim}, {Arnaud}, {Ashdown}, {Atrio-Barandela}, {Aumont}, {Baccigalupi},
  {Balbi}, {Banday}, {Barreiro}, {Bartlett}, {Battaner}, {Benabed},
  {Beno{\^\i}t}, {Bernard}, {Bersanelli}, {Bhatia}, {Bikmaev}, {Bobin},
  {B{\"o}hringer}, {Bonaldi}, {Bond}, {Borgani}, {Borrill}, {Bouchet},
  {Bourdin}, {Brown}, {Burenin}, {Burigana}, {Cabella}, {Cardoso}, {Carvalho},
  {Castex}, {Catalano}, {Cay{\'o}n}, {Chamballu}, {Chiang}, {Chon},
  {Christensen}, {Churazov}, {Clements}, {Colafrancesco}, {Colombi}, {Colombo},
  {Comis}, {Coulais}, {Crill}, {Cuttaia}, {Da Silva}, {Dahle}, {Danese},
  {Davis}, {de Bernardis}, {de Gasperis}, {de Zotti}, {Delabrouille},
  {D{\'e}mocl{\`e}s}, {D{\'e}sert}, {Diego}, {Dolag}, {Dole}, {Donzelli},
  {Dor{\'e}}, {D{\"o}rl}, {Douspis}, {Dupac}, {Efstathiou}, {En{\ss}lin},
  {Eriksen}, {Finelli}, {Flores-Cacho}, {Forni}, {Fosalba}, {Frailis},
  {Franceschi}, {Frommert}, {Galeotta}, {Ganga}, {G{\'e}nova-Santos}, {Giard},
  {Giraud-H{\'e}raud}, {Gonz{\'a}lez-Nuevo}, {G{\'o}rski}, {Gregorio},
  {Gruppuso}, {Hansen}, {Harrison}, {Hempel}, {Henrot-Versill{\'e}},
  {Hern{\'a}ndez-Monteagudo}, {Herranz}, {Hildebrandt}, {Hivon}, {Hobson},
  {Holmes}, {Hurier}, {Jaffe}, {Jaffe}, {Jagemann}, {Jones}, {Juvela},
  {Keih{\"a}nen}, {Khamitov}, {Kisner}, {Kneissl}, {Knoche}, {Knox}, {Kunz},
  {Kurki-Suonio}, {Lagache}, {L{\"a}hteenm{\"a}ki}, {Lamarre}, {Lasenby},
  {Lawrence}, {Le Jeune}, {Leonardi}, {Liddle}, {Lilje}, {L{\'o}pez-Caniego},
  {Luzzi}, {Mac{\'\i}as-P{\'e}rez}, {Maino}, {Mandolesi}, {Maris}, {Marleau},
  {Marshall}, {Mart{\'\i}nez-Gonz{\'a}lez}, {Masi}, {Massardi}, {Matarrese},
  {Mazzotta}, {Mei}, {Melchiorri}, {Melin}, {Mendes}, {Mennella}, {Mitra},
  {Miville-Desch{\^e}nes}, {Moneti}, {Montier}, {Morgante}, {Mortlock},
  {Munshi}, {Murphy}, {Naselsky}, {Nati}, {Natoli}, {N{\o}rgaard-Nielsen},
  {Noviello}, {Novikov}, {Novikov}, {Osborne}, {Pajot}, {Paoletti}, {Pasian},
  {Patanchon}, {Perdereau}, {Perotto}, {Perrotta}, {Piacentini}, {Piat},
  {Pierpaoli}, {Piffaretti}, {Plaszczynski}, {Pointecouteau}, {Polenta},
  {Ponthieu}, {Popa}, {Poutanen}, {Pratt}, {Prunet}, {Puget}, {Rachen},
  {Reach}, {Rebolo}, {Reinecke}, {Remazeilles}, {Renault}, {Ricciardi},
  {Riller}, {Ristorcelli}, {Rocha}, {Roman}, {Rosset}, {Rossetti},
  {Rubi{\~n}o-Mart{\'\i}n}, {Rusholme}, {Sandri}, {Savini}, {Scott}, {Smoot},
  {Starck}, {Sudiwala}, {Sunyaev}, {Sutton}, {Suur-Uski}, {Sygnet}, {Tauber},
  \& {Terenzi}}]{Planck2013}
{Planck Collaboration}, {Ade}, P.~A.~R., {Aghanim}, N., {et~al.} 2013, \aap,
  550, A131

\bibitem[{{Planck Collaboration} {et~al.}(2020){Planck Collaboration},
  {Aghanim}, {Akrami}, {Ashdown}, {Aumont}, {Baccigalupi}, {Ballardini},
  {Banday}, {Barreiro}, {Bartolo}, {Basak}, {Battye}, {Benabed}, {Bernard},
  {Bersanelli}, {Bielewicz}, {Bock}, {Bond}, {Borrill}, {Bouchet}, {Boulanger},
  {Bucher}, {Burigana}, {Butler}, {Calabrese}, {Cardoso}, {Carron},
  {Challinor}, {Chiang}, {Chluba}, {Colombo}, {Combet}, {Contreras}, {Crill},
  {Cuttaia}, {de Bernardis}, {de Zotti}, {Delabrouille}, {Delouis}, {Di
  Valentino}, {Diego}, {Dor{\'e}}, {Douspis}, {Ducout}, {Dupac}, {Dusini},
  {Efstathiou}, {Elsner}, {En{\ss}lin}, {Eriksen}, {Fantaye}, {Farhang},
  {Fergusson}, {Fernandez-Cobos}, {Finelli}, {Forastieri}, {Frailis},
  {Fraisse}, {Franceschi}, {Frolov}, {Galeotta}, {Galli}, {Ganga},
  {G{\'e}nova-Santos}, {Gerbino}, {Ghosh}, {Gonz{\'a}lez-Nuevo}, {G{\'o}rski},
  {Gratton}, {Gruppuso}, {Gudmundsson}, {Hamann}, {Handley}, {Hansen},
  {Herranz}, {Hildebrandt}, {Hivon}, {Huang}, {Jaffe}, {Jones}, {Karakci},
  {Keih{\"a}nen}, {Keskitalo}, {Kiiveri}, {Kim}, {Kisner}, {Knox},
  {Krachmalnicoff}, {Kunz}, {Kurki-Suonio}, {Lagache}, {Lamarre}, {Lasenby},
  {Lattanzi}, {Lawrence}, {Le Jeune}, {Lemos}, {Lesgourgues}, {Levrier},
  {Lewis}, {Liguori}, {Lilje}, {Lilley}, {Lindholm}, {L{\'o}pez-Caniego},
  {Lubin}, {Ma}, {Mac{\'\i}as-P{\'e}rez}, {Maggio}, {Maino}, {Mandolesi},
  {Mangilli}, {Marcos-Caballero}, {Maris}, {Martin}, {Martinelli},
  {Mart{\'\i}nez-Gonz{\'a}lez}, {Matarrese}, {Mauri}, {McEwen}, {Meinhold},
  {Melchiorri}, {Mennella}, {Migliaccio}, {Millea}, {Mitra},
  {Miville-Desch{\^e}nes}, {Molinari}, {Montier}, {Morgante}, {Moss}, {Natoli},
  {N{\o}rgaard-Nielsen}, {Pagano}, {Paoletti}, {Partridge}, {Patanchon},
  {Peiris}, {Perrotta}, {Pettorino}, {Piacentini}, {Polastri}, {Polenta},
  {Puget}, {Rachen}, {Reinecke}, {Remazeilles}, {Renzi}, {Rocha}, {Rosset},
  {Roudier}, {Rubi{\~n}o-Mart{\'\i}n}, {Ruiz-Granados}, {Salvati}, {Sandri},
  {Savelainen}, {Scott}, {Shellard}, {Sirignano}, {Sirri}, {Spencer},
  {Sunyaev}, {Suur-Uski}, {Tauber}, {Tavagnacco}, {Tenti}, {Toffolatti},
  {Tomasi}, {Trombetti}, {Valenziano}, {Valiviita}, {Van Tent}, {Vibert},
  {Vielva}, {Villa}, {Vittorio}, {Wandelt}, {Wehus}, {White}, {White},
  {Zacchei}, \& {Zonca}}]{Planck18}
{Planck Collaboration}, {Aghanim}, N., {Akrami}, Y., {et~al.} 2020, \aap, 641,
  A6

\bibitem[{{Ponman} {et~al.}(1999){Ponman}, {Cannon}, \& {Navarro}}]{Ponman1999}
{Ponman}, T.~J., {Cannon}, D.~B., \& {Navarro}, J.~F. 1999, \nat, 397, 135

\bibitem[{{Predehl} {et~al.}(2020){Predehl}, {Sunyaev}, {Becker}, {Brunner},
  {Burenin}, {Bykov}, {Cherepashchuk}, {Chugai}, {Churazov}, {Doroshenko},
  {Eismont}, {Freyberg}, {Gilfanov}, {Haberl}, {Khabibullin}, {Krivonos},
  {Maitra}, {Medvedev}, {Merloni}, {Nandra}, {Nazarov}, {Pavlinsky}, {Ponti},
  {Sanders}, {Sasaki}, {Sazonov}, {Strong}, \& {Wilms}}]{Predehl2020}
{Predehl}, P., {Sunyaev}, R.~A., {Becker}, W., {et~al.} 2020, \nat, 588, 227

\bibitem[{{Sarkar} {et~al.}(2021){Sarkar}, {Su}, {Randall}, {Gastaldello},
  {Trierweiler}, {White}, {Kraft}, \& {Miller}}]{MKW4}
{Sarkar}, A., {Su}, Y., {Randall}, S., {et~al.} 2021, \mnras, 501, 3767

\bibitem[{{Sasaki} {et~al.}(2015){Sasaki}, {Matsushita}, {Sato}, \&
  {Okabe}}]{Sasaki2015}
{Sasaki}, T., {Matsushita}, K., {Sato}, K., \& {Okabe}, N. 2015, \apj, 806, 123

\bibitem[{{Sekiya} {et~al.}(2014){Sekiya}, {Yamasaki}, {Mitsuda}, \&
  {Takei}}]{Sekiya2014}
{Sekiya}, N., {Yamasaki}, N.~Y., {Mitsuda}, K., \& {Takei}, Y. 2014, \pasj, 66,
  L3

\bibitem[{{Serlemitsos} {et~al.}(2007){Serlemitsos}, {Soong}, {Chan},
  {Okajima}, {Lehan}, {Maeda}, {Itoh}, {Mori}, {Iizuka}, {Itoh}, {Inoue},
  {Okada}, {Yokoyama}, {Itoh}, {Ebara}, {Nakamura}, {Suzuki}, {Ishida},
  {Hayakawa}, {Inoue}, {Okuma}, {Kubota}, {Suzuki}, {Osawa}, {Yamashita},
  {Kunieda}, {Tawara}, {Ogasaka}, {Furuzawa}, {Tamura}, {Shibata}, {Haba},
  {Naitou}, \& {Misaki}}]{SuzakuXRT}
{Serlemitsos}, P.~J., {Soong}, Y., {Chan}, K.-W., {et~al.} 2007, \pasj, 59, S9

\bibitem[{{Simionescu} {et~al.}(2017){Simionescu}, {Werner}, {Mantz}, {Allen},
  \& {Urban}}]{Simionescu2017}
{Simionescu}, A., {Werner}, N., {Mantz}, A., {Allen}, S.~W., \& {Urban}, O.
  2017, \mnras, 469, 1476

\bibitem[{{Simionescu} {et~al.}(2015){Simionescu}, {Werner}, {Urban}, {Allen},
  {Ichinohe}, \& {Zhuravleva}}]{Simionescu2015}
{Simionescu}, A., {Werner}, N., {Urban}, O., {et~al.} 2015, \apjl, 811, L25

\bibitem[{{Simionescu} {et~al.}(2011){Simionescu}, {Allen}, {Mantz}, {Werner},
  {Takei}, {Morris}, {Fabian}, {Sanders}, {Nulsen}, {George}, \&
  {Taylor}}]{Simionescu2011}
{Simionescu}, A., {Allen}, S.~W., {Mantz}, A., {et~al.} 2011, Science, 331,
  1576

\bibitem[{{Simionescu} {et~al.}(2013){Simionescu}, {Werner}, {Urban}, {Allen},
  {Fabian}, {Mantz}, {Matsushita}, {Nulsen}, {Sanders}, {Sasaki}, {Sato},
  {Takei}, \& {Walker}}]{Simionescu2013}
{Simionescu}, A., {Werner}, N., {Urban}, O., {et~al.} 2013, \apj, 775, 4

\bibitem[{{Smith} {et~al.}(2001){Smith}, {Brickhouse}, {Liedahl}, \&
  {Raymond}}]{Smith2001}
{Smith}, R.~K., {Brickhouse}, N.~S., {Liedahl}, D.~A., \& {Raymond}, J.~C.
  2001, \apjl, 556, L91

\bibitem[{{Snowden} {et~al.}(1995){Snowden}, {Burrows}, {Sanders},
  {Aschenbach}, \& {Pfeffermann}}]{Snowden1995}
{Snowden}, S.~L., {Burrows}, D.~N., {Sanders}, W.~T., {Aschenbach}, B., \&
  {Pfeffermann}, E. 1995, \apj, 439, 399

\bibitem[{{Snowden} {et~al.}(2008){Snowden}, {Mushotzky}, {Kuntz}, \&
  {Davis}}]{Snowden2008}
{Snowden}, S.~L., {Mushotzky}, R.~F., {Kuntz}, K.~D., \& {Davis}, D.~S. 2008,
  \aap, 478, 615

\bibitem[{{Sugiyama} {et~al.}(2023){Sugiyama}, {Ueda}, {Fukushima},
  {Kobayashi}, {Yamasaki}, {Sato}, \& {Matsushita}}]{Sugiyama2023}
{Sugiyama}, H., {Ueda}, M., {Fukushima}, K., {et~al.} 2023, \pasj, 75, 1324

\bibitem[{{Sun}(2012)}]{Sun2012}
{Sun}, M. 2012, New Journal of Physics, 14, 045004

\bibitem[{{Sun} {et~al.}(2009){Sun}, {Voit}, {Donahue}, {Jones}, {Forman}, \&
  {Vikhlinin}}]{Sun2009}
{Sun}, M., {Voit}, G.~M., {Donahue}, M., {et~al.} 2009, \apj, 693, 1142

\bibitem[{{Takei} {et~al.}(2012){Takei}, {Akamatsu}, {Hiyama}, {Maeda},
  {Ishida}, {Mori}, {Ishisaki}, \& {Hoshino}}]{Takei2012}
{Takei}, Y., {Akamatsu}, H., {Hiyama}, Y., {et~al.} 2012, in American Institute
  of Physics Conference Series, Vol. 1427, Suzaku 2011: Exploring the X-ray
  Universe: Suzaku and Beyond, ed. R.~{Petre}, K.~{Mitsuda}, \& L.~{Angelini},
  239--240

\bibitem[{{Tawa} {et~al.}(2008){Tawa}, {Hayashida}, {Nagai}, {Nakamoto},
  {Tsunemi}, {Yamaguchi}, {Ishisaki}, {Miller}, {Mizuno}, {Dotani}, {Ozaki}, \&
  {Katayama}}]{Tawa2008}
{Tawa}, N., {Hayashida}, K., {Nagai}, M., {et~al.} 2008, \pasj, 60, S11

\bibitem[{{Th{\"o}lken} {et~al.}(2016){Th{\"o}lken}, {Lovisari}, {Reiprich}, \&
  {Hasenbusch}}]{UGC}
{Th{\"o}lken}, S., {Lovisari}, L., {Reiprich}, T.~H., \& {Hasenbusch}, J. 2016,
  \aap, 592, A37

\bibitem[{{Uchida} {et~al.}(2016){Uchida}, {Simionescu}, {Takahashi}, {Werner},
  {Ichinohe}, {Allen}, {Urban}, \& {Matsushita}}]{Uchida2016}
{Uchida}, Y., {Simionescu}, A., {Takahashi}, T., {et~al.} 2016, \pasj, 68, S20

\bibitem[{{Ueda} {et~al.}(2022){Ueda}, {Sugiyama}, {Kobayashi}, {Fukushima},
  {Yamasaki}, {Sato}, \& {Matsushita}}]{Ueda2022}
{Ueda}, M., {Sugiyama}, H., {Kobayashi}, S.~B., {et~al.} 2022, \pasj, 74, 1396

\bibitem[{{Urban} {et~al.}(2017){Urban}, {Werner}, {Allen}, {Simionescu}, \&
  {Mantz}}]{Urban2017}
{Urban}, O., {Werner}, N., {Allen}, S.~W., {Simionescu}, A., \& {Mantz}, A.
  2017, \mnras, 470, 4583

\bibitem[{{Urban} {et~al.}(2014){Urban}, {Simionescu}, {Werner}, {Allen},
  {Ehlert}, {Zhuravleva}, {Morris}, {Fabian}, {Mantz}, {Nulsen}, {Sanders}, \&
  {Takei}}]{Urban2014}
{Urban}, O., {Simionescu}, A., {Werner}, N., {et~al.} 2014, \mnras, 437, 3939

\bibitem[{{Velliscig} {et~al.}(2014){Velliscig}, {van Daalen}, {Schaye},
  {McCarthy}, {Cacciato}, {Le Brun}, \& {Dalla Vecchia}}]{OWLS2014}
{Velliscig}, M., {van Daalen}, M.~P., {Schaye}, J., {et~al.} 2014, \mnras, 442,
  2641

\bibitem[{{Vikhlinin} {et~al.}(2006){Vikhlinin}, {Kravtsov}, {Forman}, {Jones},
  {Markevitch}, {Murray}, \& {Van Speybroeck}}]{Vikhlinin2006}
{Vikhlinin}, A., {Kravtsov}, A., {Forman}, W., {et~al.} 2006, \apj, 640, 691

\bibitem[{{Vikhlinin} {et~al.}(2009){Vikhlinin}, {Burenin}, {Ebeling},
  {Forman}, {Hornstrup}, {Jones}, {Kravtsov}, {Murray}, {Nagai}, {Quintana}, \&
  {Voevodkin}}]{Vikhlinin2009}
{Vikhlinin}, A., {Burenin}, R.~A., {Ebeling}, H., {et~al.} 2009, \apj, 692,
  1033

\bibitem[{{Voit}(2005)}]{Voit2005}
{Voit}, G.~M. 2005, Reviews of Modern Physics, 77, 207

\bibitem[{{Walker} {et~al.}(2022){Walker}, {Mirakhor}, {ZuHone}, {Sanders},
  {Fabian}, \& {Diwanji}}]{Walker2022}
{Walker}, S.~A., {Mirakhor}, M.~S., {ZuHone}, J., {et~al.} 2022, \apj, 929, 37

\bibitem[{{White} {et~al.}(1993){White}, {Navarro}, {Evrard}, \&
  {Frenk}}]{White1993}
{White}, S. D.~M., {Navarro}, J.~F., {Evrard}, A.~E., \& {Frenk}, C.~S. 1993,
  \nat, 366, 429

\bibitem[{{Wong} {et~al.}(2016){Wong}, {Irwin}, {Wik}, {Sun}, {Sarazin},
  {Fujita}, \& {Reiprich}}]{Antlia}
{Wong}, K.-W., {Irwin}, J.~A., {Wik}, D.~R., {et~al.} 2016, \apj, 829, 49

\bibitem[{{XRISM Collaboration} {et~al.}(2025){XRISM Collaboration}, {Audard},
  {Awaki}, {Ballhausen}, {Bamba}, {Behar}, {Boissay-Malaquin}, {Brenneman},
  {Brown}, {Corrales}, {Costantini}, {Cumbee}, {Diaz Trigo}, {Done}, {Dotani},
  {Ebisawa}, {Eckart}, {Eckert}, {Eguchi}, {Enoto}, {Ezoe}, {Foster},
  {Fujimoto}, {Fujita}, {Fukazawa}, {Fukushima}, {Furuzawa}, {Gallo}, {Garcia},
  {Gu}, {Guainazzi}, {Hagino}, {Hamaguchi}, {Hatsukade}, {Hayashi}, {Hayashi},
  {Hell}, {Hodges-Kluck}, {Hornschemeier}, {Ichinohe}, {Ishi}, {Ishida},
  {Ishikawa}, {Ishisaki}, {Kaastra}, {Kallman}, {Kara}, {Katsuda}, {Kanemaru},
  {Kelley}, {Kilbourne}, {Kitamoto}, {Kobayashi}, {Kohmura}, {Kubota},
  {Leutenegger}, {Loewenstein}, {Maeda}, {Markevitch}, {Matsumoto},
  {Matsushita}, {McCammon}, {McNamara}, {Mernier}, {Miller}, {Miller},
  {Mitsuishi}, {Mizumoto}, {Mizuno}, {Mori}, {Mukai}, {Murakami}, {Mushotzky},
  {Nakajima}, {Nakazawa}, {Ness}, {Nobukawa}, {Nobukawa}, {Noda}, {Odaka},
  {Ogawa}, {Ogorzalek}, {Okajima}, {Ota}, {Paltani}, {Petre}, {Plucinsky},
  {Porter}, {Pottschmidt}, {Sato}, {Sato}, {Sawada}, {Seta}, {Shidatsu},
  {Simionescu}, {Smith}, {Suzuki}, {Szymkowiak}, {Takahashi}, {Takeo},
  {Tamagawa}, {Tamura}, {Tanaka}, {Tanimoto}, {Tashiro}, {Terada}, {Terashima},
  {Tsuboi}, {Tsujimoto}, {Tsunemi}, {Tsuru}, {Tumer}, {Uchida}, {Uchida},
  {Uchida}, {Uchiyama}, {Ueda}, {Uno}, {Vink}, {Watanabe}, {Williams},
  {Yamada}, {Yamada}, {Yamaguchi}, {Yamaoka}, {Yamasaki}, {Yamauchi},
  {Yamauchi}, {Yaqoob}, {Yoneyama}, {Yoshida}, {Yukita}, {Zhuravleva},
  {Bellomi}, {Drury}, {Heinrich}, {Hlavacek-Larrondo}, {Meunier}, {Migkas},
  {Shefler}, {Stancil}, {Truong}, {Ueda}, {Vigneron}, {Zhang}, \&
  {ZuHone}}]{XRISMPerseus}
{XRISM Collaboration}, {Audard}, M., {Awaki}, H., {et~al.} 2025, arXiv
  e-prints, arXiv:2509.04421

\bibitem[{{Xrism Collaboration} {et~al.}(2025{\natexlab{a}}){Xrism
  Collaboration}, {Audard}, {Awaki}, {Ballhausen}, {Bamba}, {Behar},
  {Boissay-Malaquin}, {Brenneman}, {Brown}, {Corrales}, {Costantini}, {Cumbee},
  {Done}, {Dotani}, {Ebisawa}, {Eckart}, {Eckert}, {Enoto}, {Eguchi}, {Ezoe},
  {Foster}, {Fujimoto}, {Fujita}, {Fukazawa}, {Fukushima}, {Furuzawa}, {Gallo},
  {Garc{\'\i}a}, {Gu}, {Guainazzi}, {Hagino}, {Hamaguchi}, {Hatsukade},
  {Hayashi}, {Hayashi}, {Hell}, {Hodges-Kluck}, {Hornschemeier}, {Ichinohe},
  {Ishida}, {Ishikawa}, {Ishisaki}, {Kaastra}, {Kallman}, {Kara}, {Katsuda},
  {Kanemaru}, {Kelley}, {Kilbourne}, {Kitamoto}, {Kobayashi}, {Kohmura},
  {Kubota}, {Leutenegger}, {Loewenstein}, {Maeda}, {Markevitch}, {Matsumoto},
  {Matsushita}, {McCammon}, {McNamara}, {Mernier}, {Miller}, {Miller},
  {Mitsuishi}, {Mizumoto}, {Mizuno}, {Mori}, {Mukai}, {Murakami}, {Mushotzky},
  {Nakajima}, {Nakazawa}, {Ness}, {Nobukawa}, {Nobukawa}, {Noda}, {Odaka},
  {Ogawa}, {Ogorzalek}, {Okajima}, {Ota}, {Paltani}, {Petre}, {Plucinsky},
  {Porter}, {Pottschmidt}, {Sato}, {Sato}, {Sawada}, {Seta}, {Shidatsu},
  {Simionescu}, {Smith}, {Suzuki}, {Szymkowiak}, {Takahashi}, {Takeo},
  {Tamagawa}, {Tamura}, {Tanaka}, {Tanimoto}, {Tashiro}, {Terada}, {Terashima},
  {Trigo}, {Tsuboi}, {Tsujimoto}, {Tsunemi}, {Tsuru}, {Uchida}, {Uchida},
  {Uchida}, {Uchiyama}, {Ueda}, {Uno}, {Vink}, {Watanabe}, {Williams},
  {Yamada}, {Yamada}, {Yamaguchi}, {Yamaoka}, {Yamasaki}, {Yamauchi},
  {Yamauchi}, {Yaqoob}, {Yoneyama}, {Yoshida}, {Yukita}, {Zhuravleva}, {Kondo},
  {Werner}, {Pl{\v{s}}ek}, {Sun}, {Hosogi}, \& {Majumder}}]{XRISMCentaurus}
{Xrism Collaboration}, {Audard}, M., {Awaki}, H., {et~al.} 2025{\natexlab{a}},
  \nat, 638, 365

\bibitem[{{Xrism Collaboration} {et~al.}(2025{\natexlab{b}}){Xrism
  Collaboration}, {Audard}, {Awaki}, {Ballhausen}, {Bamba}, {Behar},
  {Boissay-Malaquin}, {Brenneman}, {Brown}, {Corrales}, {Costantini}, {Cumbee},
  {Diaz Trigo}, {Done}, {Dotani}, {Ebisawa}, {Eckart}, {Eckert}, {Eguchi},
  {Enoto}, {Ezoe}, {Foster}, {Fujimoto}, {Fujita}, {Fukazawa}, {Fukushima},
  {Furuzawa}, {Gallo}, {Garc{\'\i}a}, {Gu}, {Guainazzi}, {Hagino}, {Hamaguchi},
  {Hatsukade}, {Hayashi}, {Hayashi}, {Hell}, {Hodges-Kluck}, {Hornschemeier},
  {Ichinohe}, {Ishida}, {Ishikawa}, {Ishisaki}, {Kaastra}, {Kallman}, {Kara},
  {Katsuda}, {Kanemaru}, {Kelley}, {Kilbourne}, {Kitamoto}, {Kobayashi},
  {Kohmura}, {Kubota}, {Leutenegger}, {Loewenstein}, {Maeda}, {Markevitch},
  {Matsumoto}, {Matsushita}, {McCammon}, {McNamara}, {Mernier}, {Miller},
  {Miller}, {Mitsuishi}, {Mizumoto}, {Mizuno}, {Mori}, {Mukai}, {Murakami},
  {Mushotzky}, {Nakajima}, {Nakazawa}, {Ness}, {Nobukawa}, {Nobukawa}, {Noda},
  {Odaka}, {Ogawa}, {Ogorzalek}, {Okajima}, {Ota}, {Paltani}, {Petre},
  {Plucinsky}, {Porter}, {Pottschmidt}, {Sato}, {Sato}, {Sawada}, {Seta},
  {Shidatsu}, {Simionescu}, {Smith}, {Suzuki}, {Szymkowiak}, {Takahashi},
  {Takeo}, {Tamagawa}, {Tamura}, {Tanaka}, {Tanimoto}, {Tashiro}, {Terada},
  {Terashima}, {Tsuboi}, {Tsujimoto}, {Tsunemi}, {Tsuru}, {Uchida}, {Uchida},
  {Uchida}, {Uchiyama}, {Ueda}, {Uno}, {Vink}, {Watanabe}, {Williams},
  {Yamada}, {Yamada}, {Yamaguchi}, {Yamaoka}, {Yamasaki}, {Yamauchi},
  {Yamauchi}, {Yaqoob}, {Yoneyama}, {Yoshida}, {Yukita}, {Zhuravleva},
  {Bartalesi}, {Ettori}, {Kosarzycki}, {Lovisari}, {Rose}, {Sarkar}, {Sun}, \&
  {Tamhane}}]{XRISMA2029}
---. 2025{\natexlab{b}}, \apjl, 982, L5

\bibitem[{{Yoshino} {et~al.}(2009){Yoshino}, {Mitsuda}, {Yamasaki}, {Takei},
  {Hagihara}, {Masui}, {Bauer}, {McCammon}, {Fujimoto}, {Wang}, \&
  {Yao}}]{Yoshino2009}
{Yoshino}, T., {Mitsuda}, K., {Yamasaki}, N.~Y., {et~al.} 2009, \pasj, 61, 805

\bibitem[{{Zhu} {et~al.}(2023){Zhu}, {Kov{\'a}cs}, {Simionescu}, \&
  {Werner}}]{A133Suzaku}
{Zhu}, Z., {Kov{\'a}cs}, O.~E., {Simionescu}, A., \& {Werner}, N. 2023, \aap,
  678, A122

\end{thebibliography}

\end{document}